\documentclass[%
 reprint,
nofootinbib,
 amsmath,amssymb,
 aps,
onecolumn
]{revtex4-2}

\usepackage{inputenc}[utf8]
\usepackage{fontenc}[T1]
\usepackage{caption}
\usepackage{subcaption}
\usepackage[normalem]{ulem}
\usepackage{tabularx}
\usepackage{aas_macros}
\usepackage{amssymb}
\usepackage{graphicx}
\usepackage{xcolor}
\usepackage{multirow}
\usepackage{hyperref}

\begin{document}

\title{The Galactic Center as a laboratory for theories of gravity and dark matter}

\author{Mariafelicia de Laurentis}
\affiliation{
Dipartimento di Fisica, Universit\'a
di Napoli {}``Federico II'', Compl. Univ. di
Monte S. Angelo, Edificio G, Via Cinthia, I-80126, Napoli, Italy
}%
\affiliation{
INFN Sezione  di Napoli, Compl. Univ. di
Monte S. Angelo, Edificio G, Via Cinthia, I-80126, Napoli, Italy
}%

\author{Ivan de Martino}%
\author{Riccardo Della Monica}
\email{rdellamonica@usal.es}
\affiliation{Universidad de Salamanca, Departamento de Fisica Fundamental, P. de la Merced, Salamanca, ES}%

\begin{abstract}
    The Galactic Center of the Milky Way, thanks to its proximity, allows to perform astronomical observations that investigate physical phenomena at the edge of astrophysics and fundamental physics. As such, it offers a unique laboratory to probe gravity, where one can not only test the basic predictions of General Relativity, but is also able to falsify theories that, over time, have been proposed to modify or extend General Relativity; to test different paradigms of dark matter; and to place constraints on putative models that have been formulated as alternatives to the standard black hole paradigm in General Relativity. In this review we provide a general overview of the history of observations of the Galactic Center, emphasizing the importance, in particular on the smallest-observable scales, that they had in opening a new avenue to improve our understanding of the underlying theory of gravity in the surrounding of a supermassive compact object.
\end{abstract}

\maketitle

\tableofcontents

\section{Introduction}
\markboth{}{}

The central region of the Milky Way (MW), more commonly referred to as the Galactic Center (GC), is a challenging, yet very popular, target for astronomical observations \cite{Genzel2022}. The great technological efforts required for the development of observational facilities and techniques capable of penetrating the complex environment of absorbing and obscuring features in the GC  (and more in general in the Galactic plane, across which we perform observations) is motivated by one clear, ambitious, scientific goal: advance our understanding of physical processes at the edge of astrophysics and fundamental physics. Indeed, the GC represents a unique laboratory where measurements on very small scales, {\em i.e.} down to the size of the event horizon of the central supermassive black hole (SMBH), can be linked to the enormous wealth of information that we possess, across the electromagnetic spectrum, on the entire Galaxy itself. Its uniqueness resides in the fact that we can take advantage of its proximity ($\sim\,8$ kpc) to perform high-resolution observations of the MW nucleus, that provide much more details and information than what could be possible for any other galaxy's center. Such observations allow, on one hand, to test \emph{just around the corner} astrophysical models that have been formulated to explain observational evidences in extra-galactic scenarios, of the black hole (BH) paradigm for the active galactic nuclei, the scaling-relation between the SMBH mass and the galactic bulge velocity dispersion, or, more in general, the phenomenon of radiation emission by accretion onto a SMBH \cite{Ferrarese2005}. On the other hand, it gives the possibility to test predictions from General Relativity (GR) in a regime where it had never been tested before, opening \emph{de facto} a new era for experimental gravitation. The classical tests of GR (\emph{e.g.} the perihelion precession of Mercury's orbit, the gravitational redshift of light and the deflection of light by the Sun) whose experimental verification in the Solar System have historically allowed to test the validity of GR itself \cite{Will2014}, find, in the GC, a new test bench. The S-stars, orbiting Sagittarius A* (Sgr A*), allow to test the first two effects, the periapsis precession and the gravitational redshift, both predicted by GR, around a supermassive compact object, while Very Long Baseline Interferometry (VLBI) observations of Sgr A* allow to probe, around the same object, the deflection of light in a strong-field regime.

Past review articles on the subject have covered in great detail different experimental and theoretical aspects of the GC, offering, in their entirety, a broad overview of our current knowledge of this environment and of the gradual five-decade-long process of advancement of our understanding of the different physical regimes therein. For example, in the reviews by R. Genzel \cite{Genzel2010, Genzel2022} an overview of the wealth of experimental evidences that have allowed to identify Sgr A* as a SMBH is presented, along with a detailed characterization of the Nuclear Star Cluster (NSC) and its Paradox of Youth. The latter subject is analyzed in greater detail in the review article by M. Mapelli and A. Gualandris \cite{Mapelli2016}, where an overview of the several models of formation of the NSC that have been proposed is offered. On the other hand, the reviews by T. Alexander \cite{Alexander2005, Alexander2017} focus on the impact of the presence of a SMBH on the stellar processes in the surrounding environment, \emph{e.g.} the star formation. Finally, the recently published review by A. Bryant and A. Krabbe \cite{Bryant2021} provides with a detailed summary of the episodic and multi-scale phenomena occurring at the GC.

Differently from previous works, this review has the aim of providing a general overview of the history of observations of the GC, focusing in particular on the smallest-observable scales, \emph{i.e.} the astrometric tracking of the S-stars orbits, the event-horizon-scale radio-observation of the SMBH and, more specifically, on the impact that such observations have had in our understanding of the underlying theory of gravity in the surrounding of a massive compact object. To this purpose, this review is organized as follows: in Section \ref{sec:gc} we present an overview of the astronomical observations of the GC, with a focus on the NSC (Section \ref{sec:NSC}), on the S-stars (Section \ref{sec:s_stars}) and on the compact-radio source Sgr A* (Section \ref{sec:sgr_a}) and its SMBH nature; Section \ref{sec:gravity_tests} is devoted to the new avenue opened by observations of Sgr A* to perform tests of GR (Section \ref{sec:test_GR}), of the BH paradigm (Section \ref{sec:nature_Sgr}), on alternative theories of gravity (Section \ref{sec:alternative_theories}) and on the the different models of dark matter that can be tested at the GC (Section \ref{sec:darkmatter}). Finally in Section \ref{sec:conclusions} we report our final remarks.

\section{The Galactic Center: a populated environment}
\label{sec:gc}

The central few hundred parsecs (from here on pc) of the MW are indeed a populated environment (for a more comprehensive view we refer to \cite{Bryant2021} and references therein), characterized by a great variety of phenomena in the Interstellar Medium, embracing different stages of its evolution, is rich in molecular gas that accounts for $\sim5-10\%$ of the content of the entire galaxy and of its infrared luminosity (for a comprehensive review we refer to \cite{Mills2017}). The high density ($\sim 10^7 M_\odot$ pc$^{-3}$) such as supernova remnants \cite{Green2019}, compact star-forming regions, large populations of radio filaments \cite{YusefZadeh1987, Gray1991, Lang1999, Thomas2020} and radio arcs \cite{YusefZadeh1984, Pare2019, Heywood2019}. The Central Molecular Zone, a twisted torus-like structure, extending up to $\sim 580$ pc from the GC, that can be observed in the far-infrared \cite{Molinari2011}, \cite{Martin2004, Stark2004, Mills2018} of this region leads to increased gas temperature and pressure with respect to the surrounding environment \cite{Guesten1985, Huettemeister1993, Ginsburg2016, Krieger2017}, creating one of the most extreme and turbulent \cite{Bally1987, Shetty2012, Kauffmann2017} regions for star formation that can be directly observed, providing with a unique opportunity to study this vast range of energetic astrophysical phenomena.

\begin{figure}
    \centering
    \includegraphics[width = \textwidth]{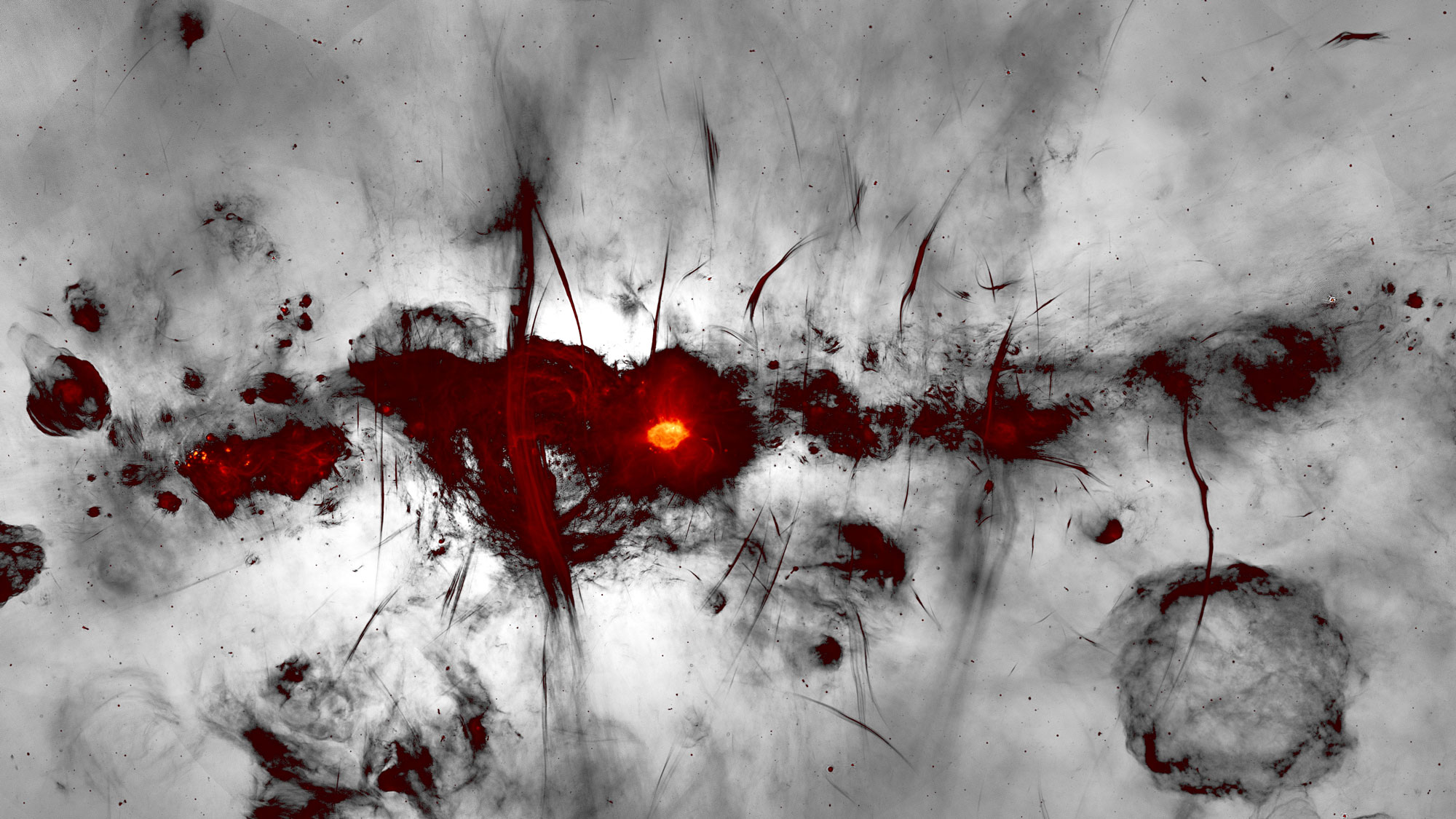}
    \label{fig:meerkat_gc}
    \caption{A recent mosaic view of the GC in radio wavelengths from South African MeerKAT radio telescope, covering 6.5 square degrees of the GC region at an angular resolution of 4 as (in physical units, the image covers a region that is $\sim300\times150$ pc). Radio observations are not affected by dust extinction and thus expose a great range of physical phenomena happening in the GC: supernova remnants \cite{Green2019}, compact star-forming regions, and a large population of mysterious radio filaments \cite{YusefZadeh1987, Gray1991, Lang1999, Thomas2020}. The prominent vertical Radio Arc corresponds to the inner part of the eastern boundary of the 430 pc radio bubbles spanning the Galactic center \cite{YusefZadeh1984, Pare2019, Heywood2019}. The images has been taken from \cite{Heywood2022}.}
\end{figure}

\begin{figure}[!t]
    \centering
    \includegraphics[width = \textwidth]{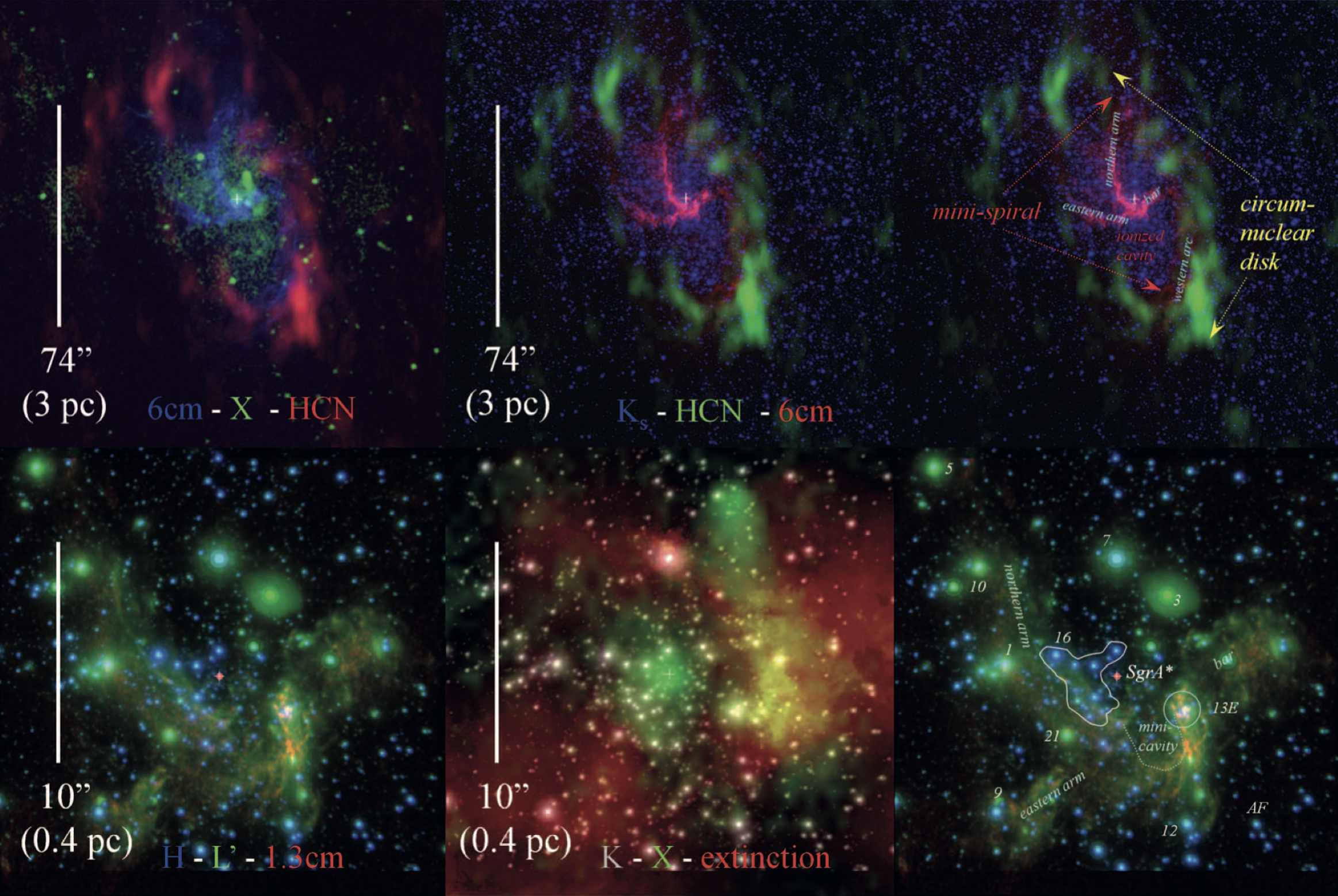}
    \caption{Multi-wavelength view of the central few pc of the GC ($\sim 3$ pc field-of-view for the top row and $\sim 0.4$ pc for the bottom one).  The cross shows the position of Sgr A* on the sky, while the galactic plane crosses the image with a position angle of $32^\circ$ in the direction SW-NE (in the images N is up and E is left). The various panels depict color-composite images from the spectral bands indicated on each box. In particular, the 6 cm and 1.3 cm observations are the radio continuum from the Very Large Array (VLA, \cite{YusefZadeh1986, Roberts1993, Zhao1998}), the HCN emission line is from Owens Valley Radio Observatory (OVRO, \cite{Christopher2005}), X-ray emission has been observed with the Chandra satellite \cite{Baganoff2003}, $K$ stands for the 2.2 $\mu$m K-band near-infrared observation at the Very Large Telescope (VLT with the instrument ISAAC, \cite{Schodel2007b}, for the top-middle panel and with the NACO adaptive optic camera \cite{Genzel2003}), $L'$ and $H$ are the 3.8 $\mu$m and 1.6 $\mu$m bands from VLT-NACO \cite{Genzel2003}. The two right-most panels are copies of the top-central and bottom-left panels respectively, where the main interstellar features have been highlighted. The image is courtesy of \cite{Genzel2010}}
    \label{fig:gc_multiband}
\end{figure}

Diving deeper into this peculiar environment, within the central few pc of the MW a range of different physical phenomena can be observed across the whole electromagnetic spectrum. The GC harbours several clusters of young and massive stars \cite{Bryant2021}. Among these, one finds the Arches \cite{Cotera1996, Espinoza2009}, the Quintuplet \cite{Nagata1990, Liermann2009} and, most importantly, the NSC within the central $\sim300$ pc around the compact radio source Sgr A* \cite{Krabbe1991, Maness2007, Lu2009, Lu2013}. We will deeply discuss the NSC, the problems to its formation and to a very peculiar group of stars (the S-stars) that has been observed therein, in Section \ref{sec:NSC}. 
Along with these star clusters, both neutral and extremely hot ionized gas have been observed \cite{Becklin1968, Genzel1987, Rieke1988, Genzel1994, Mezger1996} in the same region (see Figure \ref{fig:gc_multiband}). 
Moreover, the central pc (where most of the gas is ionized), exhibits a prominent HII region (Sgr A West) along with a bright X-ray emitting region composed of hot ($T\sim 10^6$ K) gas \cite{Baganoff2001, Baganoff2003, Muno2004}.  
 
High resolution observations of the GC, hence, offer the possibility to study the innermost regions of galaxies with much finer details and precision than is possible for other galaxies (our own GC is closer by a factor 10'000 with respect to that of the closest external galaxy \cite{Genzel2010}). This does not only allow to test the SMBH paradigm in our own Galaxy and study the impact of massive black holes (MBHs) on stellar and interstellar environments, but it also gives us the opportunity to probe gravity in a regime in which it had never been tested before, with unprecedented precision (see for example Figure 1 in \cite{Hees2017} or Figure 23 in \cite{EventHorizonTelescopeCollaboration2022f}). However, since we observe the GC from within the galactic plane, observations are challenged by the presence of interstellar dust particles in the plane of the disk, which obscure the galactic nucleus in visible wavelengths. Much of what we know today about the GC comes from observations at different wavelengths, from the infrared and microwave, to high energetic processes captured by X-ray and $\gamma$-ray facilities (see Figure \ref{fig:gc_multiband} for a multi-wavelength view of the central few pc of the GC).

\subsection{The Nuclear Star Cluster}
\label{sec:NSC}
Among the three stellar clusters that have been identified in the GC, the ‘‘NSC'' is probably the most interesting one: located right in the hearth of the MW, it lies within a few pc of Sgr A*. Its morphology and the presence of a population of very young stars so close to a SMBH pose several questions about its formation mechanism, that still remain unanswered \cite{Bryant2021}.  Early infrared observations \cite{Becklin1975, Becklin1978} of the NSC first revealed the presence of a surprisingly high number of bright stars. Many of these stars ($96\%$ of the observed population within the central pc, \cite{Genzel2010}) appeared from spectroscopic analyses to be older ($> 1$ Gyr) red giant helium-burning stars, along with a few more massive red supergiants \cite{Blum2003} and a few dozens Asymptotic Giant Branch (AGB) stars, that appear to be very bright ($L> 10^3 L_\odot$) and cooler ($\sim 2700$ K) than the other stars. These kinds of stars are naturally expected in an old NSC. Moreover, observations of the proper three-dimensional motions of these stars revealed that this population exhibits random isotropic motions, with very little bulk rotation (in the same sense of the Galaxy rotation), supporting the idea of a gravitationally relaxed system \cite{McGinn1989, Genzel1996, Trippe2008, Schodel2009}. However, between the end of the 1980s and the beginning of the 1990s, thanks to the pioneering observations of  Forrest et al. \cite{Forrest1987} and Allen \& Sanders \cite{Allen1990}, of the so-called ‘‘Allen-Forrest'' stars, an increasingly numerous population of bright early-type hot stars was discovered in the NSC \cite{Krabbe1991, Krabbe1995, Tamblyn1993, Blum1995, Libonate1995, Genzel1996, Tamblyn1996, Paumard2001, Paumard2003, Paumard2006, Tanner2006}. Spectroscopic analysis revealed that these stars are mainly post-main-sequence stars, either blue supergiants or Wolf-Rayet (WR) stars with ages below 10 Myr and zero-age-main-sequence-masses of $\sim30-100M_\odot$ \cite{Najarro1994, Najarro1997, Martins2007}. Not only members of the young population appear spectroscopically to belong to a whole different class of stars, compared to the older stellar component of the NSC, but from the study of their proper motion they also appear to posses a completely different kinematical structure. Young stars located at the south of Sgr A* appear to be red-shifted, while stars at the north of Sgr A* appear blue-shifted \cite{Genzel1996, Genzel2000, Tanner2006}, a pattern clearly indicating a coherent rotation of this population of stars. Additionally, in the 1990s with the advent of proper motion measurements \cite{Genzel2000} using the European Southern Observatory (ESO) 3m New Technology Telescope (NTT) for these stars, it became evident that most of the bright young stars in the NSC belong to an inclined Keplerian clockwise-rotating disk around Sgr A* . A small minority of such stars, on the other hand, appears to be in a counter-clockwise motion and today the young stellar population is modeled as composed by two star disks or rings rotating in opposite directions and inclined with two different angles between each other, neither of them coinciding with other known orientations from other systems in the GC.

The fact that such a population of early-type stars is located so close to Sgr A* within the central few arcsecs (from here on as) of the MW are highly surprising \cite{Genzel2010} in terms of their formation history. Their peculiar distribution into stellar disks of such estimated masses, orientations and eccentricities, as well as their inferred young ages, are challenging to explain with a unique comprehensive model, especially considering the co-presence in the same environment of a potential SMBH which would naturally prevent cluster formation. The questions, most of which still remain unanswered, related to the existence of such young stars in the GC have been named by Ghez et al. in 2003 \cite{Ghez2003}  as the ''Paradox of Youth''. 
In a usual star-forming environment, in order to trigger star formation in a molecular cloud, matter must be accumulated above a certain mass and density in order to develop Jeans instability \cite{Binney2008}, which leads to cloud collapse, fragmentation and eventually star formation. This is possible whenever the action of gravity and external environmental forces overcomes internal gas pressure. In an environment where a steep mass profile is present, on the other hand, like within 1 kpc from the GC where the compact mass of Sgr A* dominates the gravitational potential well, a strong tidal field develops shear on the collapsing cloud. The tidal forces can totally disrupt the cloud preventing star formation and only if the cloud is massive enough its self-gravity can survive the exerted shear \cite{Shu1981}. This competition between cloud collapse and disruption places a lower limit on the density required for a cloud to be able to gravitationally collapse and overcome against tidal shear, known as the Roche limit \cite{Binney2008}. For example, a cloud at a distance of 1 pc from Sgr A* would have to possess a density $\gtrsim 6\times10^7$ cm$^{-3}$  to survive tidal disruption \cite{Bryant2021} (for comparison, star forming giant molecular clouds in the MW usually possess densities in the range 1-100 cm$^{-3}$ ). This makes the GC a very inhospitable environment to host star formation, for which several models have been proposed involving either ‘‘in-situ'' or ‘‘ex-situ'' mechanisms of star formation \cite{Bryant2021}.

The latter scenario considers star cluster migration via dynamical mechanisms of an early-type stellar population formed at a more distant location where the tidal field of Sgr A* is supposed to be weaker. While the migration of single stars is a slow process to explain the relaxed and temporally stratified multi-disk population that is observed, the action of dynamical friction, within the lifetime of its early-type stars, can allow migration at a faster rate \cite{Tremaine1975, Gerhard2001, Kim2003, McMillan2003, ArcaSedda2014, Ivanov2020} (not only in the GC but in general for nuclear stellar clusters). This infall mechanism is required to be able to deposit stars within 6 Myr (which is the age of the oldest early-type stars in the disks), to resist sufficiently tidal shears in order to deposit stars at the required radii, and not to leave any trail of young stars during its migration towards the center \cite{Stostad2015}. This is possible if the cluster core can sufficiently collapse, acquiring a very high central stellar density ($\sim 10^8M_\odot$ pc$^{-3}$, see \cite{Kim2003}),before being disrupted by the Galactic tidal field \cite{PortegiesZwart2003}. In this case, the dense core, which has become rich in massive stars, manages to reach high proximity to the GC and, when it eventually dissolves, a great number of massive stars in the innermost pc of the Galactic nucleus is deposited. Recent simulations considering the infall of multiple clusters in a Milky Way-like galaxy successfully producing observational signatures in the form of age segregation and stratification of stellar disks, reflecting the infall history of the different clusters \cite{Perets2014, ArcaSedda2019}.

Alternatively, the core mass concentration in the infalling star cluster might be explained with the formation of an Intermediate Mass Black Hole (IMBH, $M\sim 10^3-10^4M_\odot$) due to runaway collisions between massive stars as a consequence of the action of dynamical friction \cite{PortegiesZwart2004, Gurkan2004}. This object might provide the sufficient core density for a stellar cluster to survive tidal shear during infall, up to galactocentric distances as short as $\sim 0.1$ pc \cite{Hansen2003}. N-body simulations of this process have been able to recreate the overall properties of the stellar population after infall but, with respect to the early-type stellar populations in the GC, the counterclockwise rotating disk does not naturally arise \cite{Berukoff2006}. This would suggest the simultaneous infall of at least two different clusters (with different eccentricities so to explain the observed eccentricity distribution) in order to form a double disk structure. Despite the failure of this mechanism to reproduce peculiar properties of the observed stellar disks, this scenario is still regarded to be a viable formation scenario for the early-type stellar disks in the GC, especially after the recent discovery of a potential IMBH in the small star cluster IRS 13E. This cluster is of special interest due to its close proximity to Sgr A*, it was estimated that in order for its stellar components to remain tightly bound and not be disrupted by tidal forces, a $10^{3}-10^4M_\odot$ BH should reside in it \cite{Maillard2004, Schodel2005, Paumard2006, Fritz2010a, Zhu2020}.
Moreover, in recent years the count of possible IMBH in the GC has increased \cite{Oka2016, Oka2017, Ballone2018, Takekawa2019a, Takekawa2019b}.

A second possible scenario for the formation of the early-type nuclear stellar cluster within the central pc of the MW considers formation of these stars at or near to their current position via an ‘‘in-situ'' mechanism. For example, a self-gravitating gaseous disk, remnant of an infalling molecular cloud, or as the result of continuous accumulation of gas (or a combination of both processes) \cite{Levin2003, Milosavljevic2004}, whose orientation is determined by the gas angular momentum, can become unstable with respect to its self-gravity, and its Roche-limit-exceeding fragments can collapse \cite{Paczynski1998} leading to formation of stars \cite{Kolykhalov1980, Shlosman1987, Collin1999}. The absence of massive stars at distances $> 0.5$ pc from Sgr A* supports this in-situ scenario \cite{Nayakshin2005b}, along with the existence of a number of candidate massive clouds within $\sim 20$ pc from the GC \cite{Solomon1972, Wardle2008}. Moreover, increasingly precise and refined numerical simulation have been able to recreate peculiar characteristics of the early-type cluster such as the mass, orientation and eccentricity distributions, the surface density, the presence of a counter-rotating disk and the age stratification in the cluster \cite{Nayakshin2005a, Alexander2007, Bonnell2008, Mapelli2008, Cuadra2008, Alig2013, Aharon2015, Trani2016}.

\begin{figure}[!t]
    \centering
    \includegraphics[width = 0.9\textwidth]{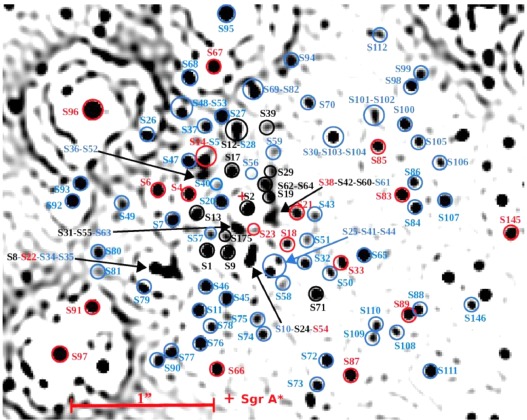}
    \caption{VLT-NACO finding chart of the S-star cluster obtained in 2018. In the plot, East direction is left, North is up. Sgr A* is denoted by a red cross. Stars in red and black circles correspond to the two separate disk populations recognized in \cite{Ali2020} and reported in Figure \ref{subfig:orientations}. Orbital elements  for all of the 29 stars are known. Blue circles, on the other hand, correspond to the stars for which only linear fit to their orbits is currently possible. Courtesy of \cite{Ali2020}.}
    \label{fig:s-stars-fieldl}
\end{figure}

\subsection{The S-stars}
\label{sec:s_stars}

\begin{figure}[!t]
\centering
\begin{subfigure}{.5\textwidth}
  \centering
  \includegraphics[width=0.9\linewidth]{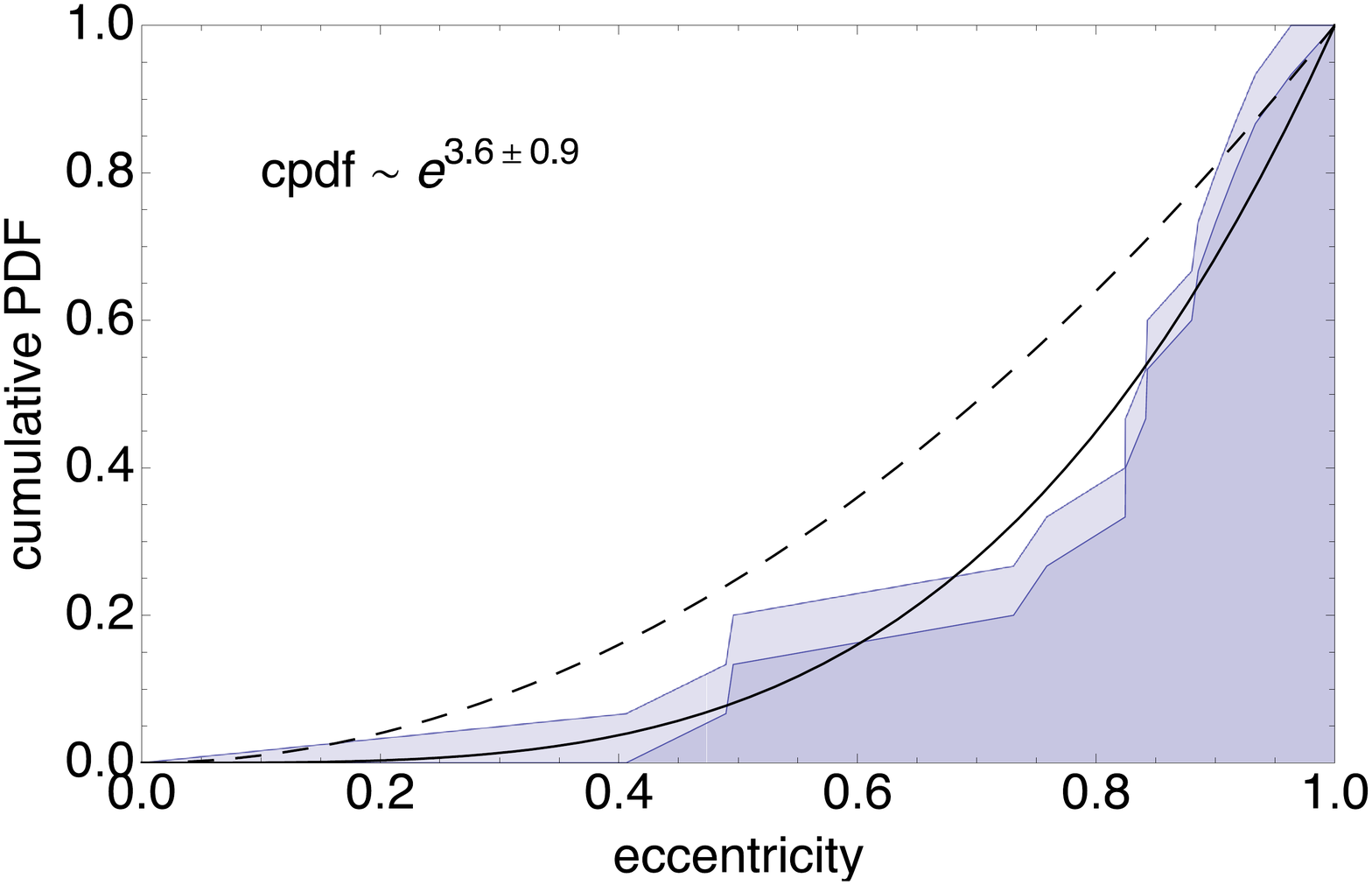}
  \caption{}
  \label{subfig:ecc}
\end{subfigure}\begin{subfigure}{.5\textwidth}
  \centering
  \includegraphics[width=\linewidth]{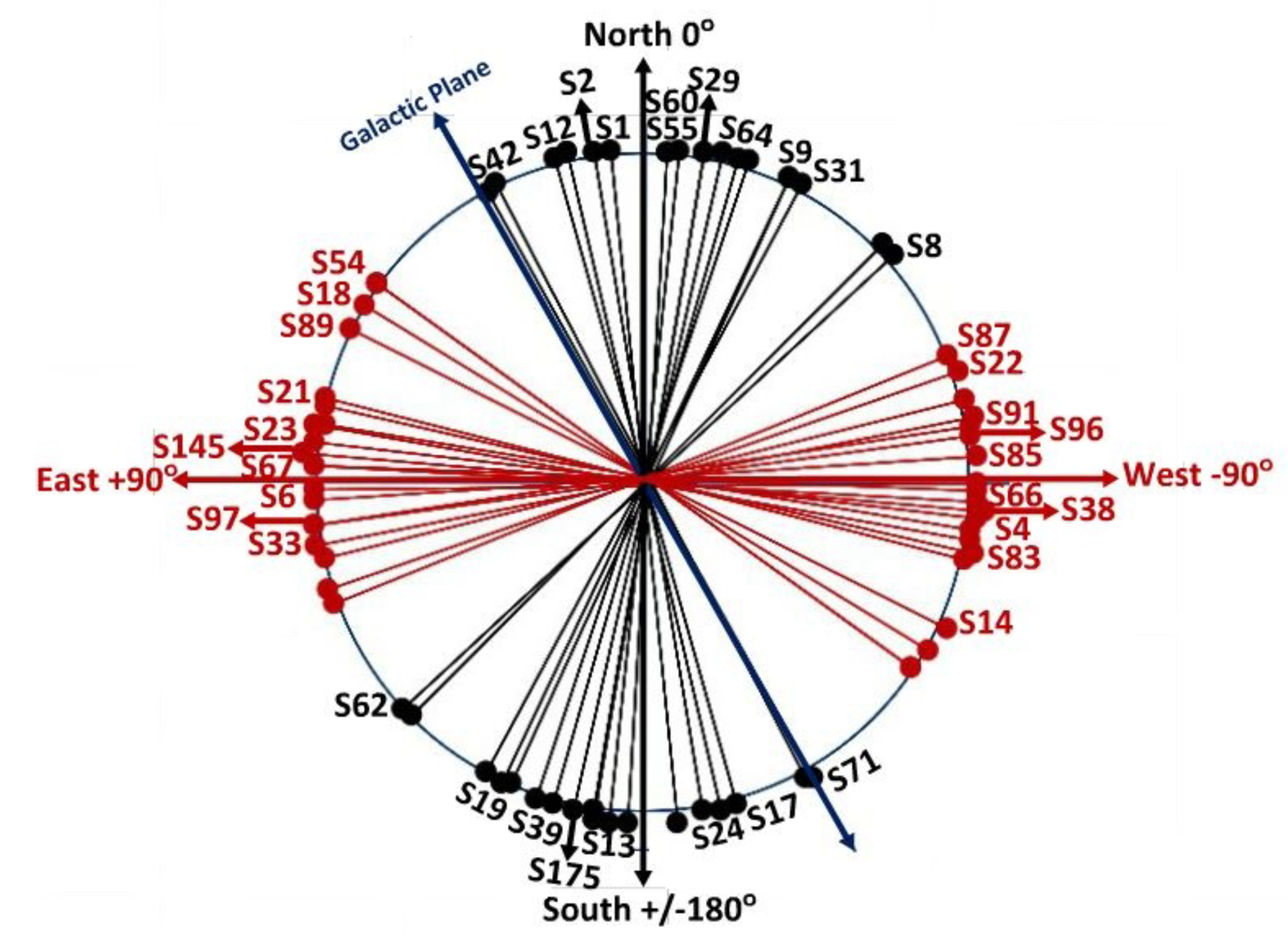}
  \caption{}
  \label{subfig:orientations}
\end{subfigure}
\caption{(a) Cumulative probability density function (PDF) for the eccentricities of the B stars in the S-stars cluster. Here, the two curves correspond to the two possible normalization factors for a cumulative PDF, with values either spanning from 0 to $(N-1)/N$ or from $1/N$ to 1. Either way, the distribution appears to be marginally compatible with a thermal isotropic model (dashed line) given by $n(e) \sim e$, while the best fitting PDF (solid black line) is given by $n(e) \sim e^{2.6\pm0.9}$ and is denoted by a more peaked probability for higher eccentricities. Image is courtesy of \cite{Gillessen2009b}; (b) angular arrangement of the disk for the 39 stars in the S-stars cluster for which orbital parameters have been estimated. In particular, the figure reports the position angles for the semi-major axes of the stars with respect to the galactic plane. From these images the demarcation between the two-disk structure, whose members have been reported with two different colors,appear clear. This image is courtesy of \cite{Ali2020}.}
\label{fig:ecc_orientations}
\end{figure}

Going even closer towards the GC, a remarkable cut-off in the disk density of the young star population is observed at a distance of around 0.04 pc form Sgr A*. This cut-off marks the separation between the young stars in the clockwise and counter-clockwise disks that we have described so far, and a third major component of the NSC, given by an impressive concentration of very fast moving dwarf B-type stars (aged 6 to 400 Myrs, \cite{Eisenhauer2005}) observed in the near-infrared within the central as around Sgr A*, called ‘‘S-stars'' \cite{Eckart1996, Ghez1998, Gillessen2009b, Gillessen2017}. The naming convention of these stars, originally proposed by Eckart and Genzel in 1997 \cite{Eckart1997} for the first 11 identified fast moving components of the cluster, is still in use today\footnote{Even though, it should be noted that the two main groups working on the subject from the Max Planck Institute for extraterrestrial Physics and University of California, Los Angeles use slightly different notations.}, while the number of recognized stars has increased to over 200 components \cite{Gillessen2009b, Gillessen2017}. Although one could naively think that the S-stars represent a low-density unbound extension of the young stellar disks, several arguments point towards a scenario where the S-stars have to be regarded as a totally different population (even though understanding the association of this cluster with the young stellar disks can shed light on the formation history of both populations). Firstly, from a spectroscopic point of view, the spectral classification of the great majority of the S-stars is of types B2-9 \cite{Eisenhauer2005}, in contradistinction to the giant and supergiant stars (OB/WR) belonging to the young disks. Furthermore, the different spectral classification, also suggests that the two populations have most likely formed at different epochs, due to the shorter average lifetime of main sequence O stars as compared to B stars.  Finally, the kinematical properties of S-stars are highly inconsistent with the coherent planar rotation of both the clockwise and counterclockwise stellar disks. The orbits of the brightest S-stars, have been tracked since the 1990s with speckle imaging methods at NTT \cite{Genzel1997, Genzel2000}, and later with the Keck telescope \cite{Ghez1998, Ghez1999, Ghez2003}, giving the possibility to measure a set of astrometric positions that allowed to constrain  stellar proper motion from the data acquired over multiple years. These studies clearly showed that these stars were characterized by randomly distributed orbits with no evidence of co-planarity with neither the clockwise nor the counterclockwise disk. The monitoring of the S-stars orbits has continued further (and still continues today), due to the fact that characterizing the motion of these stars, not only improves our understanding of this peculiar population of stars, but also serves as a probe for the gravitational field of Sgr A*, allowing to constrain its mass, its distance with unprecedented accuracy (for more details see Section \ref{sec:sgr_a}) \cite{Ghez2005b, Gillessen2009b}. These observations allowed for the first time to constrain the orbital parameters for some of the B stars in the S-stars cluster \cite{Schodel2003b, Eisenhauer2005, Ghez2005b, Gillessen2009b}, thanks to accelerated motion that some of them appeared to have undergone on the sky plane. The S-stars exhibited eccentricities of $e > 0.5$, with some going as high as $e > 0.9$ (Figure \ref{subfig:ecc}), and angular momentum pointing in random directions in the sky. This suggested that the S-stars, while exhibiting a somewhat excess of strongly non-circular orbit, was consistent with the expected probability distribution of eccentricities and angular momentum direction for a spherical, isotropic and virialized distribution of particles orbiting a central mass \cite{Schodel2003b}.

The excess of high eccentricities, on the other hand, might hint some peculiar formation mechanism that tends to favour highly non circular orbits. Moreover, since the S-stars have been proven to be a totally separate stellar component of the GC from other stellar clusters \cite{Eckart1997, Ghez1998}, their formation might be explained with a different mechanism. Such a mechanism has to be compatible with the existing early-type formation scenarios, on one hand, and on the other has to account for the observed properties of the S-star cluster.

S-stars cluster formation via an in-situ process is generally regarded to be unlikely, owing to suppression of clustering by the tidal field in the environment close to a SMBH. While migration of S-stars from the stellar disk of the MW might be possible for some of the members of the cluster owing to either Lense-Thirring precession \cite{Levin2003} (for example the S2 star might have been pulled out of the Galactic plane by this effect over its lifetime) or torques exerted by Lin–Shu-type spiral density waves on the stars at an inner Lindblad resonance \cite{Griv2010}, these mechanisms cannot account for the majority of long period eccentric orbits. Migration from the OB/WR-type stellar disks into the central as could be a possible explanation. Simulations have shown that two interacting stellar disks can result in the creation of a population of captured stars an order of magnitude closer to the central SMBH \cite{Lockmann2008}. Precession and Kozai-Lidov interaction \cite{Kozai1962, Lidov1962} on individual stars can lead them to achieve extremely high eccentricities at random orientation. Moreover, stellar binaries on such eccentric orbits could be disrupted due to close passages near the SMBH, leaving behind a single S star and ejecting the companion from the vicinity of the SMBH, providing an explanation for the observed hyper-velocity stars in the MW halo \cite{Lockmann2008}. Nevertheless, this process is still regarded to be unlikely because of two main reasons: \emph{(i)} the circumnuclear disks are populated mainly by OB/WR giant stars and only a small amount of main sequence B-type stars have been observed. In response to this, it has been proposed that a much larger population of B-stars than the observed one could exist in the disks, but that it has failed observation due to the crowded stellar environment, low spatial resolution of observations and the multi-spectral emission dominated by O-type stars \cite{Bryant2021}; \emph{(ii)} it has been demonstrated that spherical stellar bulges suppress Kozai oscillations, when the enclosed mass inside the orbit of a test body is of the order of the mass in the stellar disk \cite{Chang2009}.  This situation seems to be satisfied in the GC, here the stellar bulge is much larger than the stellar discs. This could likely suppress Kozai oscillations, making this process unable to allow S-stars migration.

This leaves with only two possible formation mechanism, both involving processes that started outside of the central as: the cluster infall scenario (that we have already seen for the early-type stars), and the binary capture scenario.

The first scenario is a direct continuation of the cluster infall mechanism for the formation of the early-type stars in the circumnuclear disks (Section \ref{sec:NSC}). Numerical simulations of this process have been able to correctly reproduce the observed properties of the S-stars population \cite{Fujii2010}. As mentioned in Section \ref{sec:NSC}, models predict that the IMBH, that forms inside the infalling cluster, deposits most of the new-born stars into the two disks. A minority of them however, keeps following the IMBH into the GC \cite{Fujii2009}, due to their orbital resonance with the IMBH itself. As the orbit of the infalling IMBH gains eccentricity, stars carried by the 1:1 resonance with the IMBH are dropped from the resonance and their orbits (with pericenter distance that can be below 0.05 pc from the SMBH) are randomized in the following 1 Myr by a chaotic Kozai mechanism, within the regime of the S-star cluster. This is able to recreate in the lifetime of the stars the observed eccentricities and apparent thermal isotropic distribution of the S-stars \cite{Merritt2009}.

In the binary capture scenario, as its name suggests, the S-stars would have arrived in the GC in the form of binary systems on highly-eccentric orbits. Gravitational interaction with the central SMBH would have resulted in the capture of one of the components of the binary in a bound eccentric orbit, and the ejection of the other star \cite{Miller2005}. In this scenario all the captured stars would have very high eccentricities ($e > 0.98$). However, relaxation occurring over 1 Myr after the capture (which lies in the lifetime of B-type stars) could eventually smooth out the distribution of eccentricities and explain the observed virialised distribution (in which an excess of high-eccentricity stars might still be surviving). In order to account for the number of S-star observed, however, this process would need a reservoir of binary systems. A possible mechanism might involve the presence of ‘‘massive perturbers'' (\emph{e.g.} molecular clouds or dense star clusters) who could result in the migration towards the SMBH of binary pairs at large radii on highly eccentric orbits \cite{Perets2007}. However, the recent kinematical analysis of the S-stars by \cite{Ali2020} has not only found a highly non-isotropic distribution of orbits, but has also implied the absence of such a perturber at least over the dynamical relaxation timescale of the two counter-rotating disks that have been observed (Figure \ref{subfig:orientations}).
\subsubsection{Observing S-stars in the Galactic Center}
\label{sec:s_stars_obs}

Two international observational groups, whose leading scientists, Reinhard Genzel and Andrea Ghez, both sharing half of the \href{https://www.nobelprize.org/prizes/physics/2020/press-release/}{2020 Nobel Prize for physics}, work at the Max Planck Institute for Extraterrestrial Physics (MPE) and at the University of California, Los Angeles (UCLA) respectively, have been tracking the orbits of S-stars in the GC for over thirty years \cite{Do2019a, Gillessen2017,Genzel2022}. The MPE group observations are performed at ESO facilities, in Chile. The UCLA GC Group, on the other hand, takes advantage of the W.M. Keck Observatory telescopes, located on the Maunakea island, in Hawaii. Despite the prominent blue emission of B-type stars in the S-stars cluster \cite{Eisenhauer2005}, optical observations of the GC are not feasible. Diffuse interstellar dust along the line of sight, towards the center of the MW, is responsible for  $\sim30$ magnitude of visible extinction \cite{Becklin1968, Becklin1978, Rieke1985, Rieke1989}. The extinction curve, however falls off steeply in the near-infrared, reaching only $\sim 3$ magnitudes of extinction toward Sgr A* in the K-band, $\lambda = 2.2 \mu$m. The drop-off goes on up to few a $\mu$m, after which obscuration by dust grains increases again. This makes the K-band an optimal spectral window to observe individual stars in the GC. Furthermore, tracking individual stars in orbit in a very crowded region like the GC requires an excellent spatial resolution and long measurement times, both in terms of single observations (long exposure times in order to recover the signal from a star) and in terms of observational campaign (following the stars over their orbital periods). In order to encounter both needs, space-based observations of the S-stars were excluded, and efforts pointed towards the improvement of ground-based facilities and analysis techniques to compensate for the effects of blurring by Earth's atmosphere, during the long observation times. A technique that has revealed to be greatly successful for near-infrared observations at the GC was the speckle imaging \cite{Fried1966, Baba1985} with the ‘‘shift-and-add'' method, which was initially adopted by both the MPE and the UCLA group. The technique is based on the fact that variations in Earth's atmosphere due to atmospheric turbulence happen on time-scales shorter than about one second, distorting astronomical images and typically limiting the angular resolution of long-exposures to $\sim 0.5 \div 1$ as, almost a factor ten times worse than the theoretical limit for large ground-based telescopes \cite{Ghez1998}. By taking numerous very short exposures (of $\sim 0.1$ s) with a sensitive detector, then shifting them spatially (in order to align the sources) and adding them in stack, a much sharper and deeper image could be produced that could reach the diffraction limit of the observing telescope. This methodology was successfully developed and implemented for the first time between 1991 and 1996 at ESO NTT by the MPE group, using the SHARP camera which was developed specifically for the project \cite{Eckart1996, Eckart1997}. This allowed them to achieve a spatial resolution of 0.15 as. However, breakthrough observations using this technique were led by the UCLA group at the W. M. Keck I 10m telescope in 1998, \cite{Ghez1998} reaching the diffraction limit (which for a 10 m telescope at 2.2 $\mu m$ wavelength - K band - correspond to 0.05 as of angular resolution), which allowed them to resolve spatial scales of 2.5 light days at the GC.

\begin{figure}[!t]
    \centering
    \includegraphics[width = \textwidth]{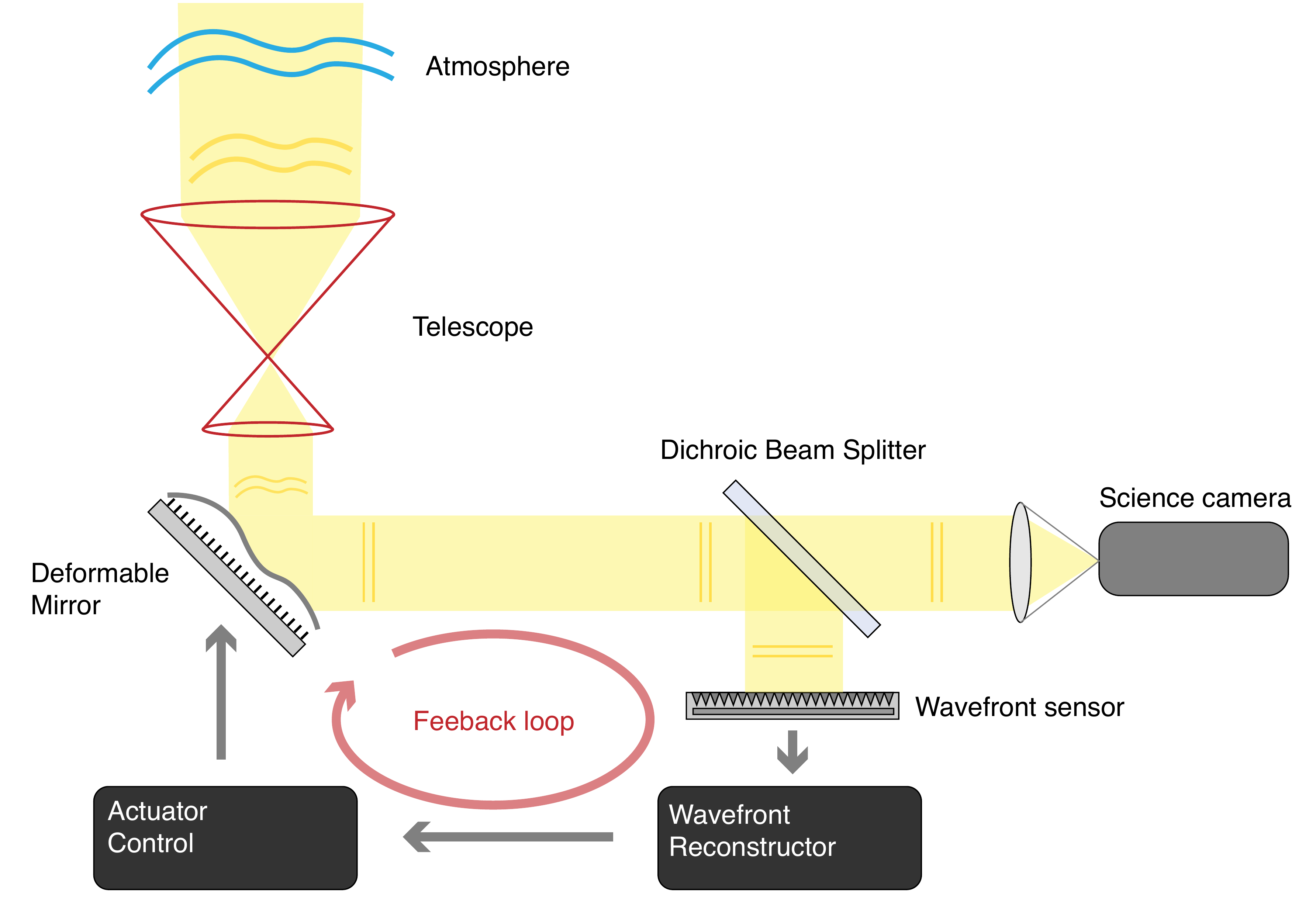}
    \caption{Schematic view of an Adaptive Optics feedback loop. A bright reference object whose optical properties are known is used as a reference to estimate the aberrations induced by Earth's atmosphere. These are then converted into the shape that a mirror should have in order to compensate for such aberrations and used to actually change the shape of a deformable secondary mirror, before reaching the sensitive surface of a science camera. Image reproduced from Figure 1 in \cite{Doble2005}.}
    \label{fig:ao_scheme}
\end{figure}

While speckle imaging allows almost effortlessly to achieve the maximum nominal resolution provided by ground-based telescopes, it has one major downside. Relying on short-time exposures, it is limited to monitoring only the brightest sources. Moreover, in order to extract reliable estimations of stars proper motion, lengthy surveys are required. To overcome these issues, developing a technique that was capable of exploiting the maximum potential of observing facilities, while performing deep enough observations to record signal from fainter stars, without being affected by Earth's atmosphere turbulence, was paramount. A ‘‘new era'' for observations at the GC began when adaptive optics (AO) techniques became available at Keck II observatory in 2000 \cite{Wizinowich2000} (however GC observation benefited from this technique only from 2004 onwards, due to the development of laser guide-star adaptive optics in 2004 on the 10 m Keck II telescope \cite{Wizinowich2006}) and, a few years later, at ESO Very Large Telescope (VLT) with the NACO instrument \cite{Rousset2003}. AO techniques were first envisioned by H. W. Babcock in 1953 \cite{Babcock1953}. However, the technology required for their application was not available until the 1990s. A schematic view of an AO system is shown in Figure \ref{fig:ao_scheme}: a bright reference object, located near on the sky to the scientific target of the observation, whose optical properties are known\footnote{Either a natural guided source - NGS-AO - like a bright star in the field-of-view of the detector, or an artificial star created with laser excitation of sodium in the upper atmosphere - LGS-AO.} is used as a reference to estimate the aberrations induced by Earth's atmosphere. These are then converted into the shape that a mirror should have in order to compensate for such aberrations and used to actually change the shape of a deformable secondary mirror \cite{Tyson2022}. This allows real-time correction of astronomical images, through a feedback loop, allowing long exposures times and diffraction-limited sharp and deep observations.

In order to coherently measure sky-projected positions for the S-stars relative to Sgr A*, an astrometric reference frame had to be established. The idea that a group of red giant stars at the GC could establish a radio-to-infrared reference frame, goes back to the work of Menten and collaborators in the mid-1990s \cite{Menten1997}. These stars were visible both in the radio (through SiO maser emission in their envelopes) and in the near-infrared, and their positions and proper motions had been characterized by radio-interferometric astrometry. However the field of view of neither Keck II telescope (10 as) nor the NACO camera at VLT (14 as) are large enough to observe simultaneously the S-stars and the masers. Thus, in order to relate observed positions of the star to the radio-to-infrared reference frame, a dither pattern of overlapping telescope-pointings had to be adopted, executing secondary-astrometric referencing (either adopting matching coordinate lists or with mosaic images). Systematic astrometric errors naturally arise in this process, caused by the geometric distortions, field dependence of the point-spread function and higher-order aberrations of the optics \cite{Yelda2010, Plewa2015, Sakai2019}.

The ability to employ AO techniques to observe the S-stars at the GC, did not only allow for better astrometry, it also allowed for the first time to perform spectrography on the single stars which, on one hand allowed for a spectral classification of the population \cite{Eisenhauer2005}, and, most crucially for the study of the orbital motion of these stars, radial velocities could be measured in addiction to sky-projected properties, providing information on the third (line-of-sight dimension). This led in 2003 to the first measurement of spectral lines for single stars in the GC \cite{Ghez2003}. At that time, the NIRC2 spectrographer was used at the Keck II telescope behind a NGS-AO system, providing for a 75 km s$^{-1}$ resolution for the K-band $\lambda \sim 2.16612\,\mu$m Bracket-$\gamma$ absorption line of hydrogen (which is a prominent spectral feature of B-type stars. \cite{Eisenhauer2005}). From that time, spectroscopic measurement combined with AO systems have improved by a great amount, with the use of specifically designed instruments. In particular, since 2005, the UCLA group makes use of the OSIRIS (OH-Suppressing Infra-Red Imaging Spectrograph) integral field spectrographer (which uses LGS-AO) at the Keck observatory which reduces uncertainties on measured radial velocities to about $10-20$ km s$^{-1}$ \cite{Larkin2006}.

\begin{figure}[!t]
    \centering
    \includegraphics[width = \textwidth]{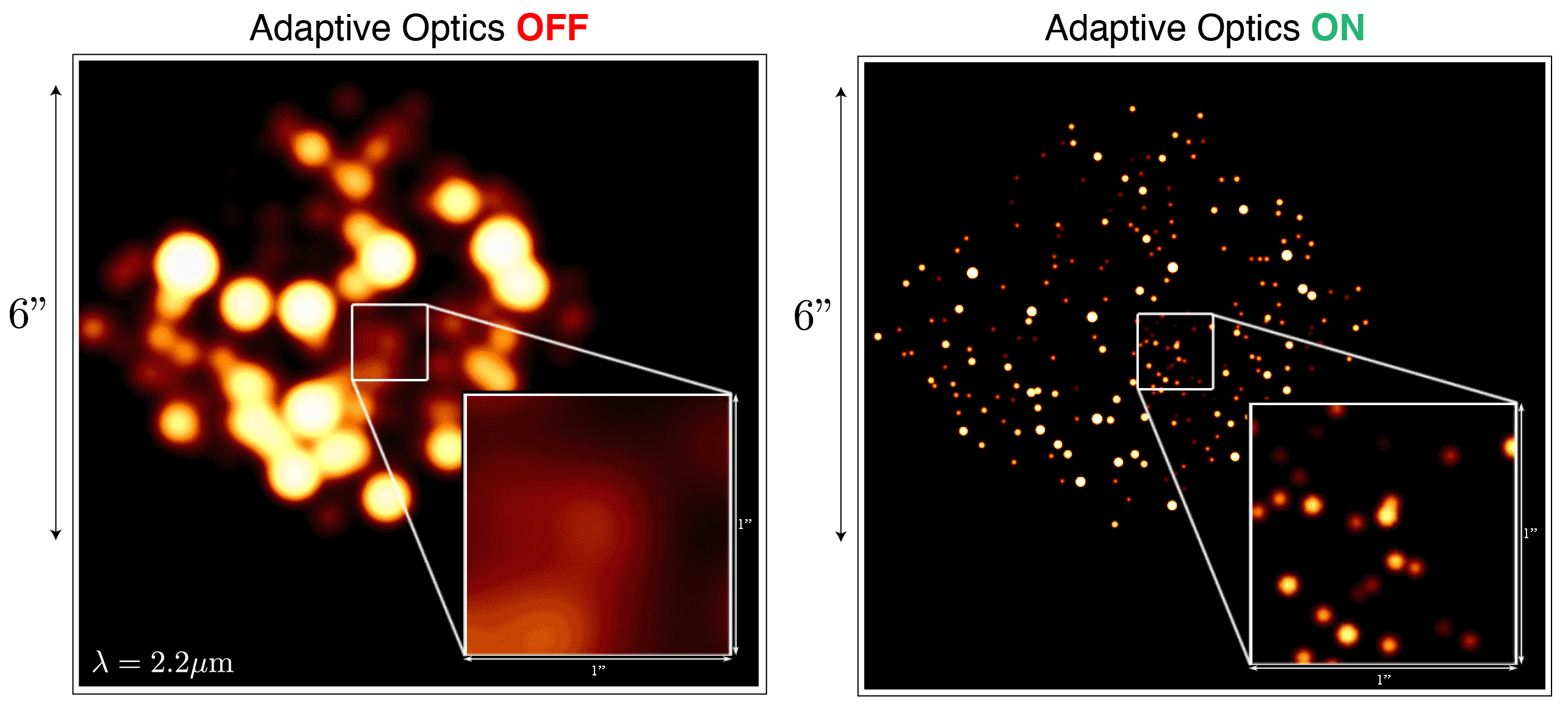}
    \caption{Observations of the $6''\times6''$ GC region with the Keck telescope in the K-band \cite{Wizinowich2006} comparing the resolution performances without (left panel) and with (right panel) the application Adaptive Optics techniques. The resolution gain is quite evident. The inset panel reports a zoomed-detail of the central region. The figure is adapted from the UCLA GC Group \href{https://galacticcenter.astro.ucla.edu/images.html}{website}.}
    \label{fig:ao_gc_comparison}
\end{figure}

\begin{figure}
    \centering
    \includegraphics[width=\textwidth]{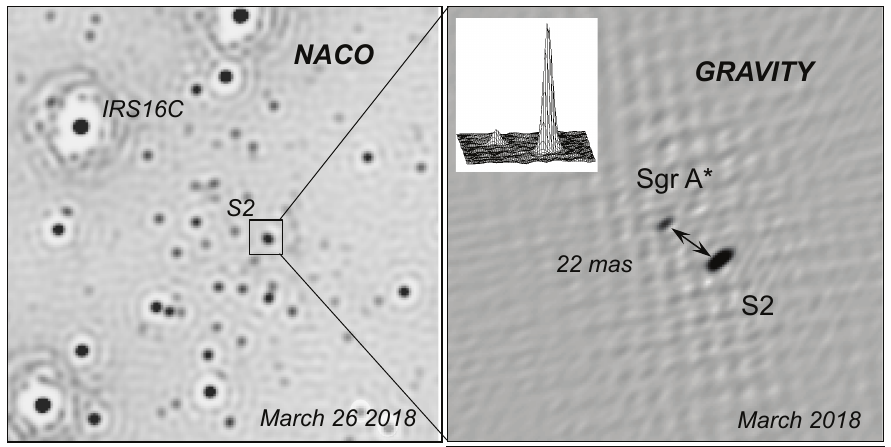}
    \caption{Observations of the S2 star in the GC using the NACO adaptive optic telescope at VLT (right panel) and the GRAVITY interferometer, which combines all four Unit Telescopes at VLT. The images have both been acquired in March 2018, a few weeks before the pericenter passage of the star around Sgr A*. The $\sim 50$-fold improvement in angular resolution achieved with GRAVITY is clearly noticeable by comparing the two images. In the NACO observation, S2 appear as being slightly elongated due to confusion with Sgr A* (the two sources are not resolved between each other, at such high proximity). In the GRAVITY observation the two sources are well separated ($\sim 20$ mas) and both distinguishable (the slightly elongated shape of S2, here, is due to shape of the interferometric clean beam). The image is adapted from \cite{GravityCollaboration2018a}.}
    \label{fig:s2_gravity}
\end{figure}

Similarly, the MPE group adopted AO-assisted spectrographic techniques to monitor the line-of-sight motion of the S-stars, starting in 2004, when the SINFONI (SINgle Faint Object Near-IR Investigation) integral field spectrographer at VLT had its first light \cite{Eisenhauer2003, Bonnet2004}. The last addition to the set of astronomical facilities employed to monitor the orbits of the S-stars is GRAVITY: an interferometric astrometry-imager that  combines all four Unit Telescopes (UTs, NACO on the contrary is composed by only one UT) of the VLT interferometer \cite{GravityCollaboration2017}. In particular, the Gravity Collaboration, with R. Genzel among the principal investigators, is the group that nowadays makes use of the exquisite astrometry provided by the GRAVITY interferometer to study the GC. The use of interferometric techniques has allowed GRAVITY to achieve an angular resolution of just about 20 $\mu$as (with a nominal instrumental goal of 10 $\mu$as), with a gain in resolution of almost a factor 100 with respect the initial SHARP speckle images \cite{GravityCollaboration2017}. This improvement is clearly shown in Figure \ref{fig:s2_gravity} were a comparison of the NACO field-of-view with reconstructed imagery from GRAVITY data for the star S2 is presented. The observations are now precise enough that the position of this particular star can be seen to change between consecutive nights \cite{GravityCollaboration2018a}. Moreover, interferometric observations of the GC are not affected by the systematic uncertainties of the reference frame. This is due to the ability of GRAVITY to directly relate the position of the observed star (mainly S2) to that of Sgr A*, being the two sources well separated. Nonetheless, other sources of systematic uncertainties, due to aberrations of the optical trains of the individual telescopes, affect GRAVITY observations \cite{GravityCollaboration2021b}. 

The ability to detect effects beyond the Keplerian motion (\emph{i.e.} post-Newtonian - PN - effects) on the S-stars orbits (see Section \ref{sec:test_GR}) heavily relies on the precision of astrometric and spectroscopic measurements. While efforts in the recent years from both the MPE and the UCLA groups have focused on improving the orbital tracking of the S2 star (the brightest, and thus easiest to observe, in the cluster), it still is not clear whether higher order effects (see Section \ref{sec:higher_order_effects}) can actually be detected on its orbit \cite{GravityCollaboration2021a}. As a matter of facts, the magnitude of these effects decreases faster with distance from the SMBH than lower order effects; moreover, stars located at a greater distance from Sgr A* are more affected by Newtonian orbital perturbations due to the stellar population itself or a dark mass concentration. A star on a closer orbit (either because of a shorter semi-major axis or because of an higher eccentricity and, thus, a closer pericenter passage), if it exists, would allow for a much greater detectability of such higher order orbital effects \cite{Alexander2005}. For this reason, the next challenge at the GC will be the successful observation of such closer stars that have so far eluded detection. These are expected to belong to the fainter end of the luminosity function of the GC stellar population, which makes their observation rather difficult \cite{GravityCollaboration2021a}. While some very close and faint stars have been recently found\footnote{However, the actual existence of these stars is still debated by the authors of the research and the Gravity Collaboration \cite{GravityCollaboration2021a, GravityCollaboration2022a, Peissker2022}.} by Peisßker and collaborators \cite{Peissker2020, Peissker2022} (see Table \ref{tab:sstars} in Section \ref{sec:test_GR}) thanks to a re-analysis of the NACO observations of the GC with a different image filtering technique, the current precision of the astrometry inferred by this analysis for such closer star is still poor and not competitive with the observational data for S2 from GRAVITY. Moreover, it is not yet clear how many faint, low-mass stars exist in the inner few mpc of the GC. Mass segregation mechanisms and collisional destruction by the massive remnants there, are expected to suppress such a population \cite{Alexander2005}. However, recently, deep observations of the GC with GRAVITY, were able to identify a slowly moving star with a K-band magnitude of 18.9 within 30 mas of Sgr A* \cite{GravityCollaboration2021a}. The observed position and proper motion of the star allowed to identify it with the previously observed S62. While passing closer than S2 on the sky, this star is at a substantially larger physical distance from Sgr A*. No other faint stars have been detected in current images of the GC with GRAVITY, and this is probably due to limitations related to the calibration of the instrument \cite{GravityCollaboration2021a}. Future upgrades of the instrument, within the GRAVITY+ project \cite{GravityCollaboration2022b}, resulting in a better calibration and removal of the point-spread-functions (PSF) from bright stars are expected to eventually lead to deeper observations of the GC and the identification of a whole new population of closer, fainter stars \cite{GravityCollaboration2021a}. Moreover, substantial progress in this direction is expected to occur with the development of greater observing facilities like the Thirty Meter Telescope (TMT) on Hawaii \cite{Weinberg2004} or the 39-meter Extremely Large Telescope (ELT) at ESO \cite{Davies2021}.

\subsection{Sgr A* and the quest for its imaging}
\label{sec:sgr_a}
The idea that compact objects like SMBHs could power bright radio sources in the center of galaxies goes back to the seminal theoretical works of Lynden-Bell and Rees \cite{LyndenBell1969, LyndenBell1971}. They proposed that most galactic nuclei, the center of the MW included, could host such compact objects, resulting in high energy emission produced by the release of gravitational energy in the form of radiation when matter accretes onto a SMBH (that in some cases can reach the luminosities and energy densities of the distant quasars). Furthermore, they provide with a set of possible observational tests to unambiguously detect the presence of a compact object in the GC. Among these, Very-Long Baseline Interferometry (VLBI) was regarded as a powerful tool to determine the size of any central BH that there may be in the Galaxy. 
The actual discovery of Sgr A* happened on the nights between 13 and 15 February 1974 \cite{Balick1974} by B. Balick and R.L. Brown. Their observation was a follow-up on previous works that had analyzed the small scale radio structure of the Sgr A* region\footnote{More than 20 year before, the Sgr A* region was initially identified by Australian astronomers Piddington and Minnet \cite{Piddington1951} as a plausible location for the center of the Galaxy} in the GC near the beginning of the 1970s \cite{Clark1966, Miley1970, Ekers1971, Downes1971, Goss2003}. All these observations were just below the resolution limit required to unambiguously resolve the presence of the compact radio source. The observation by Balick and Brown \cite{Balick1974} using the Green Bank interferometer and a single antenna at the Huntersville West Virginia site, on the other hand, achieved for the first time the resolution and u-v coverage to resolve the extended confusion from the Sgr A West complex. They remarked the unusual structure of the sub-as structure and the positional coincidence of this source with the galactic nucleus, suggesting that this structure is physically associated with the GC. 
The bright and compact radio feature was later confirmed by Ekers and collaborators \cite{Ekers1975} combining data from the Westerbork Synthesis Radio Telescope with Owens Valley Radio Observatory data at 6 cm and, for the first time, with the technique of VLBI by Lo and collaborators  \cite{Lo1975}, but it wasn't until 1982 that it was given the name by which we refer to it today, Sagittarius A*, in order to distinguish it from the extended source complex Sagittarius A \cite{Brown1982}. High-resolution VLA observations in 1981 pinpointed the location of Sgr A* to be coincident with the dynamical center of the gas streamers in the Galactic nucleus \cite{Brown1981} and a year later its radio variability was estimated for the first time by the same group \cite{Brown1982}. Sgr A*  appeared a very atypical object because of its rapid variability its high radio luminosity and its apparent compactness, supporting the compelling idea that it could be a SMBH candidate.

\subsubsection{Constraints from gas dynamics}
\label{sec:gas_dynamics}
From a dynamical standpoint, the first observations supporting the idea of a central ‘‘dark'' mass concentration in the GC emerged in the late 1970s and early 1980s \cite{Wollman1977, Lacy1980, Lacy1982}. It was around those years that a group of astronomers at University of California, Berkeley, led by Nobel laureate C.H. Townes, discovered that ionized gas (in particular, they observed the 12.8 $\mu$m line of Ne II) at the GC had radial velocities that increases to hundreds of km s$^{-1}$ in the central pc of the MW \cite{Wollman1977}. The virial analysis applied to the velocity and velocity dispersion of the ionized gas, suggested that the gravitational field was dominated by a mass of $\sim 4\times 10^6M_\odot$ concentrated in the central 1 pc of the MW. The same group led additional observations of the GC, characterizing the infrared fine-structure line emission of compact clouds of ionized gas in the Sgr A West region \cite{Lacy1980, Lacy1982}. {From analyzing the distribution of the velocities of such clouds a plausible mass distribution was derived for the region within 1 pc of the center. The most plausible model included a central point-like mass of a few $10^6M_\odot$, in addition to star clusters whose mass is a few $10^6M_\odot$.} Moreover, for the first time the idea that such a mass concentration could presumably be a massive BH was put on the table. However, due to the small number of observed clouds, there was not a definitive proof of the existence of a central point-like mass. Since then, though, the evidence for the existence of a dark mass concentration in the GC has steadily grown. The advent of increasingly more precise observations of ionized, atomic and molecular gas within the central 1 pc during the 1980s \cite{Crawford1985, Serabyn1985, Guesten1987, Serabyn1988, Schwarz1989, Lacy1991, Herbst1993, Jackson1993, Roberts1996} seemed to confirm the existence of a mass of a few million solar masses located in spatial coincidence with the compact source Sgr A*. On the other hand, evidence supported only by gas dynamics seemed unconvincing for many astronomers who argued that considering only gravitational forces would be an over-simplistic approach to appropriately model the observed gas motion while other, non-gravitational, forces (like shocks, magnetic fields, stellar winds, etc.) could heavily modify predictions \cite{Genzel1987, Zhao2009}. Moreover, a SMBH was expected to be a very luminous X-ray source, while Sgr A* appeared to be very faint in all bands except radio to sub-millimeter \cite{Falcke1998}, and this shed doubts on the actual existence of such a compact object \cite{Rees1982, Allen1986, Kormendy1995}. As a matter of fact, despite the observational improvements, gas dynamics measurements were subject to too many variables making, along with the low spatial resolution, conclusions on the nature of Sgr A* rather ambiguous and uncertain.

\subsubsection{Constraints from the stellar dynamics}
Significant progress in the field was made with the advent of stellar dynamics measurements in the GC. Spectroscopic observation of the 2 $\mu$m CO absorption bands in late-type giants and AGB stars and the 18 cm OH maser emission stars \cite{Rieke1988, McGinn1989, Sellgren1990, Lindqvist1992, Haller1996a} provided with the first stellar velocity dispersion measurements between 10 as and 3 arcmin from Sgr A*. Such observations seemed to confirm the predictions from gas dynamics of a few-million-solar-masses concentration within the GC, but the observed stars were at a too large distance (projected distances $> 10$ as) to constrain such mass to a point-like source. The evidence was further strengthened when hot ‘‘He II-stars'' blue supergiants, close to their WR stage,  were observed in the central 10 as \cite{Forrest1987, Allen1990, Krabbe1991}. The radial velocity dispersion of 35 early- and late-type stars within 1 to 12 as from Sgr A* was used to perform a virial analysis of the gravitational field, allowing to constrain the derived central dark mass of $\sim 3\times 10^6M_\odot$ within 0.14 pc of the dynamic center \cite{Krabbe1995}. In 1996, Genzel and collaborators studied for the first time the concentration $\sim$200 stars down to a distance of 0.1 pc from Sgr A* \cite{Genzel1996}. They found a statistically significant increase in projected stellar velocity dispersion towards the dynamic center (going up to 180 km s$^{-1}$ at a distance of 0.1 pc). Moreover, fitting simultaneously the observed projected surface densities and velocity dispersion, and applying Jeans analysis they derived the mass distribution between 0.1 and 5 pc. For the first time, from their analysis the evidence for a compact central dark mass of $2.5-3.2 \times 10^6 M_\odot$ was obtained at $\sim8\sigma$ significance.

\subsubsection{Constraints from the S-stars}
\label{sec:stellar_dynamics}
\begin{figure}[!t]
\centering
\begin{subfigure}{.5\textwidth}
  \centering
  \includegraphics[width=\linewidth]{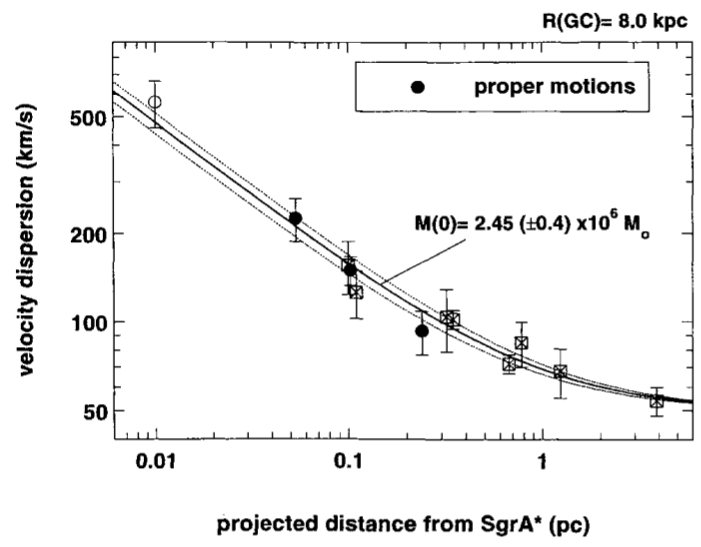}
  \caption{}
  \label{subfig:proper_motion_1}
\end{subfigure}\begin{subfigure}{.5\textwidth}
  \centering
  \includegraphics[width=\linewidth]{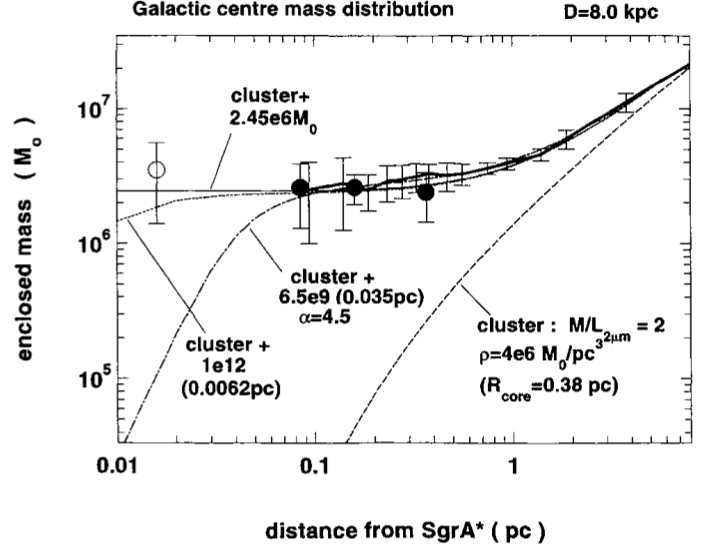}
  \caption{}
  \label{subfig:proper_motion_2}
\end{subfigure}
\caption{Mass distribution in the central pc of the GC. \textbf{(a)} Projected one-dimensional stellar velocity dispersion profile, derived from stellar proper motions (filled circles) and line-of-sight Doppler velocities (crossed squares). The solid curve represent the profile expected from theory in the case of an isotropic velocity distribution in the gravitational field generated by a point mass ($2.45\pm0.4\times10^6M_\odot$) plus an isothermal star cluster with velocity dispersion 50 km s$^{-1}$, assuming a GC distance of 8 kpc. (b); Enclosed mass as a function of distance from Sgr A* from the modelling of the stellar proper and radial motions \cite{Eckart1997}. The thick curve reports the enclosed mass profile derived from the Jeans analysis of stellar radial velocities, assuming anisotropy $b = 0$ \cite{Genzel1996}. The filled circles denote the masses estimated from an independent Jeans analysis of the proper motions for 39 stars between 0.9 and 8.8 as. The thick dashed curve represents the mass model of the stellar cluster alone, as derived from near-infrared observations. The thin continuous curve represents the sum of the stellar cluster component plus a central $2.45\times 10^6 M_\odot$ point mass. The thin dash-dotted (dotted) curves report the sum of the stellar cluster plus a dark cluster of core radius 0.035 pc (0.0062 pc) and central density $6.5\times10^{9}M_\odot/\textrm{pc}^3$ ($10^{12}M_\odot/\textrm{pc}^3$), respectively. The S-stars proper motion analysis from Eckart and collaborators in 1996 (from which the open circle is derived) seems to favour the point-like mass distribution. These images are courtesy of \cite{Eckart1997}.}
\label{fig:s_stars_proper_motion}
\end{figure}

Great advancement towards further constraints of the central mass occurred when data for the proper motion of single stars in the S-star cluster (see Section \ref{sec:s_stars}) in the central few as became available. Near-infrared observations of this stellar cluster performed by the MPE (since 1992, \cite{Eckart1996, Eckart1997, Genzel1996}) and UCLA (since 1995 \cite{Ghez1998}) groups (see Section \ref{sec:s_stars_obs}) allowed to determine the proper motion for these fast moving stars (up to 1000 km/s). The velocity $v$ of the stars, derived from the shifts in their sky-projected positions, led to a successful detection (down to a distance of 0.01 pc from Sgr A*) of the $\sigma(v)\propto r^{-1/2}$ relation that is expected for a Keplerian motion around a single massive point source (Figure \ref{fig:s_stars_proper_motion}). Nevertheless, the rather small number of recorded stellar proper motions ($\sim 40$) and the lack of information on the line-of-sight direction only allowed for a statistical approach, limiting the smallest scale to the average radius of the S-stars cluster of $\sim0.02$ pc \cite{Genzel2010}.

\begin{figure}[!t]
\centering
\begin{subfigure}{.67\textwidth}
  \centering
  \includegraphics[width=\linewidth]{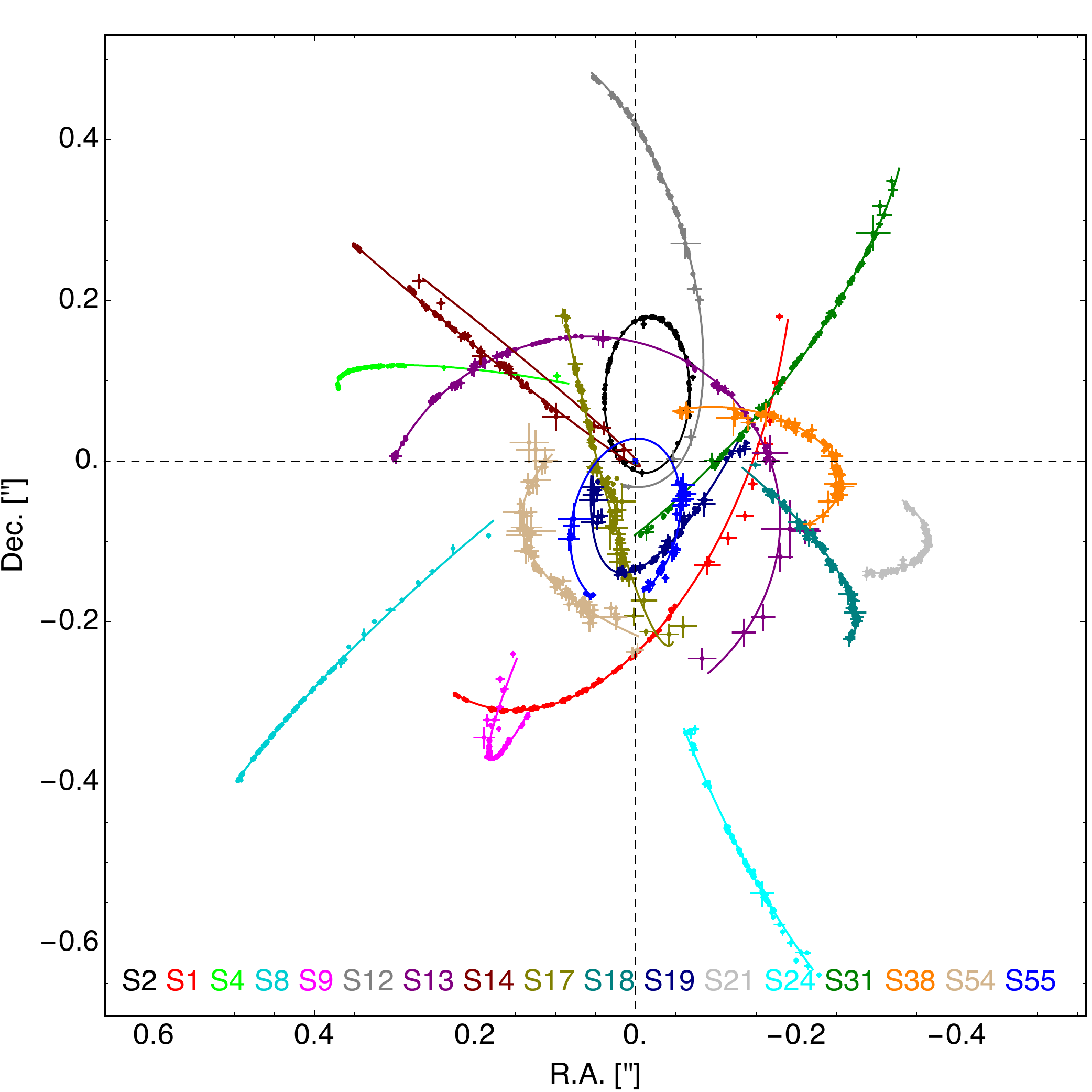}
  \label{subfig:s_stars_orbits_1}
\end{subfigure}\begin{subfigure}{.33\textwidth}
  \centering
  \includegraphics[width=\linewidth]{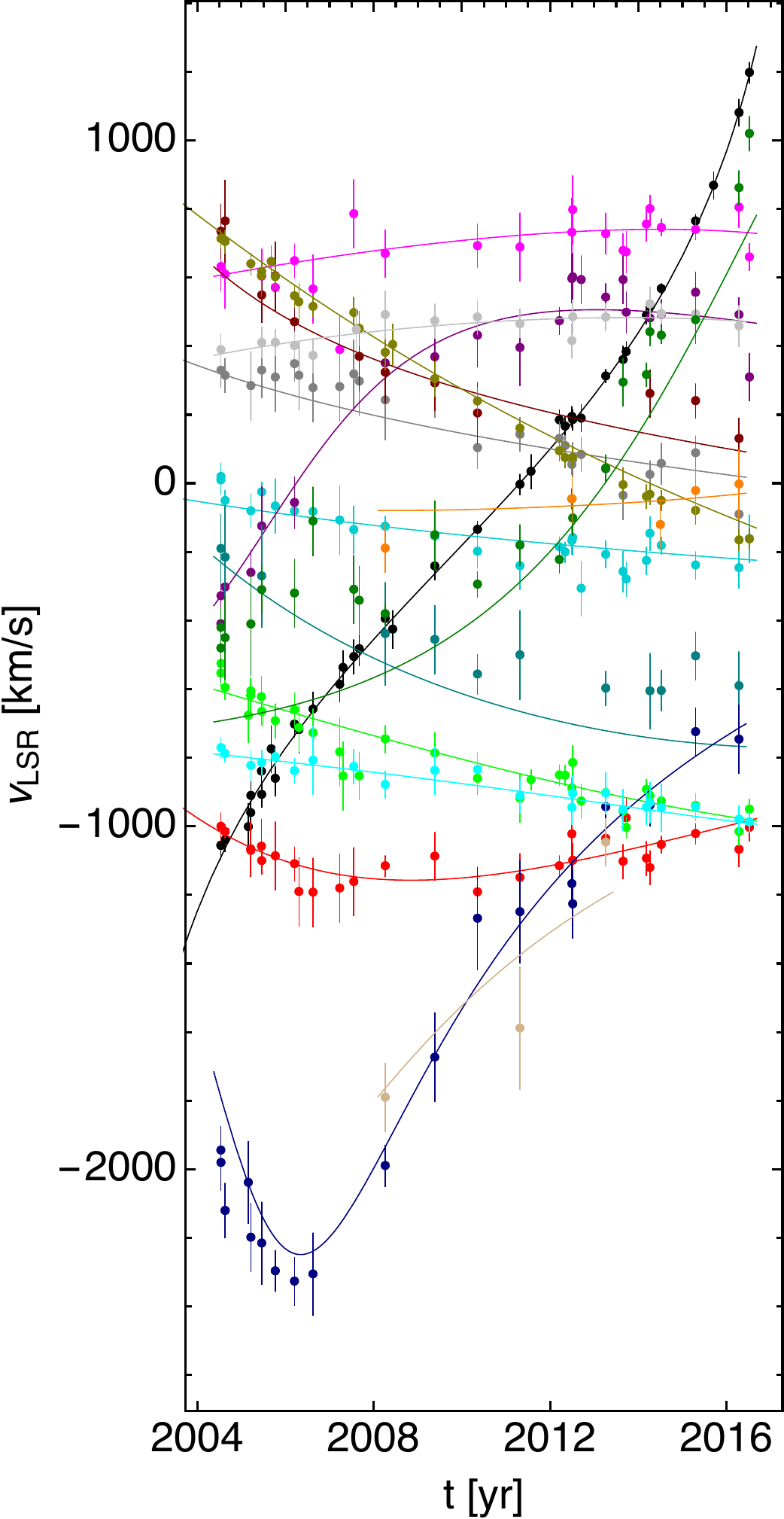}
  \label{subfig:s_stars_orbits_2}
\end{subfigure}
\caption{Astrometric (left) and radial velocity (right) data for the 17 stars used for the multi-star fit from \cite{Gillessen2017}. Solid lines represent the best-fitting orbits from the multi-star fit, and the color coding is the same for the two plots. Image is courtesy of \cite{Gillessen2017}.}
\label{fig:s_stars_orbits}
\end{figure}

The next major breakthrough happened when tracking of individual stellar orbits became possible \cite{Genzel1999, Ghez2000, Eckart2002}. Around the 2000s, the acceleration for three of the S-stars (S0-1, S0-2 and S0-4) was determined \cite{Ghez2000, Eckart2002}. These data did not only confirm a $\sim 4\times10^6M_\odot$ mass, increasing the inferred minimum mass density in the central region of the GC by a factor 10 relative to previous results (due to their proximity to the central object), but also localized the dark mass to within $0.05\pm0.04$ as of the nominal position of Sgr A* (by looking at the intersection of their acceleration vectors). Moreover, it was possible to estimate the orbital period of one of the observed stars (S-02 in the UCLA notation, or S2 in the MPE notation, the brightest in the cluster with a magnitude $m_K = 14$ in the $K$-band) that seemed to be $\sim15$ years, proving that constant monitoring of the region over multiple years could provide with orbital phase coverage spanning entire orbits. In 2002, S2 passed at its pericenter around Sgr A*, providing for the first time the opportunity to constrain its orbital parameters \cite{Schodel2002, Ghez2003}. S2 is revolving around Sgr A* every 16 years on a high-eccentricity orbit of $e\sim0.88$, indicating that at the pericenter S2 goes as close as 120 AU to Sgr A*. Thus, the $4\times 10^6M_\odot$ dark mass has to be concentrated in a radius that is necessarily smaller than the pericenter passage distance of S2. The importance of observing stellar orbits at the GC, to be used as precision probes (especially the bright short-period S2) of the gravitational field of the GC and to establish constraints on the compact dark mass associated with Sgr A*, motivated the two MPE and UCLA groups to carry on 30-year-long observational campaigns that still continue today. Over the years, advancement of the observing techniques and data analysis methods allowed the number of single orbits tracked to increase and the characterization of S-star orbital parameters to improve by a great amount \cite{Ghez2008, Gillessen2009b, Boehle2016, Gillessen2017, GravityCollaboration2018a, Do2019a, GravityCollaboration2019b}. In 2017, the MPE group presented the most updated and complete catalogue of orbital parameters, astrometric and radial velocity data of the S-stars \cite{Gillessen2017}, where orbits for 40 stars in the GC have been constrained and a multi-star fit (Figure \ref{fig:s_stars_orbits}) of the gravitational field has been performed incorporating updated systematic analysis from \cite{Plewa2015}.

The position of Sgr A* with respect to the background quasars has been monitored since 1981, initially with the VLA \cite{Backer1999} and later with the Very Long Baseline Array (VLBA) \cite{Reid2004, Reid2009c}. These observations have demonstrated that the radio source, once corrected for the Solar System motion around the center of the Galaxy, is stationary. In particular, Sgr A* appears to have small residual motion of $-7.2 \pm 8.5$ km/s on the Galactic Plane \cite{Reid2009c} (the high uncertainty is mainly due to imprecise estimation of the Sun's motion around the GC between 220 km/s and and 255 km/s) and a much smaller value for the motion perpendicularly to the Galactic plane, $-0.4 \pm 0.9$ km/s \cite{Reid2004}.  Consequently Sgr A* moves hundreds to thousands of times slower than the S-stars orbiting around it (whose orbital speed have been estimated to reach up to $10^4$ km/s, and whose mass are supposed to be in the $10-15M_\odot$ range). This is an other evidence in favour of the conclusion that the compact radio source contains a significant fraction of the dynamical mass derived from the S-stars dynamics. As a matter of fact, owing to equipartition of kinetic energy between a central SMBH and a population of stars orbiting around it, both analytical and numerical results allow the determination of a lower limit on the mass of Sgr A*, $\gtrsim4\times10^5M_\odot$, only due to its small proper motion
\cite{Chatterjee2002, Dorband2003, Reid2004, Merritt2007}.

With improved astrometry from interferometric observations of the GC from the Gravity Collaboration using VLT/GRAVITY, the orbital parameters (especially for the S2 star) have been measured with increasingly high precision, allowing to improve accordingly the constraints on the properties of the central compact object (\emph{i.e.} its mass $M_0$ and distance $R_0$ from Earth) and to detect relativistic effects on the orbit of the star (see Section \ref{sec:test_GR}). The most updated values for the mass and distance of Sgr A* derived by the Gravity Collaboration in 2022 using NACO and GRAVITY data for the four stars S2 (which yields most of the constraining power \cite{Gillessen2017}), S29, S38 and S55 (which are among the brightest and shortest-period known stars in the GC) are \cite{GravityCollaboration2022a}:
\begin{align*}
    &\textrm{Gravity Collaboration:}& M_{0} &= 4.297\pm0.012\times10^6M_\odot,& R_0 &= 8.277\pm0.009\textrm{ kpc},
\end{align*}
which represent a more than 10-fold improvement over the first measurements of the dark mass concentration at the GC from the early studies on the subject \cite{GravityCollaboration2021b}.

The latest published result from the UCLA GC Group on the S-stars are from 2019 and, apart from reporting the detection of relativistic gravitational redshift on the observed radial velocity of S2 (see Section \ref{sec:test_GR}), provide with the most updated values for the mass and distance of Sgr A* derived with Keck observations \cite{Do2019a}:
\begin{align*}
    &\textrm{UCLA:}& M_{0} &= 3.984\pm0.058\times10^6M_\odot,& R_0 &= 7.971\pm0.059\textrm{ kpc}.
\end{align*}

\begin{figure}[!t]
\centering
\includegraphics[width = \textwidth]{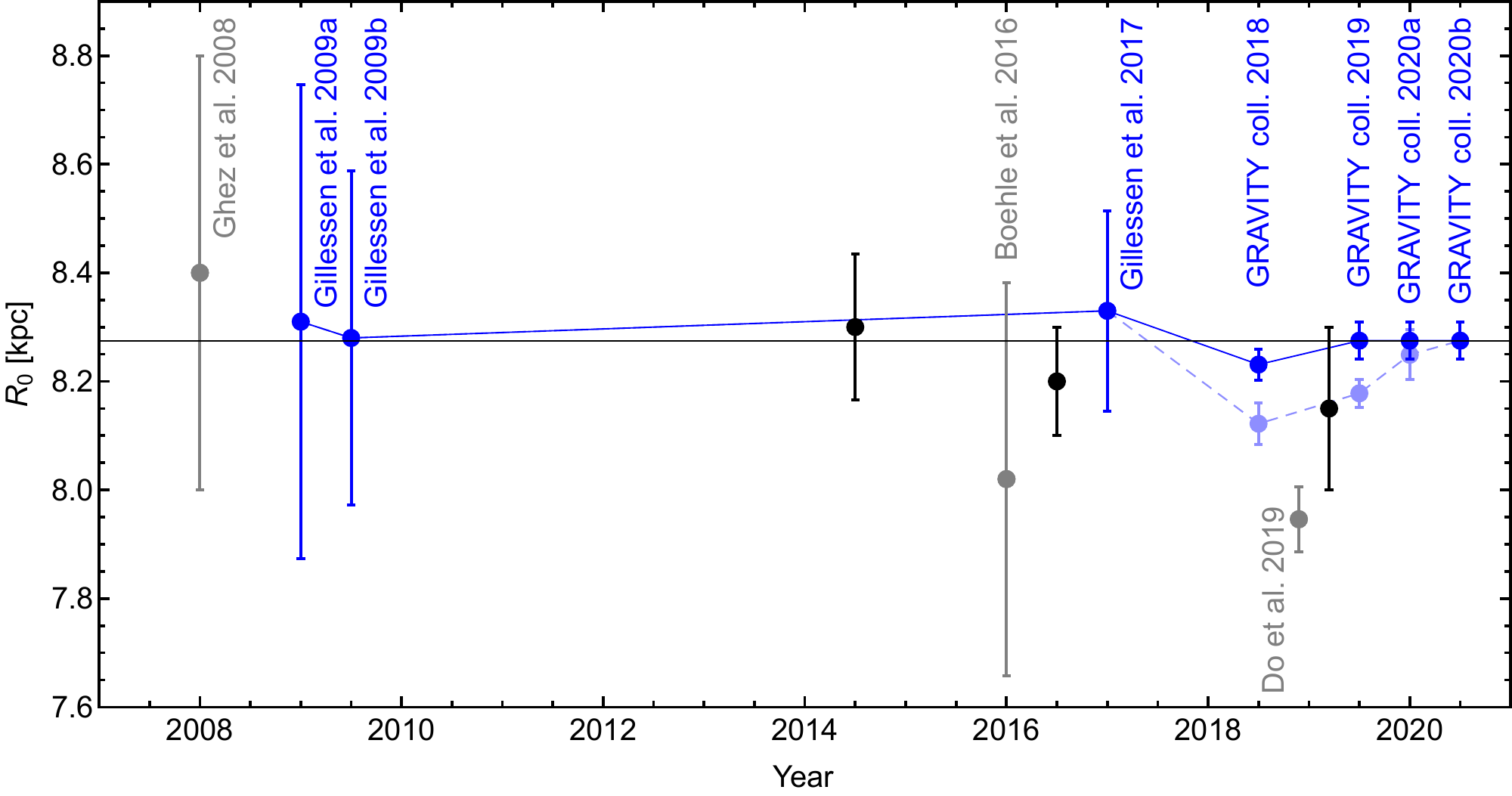}
\caption{Timeline of different measurements of the GC distance. Blue points show results obtained with the SINFONI, NACO, and GRAVITY data of the S2 star orbit with (dark blue) and without (light blue) application of the aberration corrections \cite{GravityCollaboration2021b}. Gray data points determinations are based on data from the Keck observatory. Black data points report measurements of the GC distance based on different methods not involving the S-stars: in $\sim$2015 using statistical parallax of the NSC \cite{Chatzopoulos2015}; in $\sim2016$ with a combination of different measurements \cite{BlandHawthorn2016}; and in $\sim$2019 from observations of molecular masers \cite{Reid2019}. Image is courtesy of \cite{GravityCollaboration2021b}.}
\label{fig:r_0_improvement}
\end{figure}

The values of $R_0$ measured by the UCLA group in 2008 and 2019 are reported in Figure \ref{fig:r_0_improvement} along with the values from the MPE/Gravity Collaborations. While in early studies, due to higher uncertainties on the measurements, the values derived by the two groups were compatible between each other, improved astrometry over the years led to a more than $5\sigma$ tension, $\Delta R_0 = 340\pm45$ pc, between the values from the two analysis (apart from $R_0$, the semi-major axis of the orbit of S2, its inclination and its longitude of the ascending node also seem to yield different values). The source of this discrepancy was studied in 2021 by the Gravity Collaboration \cite{GravityCollaboration2021b} and it seems to reside in \emph{(i)} a possible offset, drift and different orientations of the two reference frames (which was first observed in \cite{Gillessen2009b} by simply overlapping the datasets on the same plot); \emph{(ii)} a difference between the measured radial velocities of the stars from $\sim 2011$ on (especially noticeable for high values of the line-of-sight velocity, reached just before and right after the pericenter passage) which partly owes to the algorithm for radial velocity extraction from the spectrum used by the UCLA group (though a significant fraction of difference in the radial velocity has no conclusive explanation) and \emph{(iii)} a possible non-zero offset on the declination direction for measurements from the Keck telescope which ultimately yields a greater value for the semi-major axis, affecting the entire analysis.

\subsubsection{Direct observation of Sgr A*}
\label{sec:eht}
Despite possible systematics and uncertainties on the mass and distance of Sgr A*, observations of the S-stars provide with a clear evidence of the the presence of a dark compact mass in the GC beyond any reasonable doubt. Nonetheless, a definitive and direct proof for the existence of a SMBH in the GC requires a direct observation of peculiar features that uniquely belong to these objects. The characteristic size associated with a BH is its Schwarzschild radius \cite{Wald1984}
\begin{equation}
    r_s \equiv \frac{2GM}{c^2}\,,
    \label{eq:sch_radius}
\end{equation}
where $G$ is the gravitational constant, $c$ is the speed of light and $M$ is the BH mass. This quantity defines the radial coordinate where its event horizon is located. This surface is described as the boundary of the region of the space-time within which the escape velocity is greater than $c$ \cite{Wald1984}. The existence of an event horizon and effects of strong lensing (deflection) of photons due to the extreme gravitational field near a BH are assumed to produce distinctive features in the observed images of BHs \cite{Cunningham1973, Luminet1979, Jaroszynski1997, Falcke2000a, Melia2001}. Since the first pioneering simulations of the optical appearance of a rotating BH endowed with a thin accretion disk by J. Luminet in 1979 \cite{Luminet1979}, a noticeable feature of BH images stood out: a central brightness depression encircled by a bright emission ring, whose shape has been described as a ‘‘crescent'' because of fast rotation and relativistic beaming \cite{Falcke2000a, Bromley2001, Noble2007, Broderick2009a, Broderick2009b, Broderick2009c, Kamruddin2013, Lu2014}.

The central depression is what we refer to as the BH ‘‘shadow'' \cite{Falcke2000a}. By definition, this feature is a purely geometric property of space-time around a BH and does not depend on astrophysical effects. As a matter of fact, the boundary of the shadow on the image plane of a distant observer represent a critical curve, identified by points corresponding to photons, which, when traced back toward the BH, are tangent to the spherical surface where photons sit on unstable spherical orbits close to the horizon (the so-called photon sphere) \cite{Bardeen1973, Bardeen1974, Chandrasekhar1983}. When a photon reaches a distant observer with an impact parameter that is inside the boundary of the shadow, it necessarily has to belong to a path that crosses the horizon and consequently it possesses a shorter optical path through the space-time, if compared to photons that are outside of this boundary. This is what ultimately produces the brightness depression in the center of a BH image 
\cite{Jaroszynski1997, Johannsen2010, Narayan2019, Ozel2021, Bronzwaer2021, Kocherlakota2022}.

For the BH shadow to be observable, two conditions are required \cite{EventHorizonTelescopeCollaboration2022f}. Firstly, a bright source of photons is needed close to the BH, in order to expose the radiation field to the near-horizon strong gravitational lensing which produces the distinctive observational features. Secondly, the bright source is required to be optically thin at the observing wavelength, in order for the shadow not to be completely hidden by the material generating this radiation. Both radiatively efficient and inefficient accretion flows around BHs can satisfy both these conditions \cite{Luminet1979, Falcke2000a}.

Because of a fortunate cancellation of effects between the frame dragging and the quadrupole moment of the space-time, the shape and size of the shadow have a very weak dependence on the angular momentum of a BH and on the inclination of the observer \cite{Zakharov2005, Johannsen2010}. What ultimately determines the angular size of the shadow are predominantly the mass of the object and distance of the observer. More precisely the angular diameter $d_{\rm sh}$ of the shadow for a non-rotating BH results to be directly proportional to the mass-over-distance, $M/R$, ratio \cite{Chandrasekhar1983, Psaltis2015}
\begin{equation}
    d_{\rm sh} = 6\sqrt{3}\frac{GM}{c^2R}.
    \label{eq:shadow_diameter}
\end{equation}
The $M/R$ ratio for Sgr A* is precisely known from the three-decade-long effort of tracking S-stars orbits (Section \ref{sec:s_stars}) making its shadow a direct probe of the metric properties \cite{Psaltis2015}. Particularizing the expression of the angular diameter for Sgr A*, one obtains:
\begin{equation}
    d_{\rm sh} = 51.195\;\mu\textrm{as}\left(\frac{M_0}{4\times 10^6 M_\odot}\right)\left(\frac{D_0}{8\textrm{ kpc}}\right)^{-1}
\end{equation}

It is clear that the scales probed by near-infrared imaging, {while accurate enough to provide exquisite astrometric precision (even up to distances as small as the last stable circular orbit, \emph{e.g.} $\sim 0.12$ AU, for Sgr A* \cite{GravityCollaboration2018b})}, are much larger than the apparent size of Sgr A* and only much higher-resolution observations can be able to resolve features and perform imaging on the scale of the event horizon for Sgr A*. VLBI delivers the highest resolution images and sub-mas astrometry and for this reason Sgr A* has been observed with millimeter VLBI for over 25 years. As its name suggests, VLBI is an interferometric technique which takes advantage of radio-antennas placed at different locations on Earth (the spatial distance between each pair of telescopes corresponds to a baseline) which simultaneously collect signals coming from the same source. The different signals are then combined with a signal processing technique called aperture synthesis, to create a high resolution image of the source, whose angular resolution is, nominally, the same as that of a single dish with the size of the entire collection of antennas \cite{Jennison1958, Burke2009}. Sgr A* was observed via VLBI techniques at progressively shorter wavelengths. For wavelengths on the centimeter regime, the source size is entirely determined by scatter broadening by the ionized interstellar medium. This process has a know scaling relation with wavelength given by $\propto\lambda^2$. Going down to the millimeter regime, the observed angular size of the source starts to deviate from the scatter broadening dependence and at 7 mm and below the imprint of the intrinsic size of the source starts to became noticeable \cite{Rogers1994, Lo1998, Doeleman2001, Bower2004, Shen2005, Bower2006}. The first pioneering detection of the intrinsic structure of Sgr A* with VLBI happened in 1995 using a 980 km projected baseline between a 30 m telescope on Pico Veleta in Spain and a single antenna in France, at the Plateau de Bure Interferometer \cite{Krichbaum1997}. This observation at 1.4 mm wavelength detected a compact source, whose angular size was estimated to be $110\pm60\;\mu$as \cite{Krichbaum1998}, the large error being due to large calibration uncertainties. Successively, event-horizon-scale structures in Sgr A* were unambiguously detected for the first time with VLBI experiments in 2008 at a wavelength of 1.3 mm. This was possible with a wider recorded bandwidth and a three-station VLBI array consisting of the Arizona Radio Observatory 10-m Sub-millimetre Telescope (ARO/SMT) on Mount Graham in Arizona, one 10-m element of the Combined Array for Research in Millimeter-wave Astronomy (CARMA) in Eastern California, and the 15-m James Clerk Maxwell Telescope (JCMT) near the summit of Mauna Kea in Hawaii. The longer baselines of the experiment provided a tighter estimate of the source size: $43\pm14 \mu$as \cite{Doeleman2008}. Moreover, the 1.3 mm VLBI data could also be fit by a uniform thick bright ring of inner diameter $35\;\mu$as and outer diameter of $80\;\mu$as plus a shadow feature in the center, convolved with the interstellar scattering. This distribution was motivated by early imaging simulations of the Sgr A* accretion region with general relativistic ray tracing by Falcke and collaborators \cite{Falcke2000a}, who predicted that sub-mm VLBI could directly image the brightness depression related to the BH shadow {(see also \cite{Zakharov2005})}. 

\begin{figure}[!t]
    \centering
    \includegraphics[width = 0.7\textwidth]{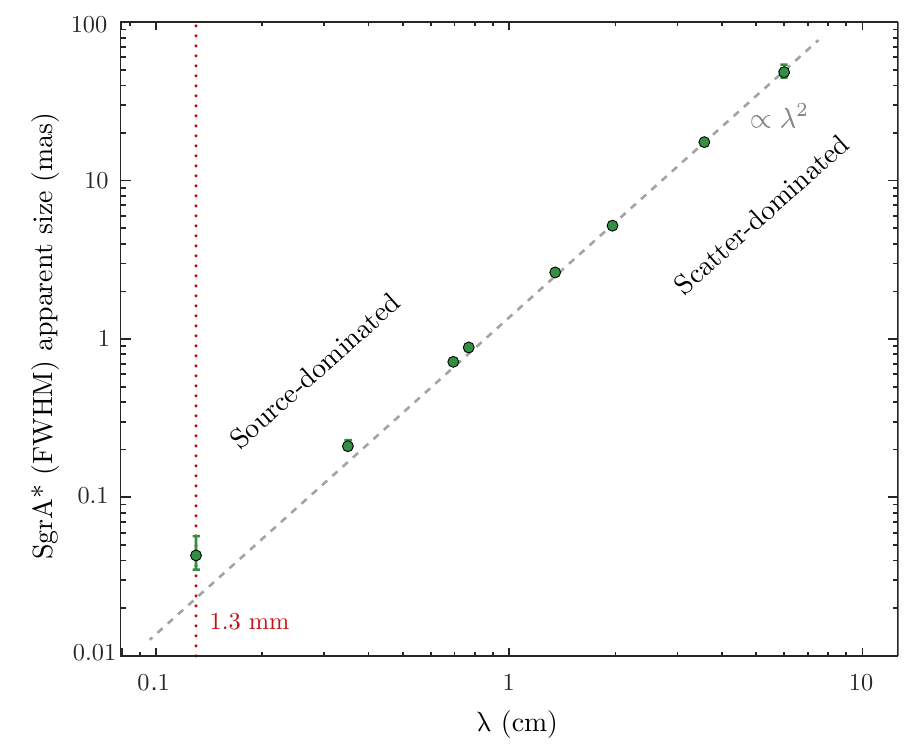}
    \caption{Determinations of the observed radio-size of Sgr A* as a function of wavelength across the cm- and mm-regimes. Various points correspond to progressively shorter-wavelengths experiments between 1997 and 2007 from \cite{Bower2004, Bower2006, Shen2005, Doeleman2008}. A break is evident in the mm regime (in correspondence with the 1.3 mm dotted red vertical line), where the observed size of Sgr A* stops being dominated by interstellar scattering (the dashed gray line is the $\lambda^2$ scattering broadening dependence, hence measurements that fall below or on this line are scatter-dominated) and observations measure the intrinsic size of the source at that wavelength (source-dominated regime).}
    \label{fig:sgra_vlbi}
\end{figure}

Observations of horizon-scale structures in Sgr A*, on one hand gave important constraints for theoretical models \cite{Broderick2009b, Dexter2009, Moscibrodzka2009}, on the other showed the feasibility of 1.3 mm VLBI techniques to peer into the innermost region of Sgr A*. This motivated theoretical advancement that allowed increasingly sophisticated general-relativistic simulations of accretion flows and radiative transfer of SMBH, reproducing realistic images for a great variety of near-horizon emission models, alternative theories of gravity (see Section \ref{sec:alternative_theories}) and alternative to BHs \cite{Broderick2006b, Moscibrodzka2009, Bambi2009, Dexter2012, Dibi2012, Grenzebach2014, Chan2015, Moscibrodzka2016, Younsi2016, Vincent2016a, Porth2017, Mizuno2018, Chael2018a, Ryan2018, Davelaar2018, Olivares2020,Roder2022IP,Roder2022,Zakharov2022b}. In parallel, VLBI studies of Sgr A* at 1.3 mm wavelength with progressively enhanced arrays, allowed a better characterization of the compact emission, its variability, significant polarization (which suggest the presence of intense magnetic field in the emitting region) and asymmetry in the image structure \cite{Fish2011, Johnson2015a, Lu2018}.

\begin{figure}[!t]
    \centering
    \includegraphics[width = 0.9\textwidth]{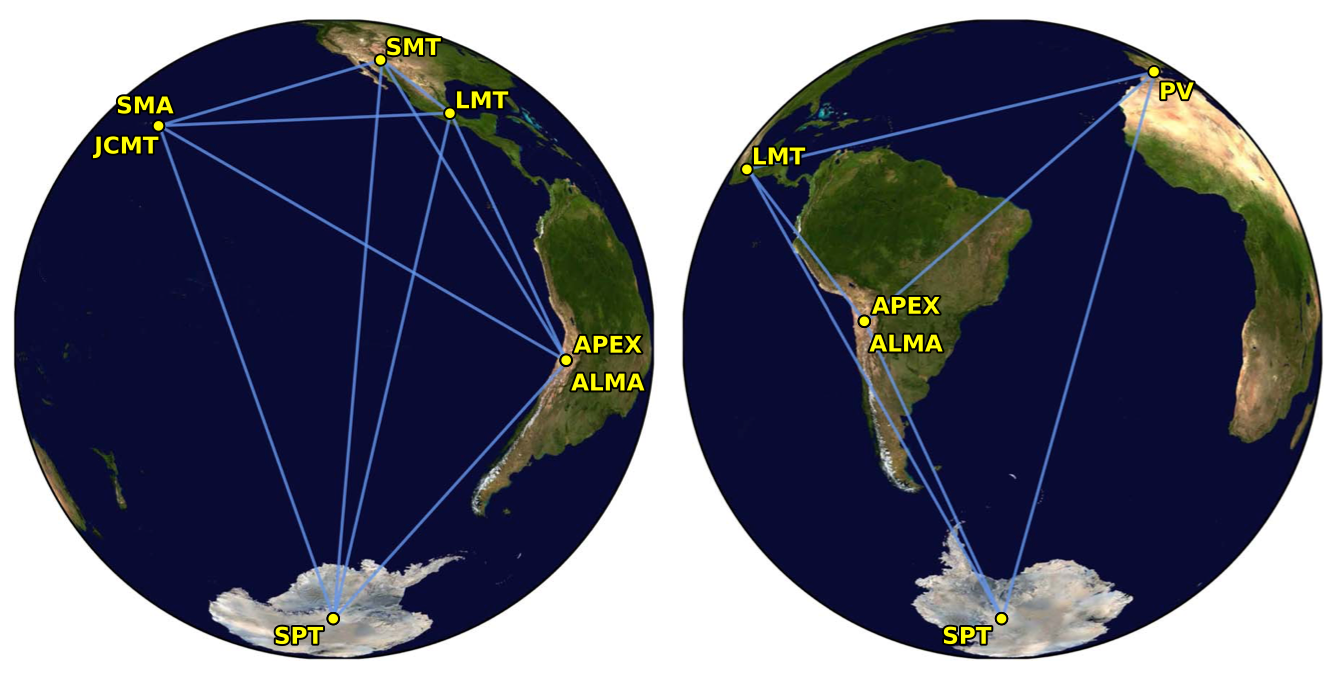}
    \caption{The EHT global VLBI array includes the Atacama Large Millimeter/submillimeter Array (ALMA) \cite{Matthews2018, Goddi2019} and the Atacama Pathfinder Experiment telescope (APEX) \cite{Wagner2015}, both in Chile; the Large Millimeter Telescope Alfonso Serrano (LMT) in Mexico \cite{OrtizLeon2016}; the IRAM 30 m telescope on Pico Veleta (PV) in Spain \cite{Greve1995}; the Submillimeter Telescope Observatory in Arizona (SMT) \cite{Baars1999}; the James Clerk Maxwell Telescope (JCMT) and the Submillimeter Array (SMA) in Hawaii \cite{Doeleman2008,Primiani2016, Young2016}; and the South Pole Telescope (SPT) in Antarctica \cite{Kim2018b}.}
    \label{fig:eht_network}
\end{figure}

\begin{figure}[!t]
    \centering
    \includegraphics[width=\textwidth]{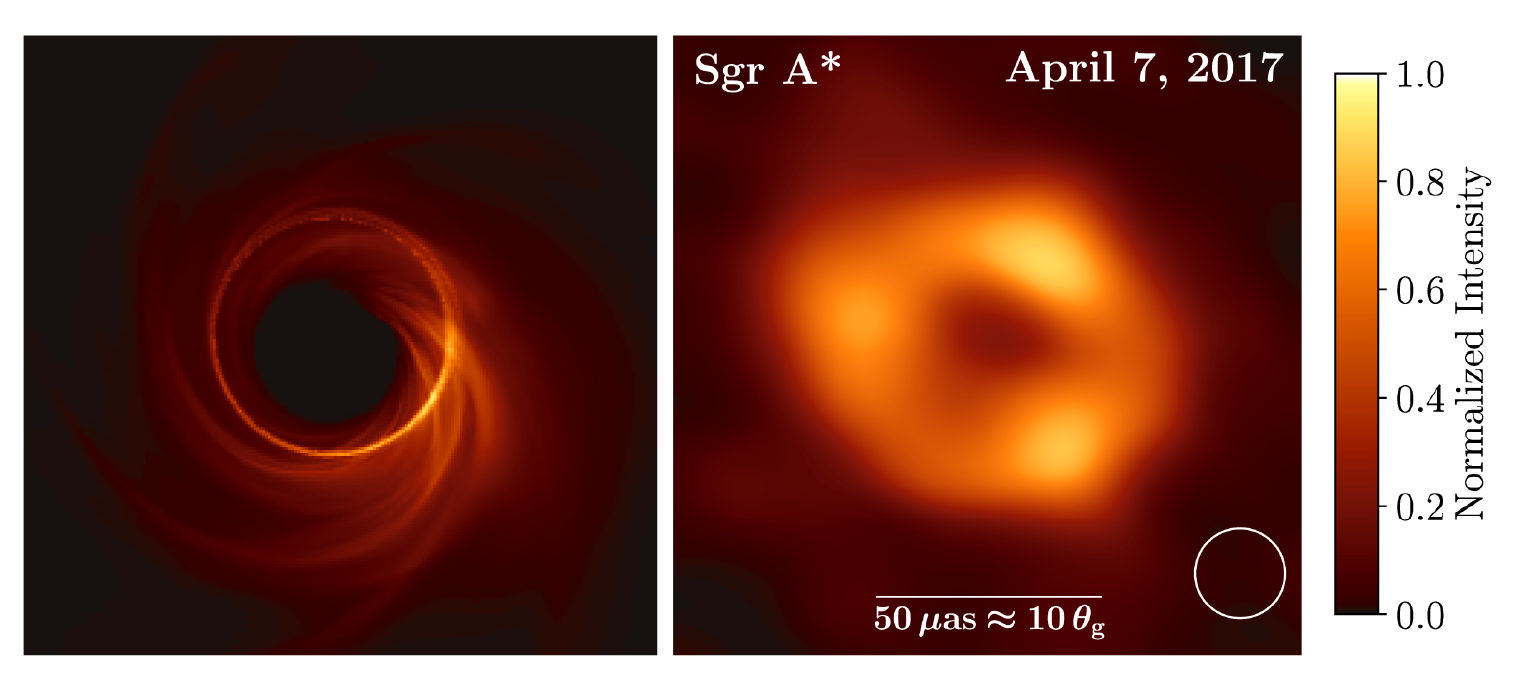}
    \caption{\emph{Left panel:} example from the library of synthetic images of Sgr A* resulting from state-of-the art general-relativistic magneto-hydrodynamics numerical simulations \cite{EventHorizonTelescopeCollaboration2022e}; \emph{Right panel:} Representative EHT image of Sgr A* from observations on April 7, 2017 \cite{EventHorizonTelescopeCollaboration2022a}, resulting from an average over different reconstruction methodologies and reconstructed morphologies. Color denotes the specific normalized intensity of the source. The reconstructed image shows a prominent ring morphology, but while the diameter and thickness of the ring is consistent across reconstructions from different methodologies, the azimuthal structure is poorly constrained. Courtesy of \cite{EventHorizonTelescopeCollaboration2022a}.}
    \label{fig:eht_obs}
\end{figure}

Starting in 2009, the Event Horizon Telescope (EHT) collaboration was established with the goal to build a global 1.3 mm VLBI network of antennas for achieving the angular resolution required to resolve the shadow of Sgr A* (and of M87* the SMBH in the center of the active galaxy M87 which being much more massive, but at much greater distance, happens to has a similar $M/R$ ratio to Sgr A* and thus the same size on the sky). The EHT directly measures Fourier components (each baseline corresponds to a specific component in Fourier space) of the radio brightness distribution on the sky. Moreover, as Earth rotates, telescope pairs samples a range of spatial frequencies. An array of VLBI antennas is endowed with a nominal angular resolution given by the $\lambda/L$ ratio, where $\lambda = 1.3$ mm is the observing frequency and $L$ corresponds to the maximum projected baseline in the network \cite{Thompson2017}, thus allowing EHT to create a virtual telescope whose size is nearly the full diameter of the Earth. {Remarkably, this idea was first discussed in the 1970s by Matveenko in a US-USSR combined effort to set up a VLBI experiment (see for example discussion in \cite{Matveenko2007a, Zakharov2022a}).} In order to achieve this goal, telescopes at large distances need to simultaneously sample the radiation field from the source and record it coherently. Temporal synchronization over the entire network is paramount for the success of the observation. For this reason hydrogen maser frequency standards at all EHT sites combined with Global Positioning System (GPS) synchronization were required to achieves temporal alignment of the recordings within tens of nanoseconds. After observations, recordings are stored on hard drives and physically shipped at a central location, where the data analysis (temporal alignment and cross-correlation for signals from each telescope-pair) is performed \cite{EventHorizonTelescopeCollaboration2019b}. The great challenge faced by the EHT Collaboration was the extension of known and established VLBI techniques at cm wavelengths down to 1.3 mm. Observations at shorter wavelengths are more affected by noise in the electronics of radio receivers, a decreased efficiency and size of radio telescopes in the mm and sub-mm bands and, finally, a greater impact of Earth's atmosphere on the observations, both in terms of a higher opacity and increased phase fluctuations due to atmospheric turbulence \cite{Doeleman2009}.  The baselines range and coverage, along with the overall collecting area of EHT were increased by upgrading radio-observations facilities around the globe: the Atacama Large Millimeter/submillimeter Array (ALMA) \cite{Matthews2018, Goddi2019} and the Atacama Pathfinder Experiment telescope (APEX) \cite{Wagner2015}, both in Chile; the Large Millimeter Telescope Alfonso Serrano (LMT) in Mexico \cite{OrtizLeon2016}; the IRAM 30 m telescope on Pico Veleta (PV) in Spain \cite{Greve1995}; the Submillimeter Telescope Observatory in Arizona (SMT) \cite{Baars1999}; the JCMT and the Submillimeter Array (SMA) in Hawaii \cite{Doeleman2008,Primiani2016, Young2016}; and the South Pole Telescope (SPT) in Antarctica \cite{Kim2018b} (Figure \ref{fig:eht_network}). This effort allowed a $\sim$30-fold improvement in sensitivity achieved by the EHT with respect to the early experiments at 1.3 mm wavelenght \cite{Doeleman2008, Doeleman2012, Akiyama2015, Johnson2015a, Fish2016, Lu2018}. In 2017, the EHT was first operated as a VLBI array of eight stations over six geographical locations, with baseline ranging from 160 m to 10700 km, achieving a theoretical diffraction-limit resolution of $\sim25\;\mu$as, with main scientific goal of resolving the shadows of M87* and Sgr A*. While the results for M87* were published only two years later in 2019 \cite{EventHorizonTelescopeCollaboration2019a,EventHorizonTelescopeCollaboration2019b,EventHorizonTelescopeCollaboration2019c,EventHorizonTelescopeCollaboration2019d,EventHorizonTelescopeCollaboration2019e,EventHorizonTelescopeCollaboration2019f}, the data analysis for the Sgr A* represented a much greater challenge owing mainly to the difference in masses of the two sources which is directly related to the physical size of the event horizon and consequently to their variability timescales. As a matter of facts, the period of the innermost stable circular orbit (ISCO) gives an approximate of dynamical timescale for matter orbiting and accreting onto the BH. For M87*, this time lies in a range from 5 days to 1 month (depending on the orientation of the revolution, either prograde or retrograde \cite{Chandrasekhar1983}), so the source emission is expected to be constant over the course of an observation night. However, for Sgr A*, the range is only 4–30 minutes, so the source structure can evolve within a single night. After a great scientific effort, in 2022 the EHT collaboration published the much anticipated imaging results of the SMBH in the center of the MW \cite{EventHorizonTelescopeCollaboration2022a, EventHorizonTelescopeCollaboration2022b, EventHorizonTelescopeCollaboration2022c, EventHorizonTelescopeCollaboration2022d, EventHorizonTelescopeCollaboration2022e, EventHorizonTelescopeCollaboration2022f}. The EHT data resolve a compact emission region with intra-hour variability. All imaging and modeling analyses detect evidence of a source dominated by a bright, thick ring with a diameter of $51.8\pm2.3\;\mu$as with an asymmetric azimuthal brightness and a central brightness depression (Figure \ref{fig:eht_obs}). Comparison with the results from a large suite of general-relativistic magneto-hydrodynamics numerical simulations \cite{EventHorizonTelescopeCollaboration2022e}, provides with strong evidence towards a source consistent with a Kerr BH of four-million-solar-masses BH at a distance of $\sim8$ kpc. This breakthrough observation allowed for the first time to connect the predictions from dynamical measurements of S-stars on scales of $10^3-10^5$ gravitational radii (and at even larger scales from gas dynamics) to images on the event-horizon-scale. 

In parallel to VLBI increasingly-precise observations, Sgr A* was also being extensively studied at other wavelengths. Sgr A* is characterized by a flat/inverted radio spectrum, which is usually associated with compact jet emission \cite{Blandford1979} in low-luminosity active galactic nuclei (LLAGNs) \cite{Ho1999, Nagar2000}. X- and gamma-ray observations of the GC with the instruments ROSAT and Sigma/GRANAT respectively \cite{Goldwurm1994, Predehl1994} could not identify a bright central source, indicating that Sgr A* is either obscured or very faint, if compared to other known LLAGNs. Subsequent radio, millimeter, infrared, and X-ray (in which Sgr A* was finally observed in 2001 with Chandra \cite{Baganoff2001}) observations allowed to place very low upper limits on the Sgr A* bolometric-to-Eddington luminosity ratio of $L/L_{\rm Edd} \sim 10^{-9}$ \cite{Genzel2010} and on its mass accretion rate of $\dot{M}\sim10^{-9}-10^{-7} M_\odot \textrm{yr}^{-1}$ \cite{Baganoff2003, Marrone2006, Marrone2007, Shcherbakov2012, YusefZadeh2015}. In particular, in the X-ray Sgr A* appears as a persistent, albeit very feeble, source whose emission mechanism has been identified to be thermal bremsstrahlung originating from hot plasma near the Bondi radius \cite{Quataert2002, Baganoff2003, Yuan2003, Liu2004, Wang2013}. The particularly faint X- and gamma-ray emission poses several question on the nature of Sgr A* accretion properties. Two possible scenarios have been proposed: \emph{(i)} the emission originates in inflow of accreting matter. In this case either the plasma is strongly radiatively inefficient \cite{Narayan1995a} or the accretion rate has to be very small compared to what is captured, or perhaps a combination of the two effects \cite{Blandford1999, Narayan2000, Quataert2000a, Yuan2014}; \emph{(ii)} the emission is given rise in the plasma outflow, which would be favoured by a small accretion rate \cite{Falcke1993}, even during flares \cite{Markoff2001}. Today, the emitting region of Sgr A* is described as a weakly bound, magnetized accretion flow that is so diffuse that the electron and ion temperatures are unable to remain strongly coupled. This picture is consistent with its low luminosity, low radiative efficiency, and its weak Faraday rotation \cite{EventHorizonTelescopeCollaboration2022a}.

Finally, Sgr A* exhibits flares at most wavelengths and continuous variability in the millimeter band. In particular, X-ray flares are observed about once per day and are characterized by non-thermal emission centered on the central source \cite{Neilsen2013} and are associated with flares in the near-infrared \cite{Eckart2004} whose emission peaks are observed more frequently than in the X-ray \cite{Genzel2003, Ghez2004, GravityCollaboration2020b}. Near-infrared flares have also been tracked astrometrically with the GRAVITY interferometer \cite{GravityCollaboration2021b}. Variability of the X-ray and near-infrared emission is detected on timescales of hours, suggesting an origin of the flaring emission from within $\sim5r_s$ of the SMBH, consistent with emission originating near the ISCO of the BH. Thus, multiwavelength observations during the EHT campaign provide with an unprecedented opportunity to connect the variability of the source with changes observed at horizon scales.


\section{Probing gravity in the Galactic Center}
\label{sec:gravity_tests}

\begin{figure}[!ht]
    \centering
    \includegraphics[width = 0.7\textwidth]{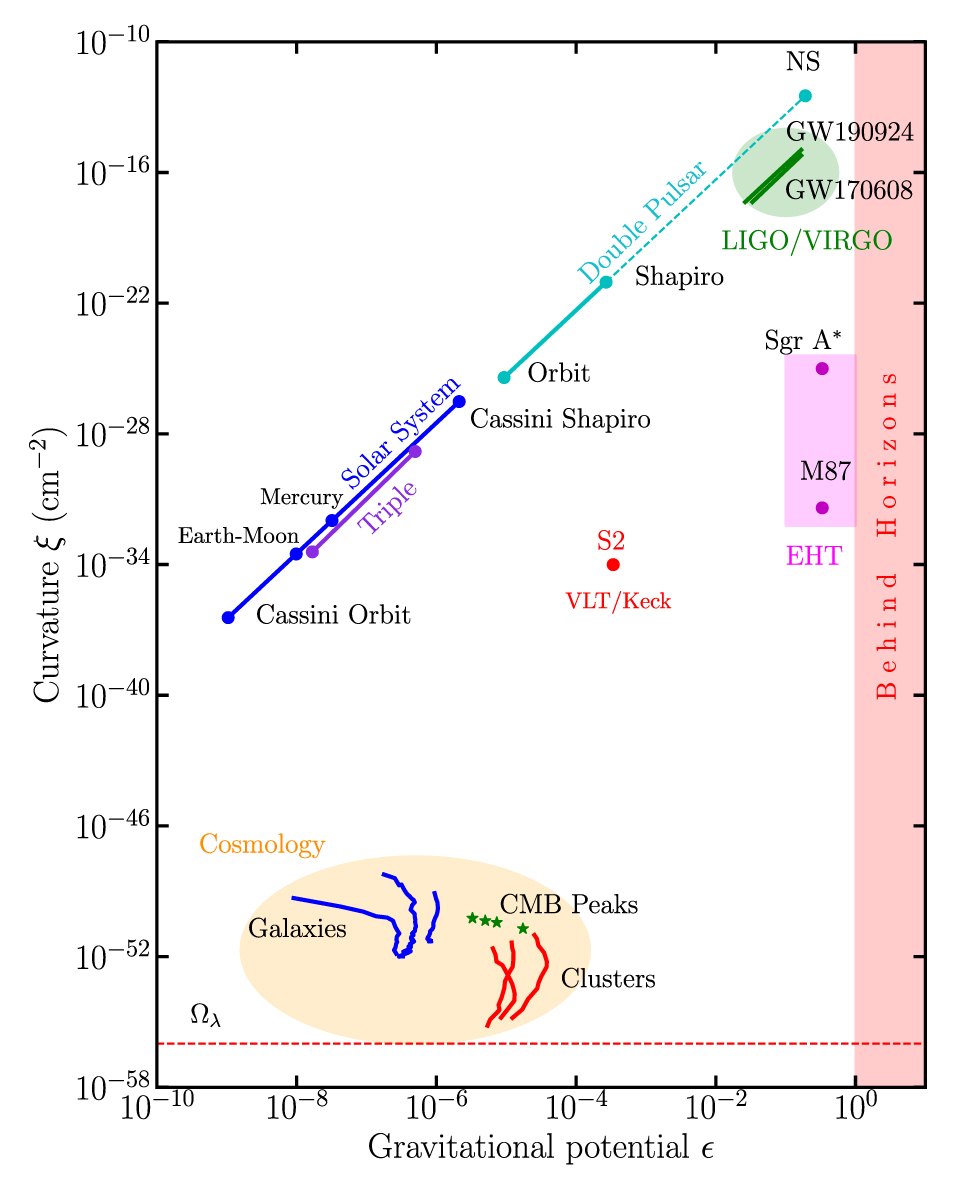}
    \caption{\small A parameter space of tests of gravity with astrophysical and cosmological systems. In particular, the x-axis reports the Gravitational potential $\epsilon = GM/rc^2$ evaluated at its maximum for each gravitational test, and the y-axis the maximum corresponding curvature for each probe is reported. The curvature is quantified by the Kretschmann scalar $\xi = (R^{\alpha\beta\gamma\delta}R_{\alpha\beta\gamma\delta})^{1/2} = \sqrt{48}GM/r^3c^2$ particularized for a Schwarzschild metric \cite{Baker2015}. For the Solar System and pulsar tests, the straight lines connect the range of gravitational fields that could affect, in principle, the outcome of each test, from the location of the outermost probe to the location of the central massive object; the dashed region of the cyan line indicates that the connection to the largest curvatures is theory specific. The green lines connect the range of gravitational fields probed by two gravitational-wave tests with BH inspirals. Filled areas show the typical range of gravitational fields probed by cosmological (orange), gravitational- wave (green), and BH (magenta) imaging tests. Even though different tests explore, in principle, different aspects of the gravitational theory, they also probe vastly different scales. In particular, the horizon-scale images of Sgr A* and the S2 star orbiting around it at its pericenter prove a region of the parameter space that is not explored by other probes. Image is courtesy of \cite{EventHorizonTelescopeCollaboration2022f}.}
    \label{fig:eht_gravitational_probes}
\end{figure}

\subsection{Tests of General Relativity around Sgr A*}
\label{sec:test_GR}

Since the first measurements of the proper motions of the S-stars orbiting Sgr A* in the GC, it became clear that tracking the orbits of such stars could be an efficient way to probe the BH paradigm and to constrain the distribution of dark matter around Sgr A* \cite{Jaroszynski1998}. The high eccentricities of these stars, lead to very close pericenter passages (up to a few hundreds of gravitational radii of Sgr A*) and extreme orbital velocities at apocenter ($\sim 10^4$ km/s), that can possibly provide a groundwork for interesting tests of GR, the greater the observational accuracy the more stringent such tests. The first studies where such low-order GR effects of transversal Doppler shift or the gravitational redshift (section \ref{sec:grav_redshift}) and pericenter advance (section \ref{sec:orbital_precession}) were taken into account date back to late 1990s - early 2000s \cite{Jaroszynski1998, Fragile2000, Rubilar2001, Weinberg2005a, Zucker2006} when the observational accuracy at the GC was still poor and such effects could only be theoretically estimated. Moreover, the effects that were considered were the ones affecting the sky-projected proper motions of the stars, due to lack of radial velocities observations at the time. Subsequent theoretical studies have considered high-order relativistic effects (section \ref{sec:higher_order_effects}) such as the frame-dragging effect of Lense-Thirring for a rotating BH, effects related to the BH quadrupole moment  \cite{Kraniotis2007, Will2008, Kannan2009, Merritt2010b, Angelil2010a, Angelil2010b}, affecting both the astrometry and the spectroscopic observables, and the detection of gravitational lensing by Sgr A* \cite{Bozza2004, Bozza2005, Bozza2012}. In Table \ref{tab:GR_literature} we report a complete chronologically-ordered summary of all theoretical studies where relativistic effects have been taken into account on the orbits of the S-star.

\begin{table*}
    \small
        \newcolumntype{C}{>{\centering\arraybackslash}X}
        \renewcommand{\arraystretch}{1.7}
        \resizebox{\textwidth}{!}{
        \parbox{1.4\linewidth}{
        \begin{tabularx}{\linewidth}{|l|l|XX|r|} 
            \hline
            Metric & GR effects & Other effects & Observables & Ref. \\\hline
                        Kerr &  RP, LT & lensing     & time averaged orbital effects & \cite{Jaroszynski1998} \\ 
                        Schwarzschild & GZ, RP, LT  &      & astrometric orbital fitting, time averaged orbital effects, spectroscopic  & \cite{Fragile2000} \\
                        PN & RP & NP from extended mass & astrometric orbits  &\cite{Rubilar2001} \\
                        Schwarzschild, Kerr & GL & comparison of full-GR GL with retro-lensing and strong-field-lensing approximations & astrometric shift and magnification &  \cite{Bozza2004} \\
                        Schwarzschild & GL & & secondary and relativistic lensed images &  \cite{Bozza2005} \\
                        PN & RP, LT & NP from extended mass, stellar remnants & time averaged orbital effects  &\cite{Weinberg2005a} \\
                        PN & GZ  &      & astrometric orbits and spectroscopic  &\cite{Zucker2006} \\
                        Kerr & RP, LT & Kerr-de Sitter gravitational field     & time averaged orbital effects &\cite{Kraniotis2007} \\ 
                        PN & RP, LT, NP &stellar cluster     & pericentre shift &\cite{Nucita2007} \\ 
                        PN & RP, LT, QP & testing ‘‘no-hair'' theorem & time averaged orbital effects & \cite{Will2008} \\ 
                        Kerr & RP, LT &   & integrated orbital effects, spectroscopic & \cite{Kannan2009} \\ 
                        PN & RP, LT &NP from extended mass   &general astrometric, spectroscopic & \cite{Preto2009} \\ 
                        PN & GZ, RP, LT  & light-path effects  &spectroscopic & \cite{Angelil2010a} \\ 
                        PN & GZ, LT, QP  & pulsar timing  &spectroscopic & \cite{Angelil2010b} \\ 
                        PN & RP, LT, QP & NP from extended mass  & astrometric & \cite{Merritt2010b} \\ 
                        PN & RP, LT, QP & NP from extended mass  & spectroscopic & \cite{Iorio2011a} \\ 
                        PN & RP, LT, QP & gravitational waves  & time averaged orbital effects & \cite{Iorio2011b} \\ 
                        PT& LT, QP & perturbing effects of cluster stars  & time averaged orbital effects & \cite{Sadeghian2011} \\
                        Schwarzschild & GL & & astrometric shift &  \cite{Bozza2012} \\
                        Kerr  &full GR & full lensing    &astrometric, spectroscopic, BH spin measurement  & \cite{Zhang2015}  \\    
                        PN & RP, LT, QP  & NP, EHT, pulsars, ‘‘no-hair'' theorem    & astrometric  & \cite{Psaltis2016}  \\
                        Kerr & full GR  & full lensing    & astrometric, spectroscopic, BH spin inclination effects  & \cite{Yu2016}  \\     
                        Kerr   & full GR & lensing primary & astrometric, spectroscopic & \cite{Grould2017b}  \\       
                        PN & RP, LT, QP  & effect of orbital eccentricity  & periods evolution &\cite{Iorio2017a}  \\
                        PN & RP, LT  &  &spectroscopic, time averaged orbital effects  &\cite{Iorio2017b}  \\
                        PN & RP &&astrometric&\cite{Parsa2017}\\
                        Kerr & full GR & full lensing, Newtonian perturbations  & astrometric, spectroscopic, time averaged orbital effects &\cite{Zhang2017a}\\
                        \hline
        \end{tabularx}
        }}

    \caption{Timeline of works in literature in which GR and other effects on the orbits of the S-stars have been taken into account. The table is inspired by Table 1 in \cite{GravityCollaboration2019b}. We employ the following abbreviations: post-Newtonian (PN), orbital perturbation theory (PT), gravitational redshift (GZ), Newtonian precession (NP), relativistic precession (RP), Lense-Thirring precession (LT), quadrupole precession (QP), gravitational lensing (GL).}\label{tab:GR_literature}
\end{table*}

When the S2 star passed for the last time at its pericenter in 2018, astronomers at the UCLA group and MPE/Gravity Collaboration (see Section \ref{sec:s_stars_obs}) were prepared with technologically advanced instruments to detect signals up the lowest order gravitational effects like the gravitational redshift \cite{Do2019a, GravityCollaboration2018a} and relativistic orbital precession \cite{GravityCollaboration2020c}. The chances of successfully detecting higher order effects will be much improved if with future observations (for example the 39-meter ELT \cite{Davies2021} or TMT \cite{Weinberg2004}) we could be able to detect fainter stars on even tighter and more eccentric orbits. As a matter of fact, the tighter the orbit, the larger PN effects are, and the smaller the uncertainties of the unknown amount of extended mass around the SMBH gets.
A useful parameter to quantify how relativistic effects are evident on the orbit of a a celestial body, is the relativistic parameter $\Gamma \equiv r_{s}/r$ where $r_s$ is the gravitational radius of a massive object given in Equation \eqref{eq:sch_radius}, and $r$ is the distance of a body orbiting around it. The greater the relativistic parameter, the more pronounced deviations from Newtonian dynamics get. The relativistic parameter changes along the orbit and it is maximized at pericenter, $r_p$  (the closest point along an orbit), where the object dives deeper into gravitational field of the central massive object and is also faster (due to Kepler's second law), thus exhibiting special relativistic effects. 
The S2 star orbiting around Sgr A* has a relativistic parameter of $\Gamma \sim 7\times10^{-4}$ (we report the relativistic parameters for several S-stars in Table \ref{tab:sstars}). For comparison, the relativistic parameter on the surface of a white dwarf
is $\sim10^{-4}$ and on a neutron star $\sim10^{-1}$. This hints the fact that S-stars at the GC are supposed to exhibit PN effects. When considering a mildly relativistic orbit, \emph{i.e.} when the dynamical effects of deviation from a Keplerian orbit are very small compared to the orbital motion itself, one can express the relativistic parameter in terms of energy conservation, by applying Kepler's equation:
\begin{align}
        \frac{v^2}{c^2}-\frac{2GM}{c^2r} &= -\frac{GM}{c^2a}\qquad\Rightarrow\qquad
    \beta^2-\Gamma = \Gamma_0\qquad\Rightarrow\qquad \Gamma = \beta^2+\Gamma_0
    \label{eq:kepler_eq}
\end{align}
where $\beta$ is the Lorentz parameter $\beta\equiv v/c$ and $\Gamma_0$ is the relativistic parameter evaluated at the semi-major axis $a$ of the orbit. For highly eccentric orbits like the S-stars (we refer to Section \ref{sec:s_stars} for more details), the star at pericenter is much closer to the central object than its semi-major axis, thus $\beta^{2}\gg\Gamma_{0}$
and, hence, $\Gamma\sim\beta^{2}$ implying that the star's orbit is mostly perturbed by its nearly-Newtonian motion. When $\Gamma\!\ll\!1$ one can expand quantities in a series of power of $\beta$, each term in the series corresponding to progressively high PN-order effects.

\begin{table*}[!t]
    \small
    \setlength{\tabcolsep}{5pt}
    \renewcommand{\arraystretch}{1.5}
    \resizebox{\textwidth}{!}{
    \begin{tabular}{|r|ccccc|cc|ccr|}
        \hline
        Star & $T$ (yr) & $a$ (AU) & $e$ & $r_p$ (AU) & $r_A$ (AU) & $\Gamma$ ($10^{-4}$) & $v_p$ (km/s) [$\beta$] & c$\beta_r$ (km/s) & $\Delta\omega$ (arcmin) & $\Delta p_a$ (mas)\\ \hline
        S4714 & 12.00 & 841 & 0.985 & 13 & 1670 & 66.67 & 24386 [0.081] & 1984 & 108.82 & 6.409\\
        S62 & 9.90 & 746 & 0.976 & 18 & 1475 & 46.96 & 20421 [0.068] & 1391 & 77.00 & 4.006\\
        S14 & 55.30 & 2361 & 0.976 & 56 & 4666 & 14.91 & 11505 [0.038] & 442 & 24.44 & 4.022\\
        S4716 & 4.00 & 400 & 0.756 & 98 & 702 & 8.62 & 8247 [0.028] & 227 & 15.90 & 0.394\\
        S2 & 16.05 & 1031 & 0.885 & 119 & 1944 & 7.07 & 7738 [0.026] & 200 & 12.16 & 0.833\\
        S4711 & 7.60 & 619 & 0.768 & 143 & 1094 & 5.86 & 6825 [0.023] & 155 & 10.74 & 0.414\\
        S12 & 58.90 & 2463 & 0.888 & 275 & 4651 & 3.06 & 5093 [0.017] & 87 & 5.25 & 0.861\\
        S8 & 92.90 & 3337 & 0.803 & 657 & 6018 & 1.28 & 3221 [0.011] & 35 & 2.30 & 0.488\\
        S13 & 49.00 & 2178 & 0.425 & 1252 & 3104 & 0.67 & 2074 [0.007] & 14 & 1.53 & 0.167\\
        S1 & 166.00 & 4907 & 0.556 & 2179 & 7635 & 0.39 & 1643 [0.005] & 9 & 0.80 & 0.217\\ \hline
    \end{tabular}
    }
    \caption{Main orbital parameters for some of the S-stars, along with the corresponding relativistic parameters and effects. In particular, columns 2 to 6 report the orbital period, semi-major axis, eccentricity and corresponding pericenter and apocenter distances, respectively. Columns 7 and 8 report the relativistic parameter $\Gamma = r_s/r_p$ and the pericenter velocity (with the corresponding Lorentz factor $\beta = v/c$). Columns 9 to 11 report the corresponding special + general relativistic additional redshift contribution at pericenter, the rate of orbital precession per revolution and the corresponding astrometric shift computed at apocenter, respectively. Orbital data for the stars S62, S4711, S4716 and S4717 result from independent analyses by Peißker and collaborators \cite{Peissker2020, Peissker2022} with near-infrared observations at VLT/NACO. All the other stars have been observed by the MPE/Gravity Collaboration group \cite{Gillessen2017, GravityCollaboration2022b}}
    \label{tab:sstars}
\end{table*}

An effect that is important to take into account when modeling the orbital motion of the S-stars around Sgr A* is the Rømer delay. It is a classical effect (not a consequence of GR) caused by the fact that the time it takes for light to travel on a straight line from the star to an observer changes along the orbit, when the orbital plane is not observed face-on, \emph{i.e.} when the orbit possesses an inclination. The precise amount of delay related to this effect depends on the orbital parameters and on the orbital phase, and leads to apparent deviations from a Newtonian orbit \cite{Loeb2003}. A simplified calculation considering a circular edge-on orbit can give an idea of the maximum contribution and of the $\beta$-order of this effect. In particular, suppose that the emitting object and the observer are in opposition with respect to the central object. This means that the light emitted comes with a delay given by the light crossing time of the diameter of the orbit. This corresponds to:
\begin{equation}
    \Delta t_{\textrm{Rømer}} \sim \frac{2a}{c}\,.
\end{equation}
This means that observations performed at a time $\bar{t}$ would show the star at the position and velocity it was at time $\bar{t}-\Delta t_{\textrm{Rømer}}$. Consequently, given that the star's Newtonian acceleration is given by $\dot{v}\sim GM/a^2$, the shift observed in the velocity can be approximated by
\begin{equation}
    \Delta v \sim = \dot{v}\Delta t_{\textrm{Rømer}} = \frac{2GM}{ac},
\end{equation}
whose corresponding redshift difference is 
\begin{equation}
    \beta_r \sim \frac{\Delta v}{c} = \frac{2GM}{ac^2} \sim 2\beta^2,
\end{equation}
where Kepler's Equation \eqref{eq:kepler_eq} has been applied to a circular orbit. This calculation shows that the Rømer time delay affect spectroscopic measurements  of the star's radial velocity at order $\mathcal{O}(\beta^{2})$
with a pre-factor $B_{R}$ (2 in our simple calculation) that is orbit- and phase-depending. This effect is typically larger than that of either the transverse Doppler shift or the gravitational redshift (Section \ref{sec:grav_redshift}) and thus it is important to properly quantify Rømer delay if one wants to the detect the signal from the gravitational redshift.
The Rømer delay also affect the observed sky-projected position by an amount
\begin{equation}
    \Delta s = v\Delta t_{\textrm{Rømer}} = 2a\frac{v}{c} \propto \beta,
\end{equation}
determining a much stronger (first $\beta$-order) impact.

\subsubsection{The gravitational redshift}
\label{sec:grav_redshift}

With the advent of AO-assisted near-infrared spectroscopy of the S-stars in the GC (see Section \ref{sec:s_stars_obs}), the measurement of the radial velocities of the star has become the main candidate to exhibit PN effects on the orbital motion of the stars. The apparent radial velocity $v_r$ of the S2 star is measured by spectroscopic observations of the photon's frequency shift
\begin{equation}
	\beta_r= \frac{v_R}{c}=\frac{\Delta \nu}{\nu} = \frac{\nu_{\rm em}-\nu_{\rm obs}}{\nu_{\rm obs}}\,.
\end{equation}
The main sources of this shift are the special relativistic Doppler effect, $\beta_D$,  and the gravitational redshift $\beta_{G}$. This two effects are produced by the high relative velocity between the star and a distant observer (which for S2 can be as high as $\approx 7650$ km/s \cite{GravityCollaboration2018a} at pericenter), and by space-time curvature in the vicinity of a massive compact object. In particular, assuming that \emph{(A)} the observer is in a region where the space-time curvature is negligible (which is applicable to an infinitely distant observer in an asymptotically flat space-time), and \emph{(B)} the relative motion between the observer and the star is totally ascribable to the motion of the star around the central object (\emph{i.e.} proper motion of the observer is negligible), the total redshift can be expressed as:
\begin{align}
	\beta_r +1 = \underbrace{\frac{1}{\sqrt{-g_{00}(t_{\rm em},\vec{x}_{\rm em})}}}_{\beta_{G}}\cdot\underbrace{\frac{1-\vec{k}\cdot\vec{v}(t_{\rm em})}{\sqrt{1-\frac{v^2(t_{\rm em})}{c^2}}}}_{\beta_{D}}\,,
\end{align}
where $g_{00}$ is the time component of the space-time metric, $\vec{v}$ is the spatial velocity of the star, and $\vec{k}$ is the spatial part of the wave vector of the observed light ray (the scalar product $\vec{k}\cdot\vec{v}$ thus coinciding with the projection of $\vec{v}$ along the line of sight, \emph{i.e.} $-\beta \cos\theta$, being $\theta$ the angle between the velocity vector $\vec{v}$ and the line of sight). Since all quantities are evaluated at $t=t_{\rm em}$, Rømer delay is naturally accounted for in the spectroscopic observables.

The observed radial velocity $\beta_{r}$ can be expanded in terms of the 3D stellar velocity $\beta=v/c$ of the star, as \cite{Alexander2005, GravityCollaboration2018a}

\begin{equation}
\beta_{r}=B_{0}+B_{1}\beta+B_{2}\beta^{2}+{\cal {O}}(\beta^{3})\,.\label{e:br}
\end{equation}

where $B_0$, $B_1$, $\dots$, are the coefficient of the expansion (that depend on the specific effect considered), and one can continue the PN expansion up to the desired order. The special relativistic Doppler shift can be expanded to second order as
\begin{equation}
    \beta_{D}=\frac{1+\beta\cos\theta}{\sqrt{1-\beta^{2}}}\simeq1+\beta\cos\theta+\frac{1}{2}\beta^{2}.
\end{equation}
Thus the special relativistic Doppler shift contributes to the total redshift with two terms: the first is the classical Doppler effect due to the relative longitudinal motion between observer and emitter along the line of sight, which is a first order ($\propto\beta)$ effect, and transverse relativistic Doppler shift, due to the time dilation experienced by a fast moving object with respect to a stationary observer, which is a second order effect ($\propto \beta^2$).

The gravitational redshift, on the other hand, can be expanded considering the weak field approximation of the metric tensor for the space-time around a point mass $M$. At a distance $r$ from the source, the $g_{00}$ component can be expressed in terms of the classical gravitational potential $\Phi$. Namely, \cite{Poisson2014}
\begin{equation}
    g_{00}(r) \simeq \left(1-\frac{2\Phi(r)}{c^2}\right)= \left(1-\frac{2GM}{c^2r}\right)=(1-\Gamma).
\end{equation}
The gravitational contribution to the redshift, can hence be expanded in PN orders by considering the mildly-relativistic orbit approximation in Equation \eqref{eq:kepler_eq},
\begin{equation}
    \beta_G = \left(1-\Gamma\right)^{-1/2}\simeq1+\frac{1}{2}\Gamma(r)= 1+\frac{1}{2}(\Gamma_0+\beta^2).
\end{equation}
The space-time curvature thus contributes a constant term, related to the mean gravitational field of the central massive object, and a second order ($\propto\beta^2$) term related to the fact that the star during its orbital motion approaches closer distances to the central SMBH and thus experiences higher gravitational time dilation. It is worth to notice that the special relativistic transverse Doppler effect and the gravitational redshift have the same magnitude ($\beta^2/2$) and are completely degenerate in observation, as spectroscopic data are only sensitive to the sum of the two contributions \cite{Do2019a}. Moreover, since the two terms are always positive, they correspond to a redshift along the entire orbit (albeit being very small except around pericenter). On the contrary, the first-order longitudinal Doppler effect can be either positive or negative depending on the approach/recession of the observed star. Finally, in the second order term the Rømer delay contribution ($B_R\beta^2$) has to be taken into account, which, as shown previously, can be either positive or negative depending on the orbital phase of the star. By summing all contributions one obtains the final expression for the observed radial velocity of a star at the GC
\begin{equation}
    \beta_{r}=\left(\beta_{\odot}+\beta_{z,\mathrm{gal}}+\beta_{z,\star}+\frac{1}{2}\Gamma_{0}\right)+\cos\vartheta\beta+(1+B_{R})\beta^{2}+\mathcal{O}(\beta^{3}),
    \label{eq:beta_r}
\end{equation}
where in the zeroth-order constant term several additional contributions have been taken into account \cite{Alexander2005}:
\begin{enumerate}
    \item  the constant velocity shift $\beta_{\odot}$ due to the compound motion of the Sun and the Earth relative to the GC as well as the blueshift due to the potential wells of the Sun, Earth and planets;
    \item the constant velocity shift $\beta_{\rm gal}$ due to redshift by the potential of the Galaxy;
    \item the gravitational redshift $\beta_{\star}$ due to the star's potential.
\end{enumerate}
However, all these constant terms $\beta_{\odot}$, $\beta_{\mathrm{gal}}$ and $\beta_{z,\star}$ are very small compared to the $\beta^{2}$ term \cite{Alexander2005}. In particular, the peculiar radial motion of the Solar System relative to the GC is estimated to be $<\!10\,\mathrm{km\, s^{-1}}$ whose corresponding transverse Doppler redshift is $\sim0.08$ km/s. The constant gravitational redshift $\beta_{\rm gal}$ due the the Galactic potential is of order 0.16 km/s, while the gravitational blueshift due to the proximity of the Sun is only $-0.003$ km/s. Similarly, the redshift due to the star's potential and the blueshift due to the Earth's potential are only 1 km/s and -0.0002 km/s, respectively. Moreover, in the orbital fitting of the S-stars motion, the radial velocity is always corrected for the Local Standard of Rest which introduces a constant offset $v_{z,0}$ that is left as a free parameter in the fits. Thus, all constant terms are completely degenerate with such offset and it is impossible to measure them separately.

\begin{figure}[!t]
\centering
\begin{subfigure}{.56\textwidth}
  \centering
  \caption{}
  \includegraphics[width=\linewidth]{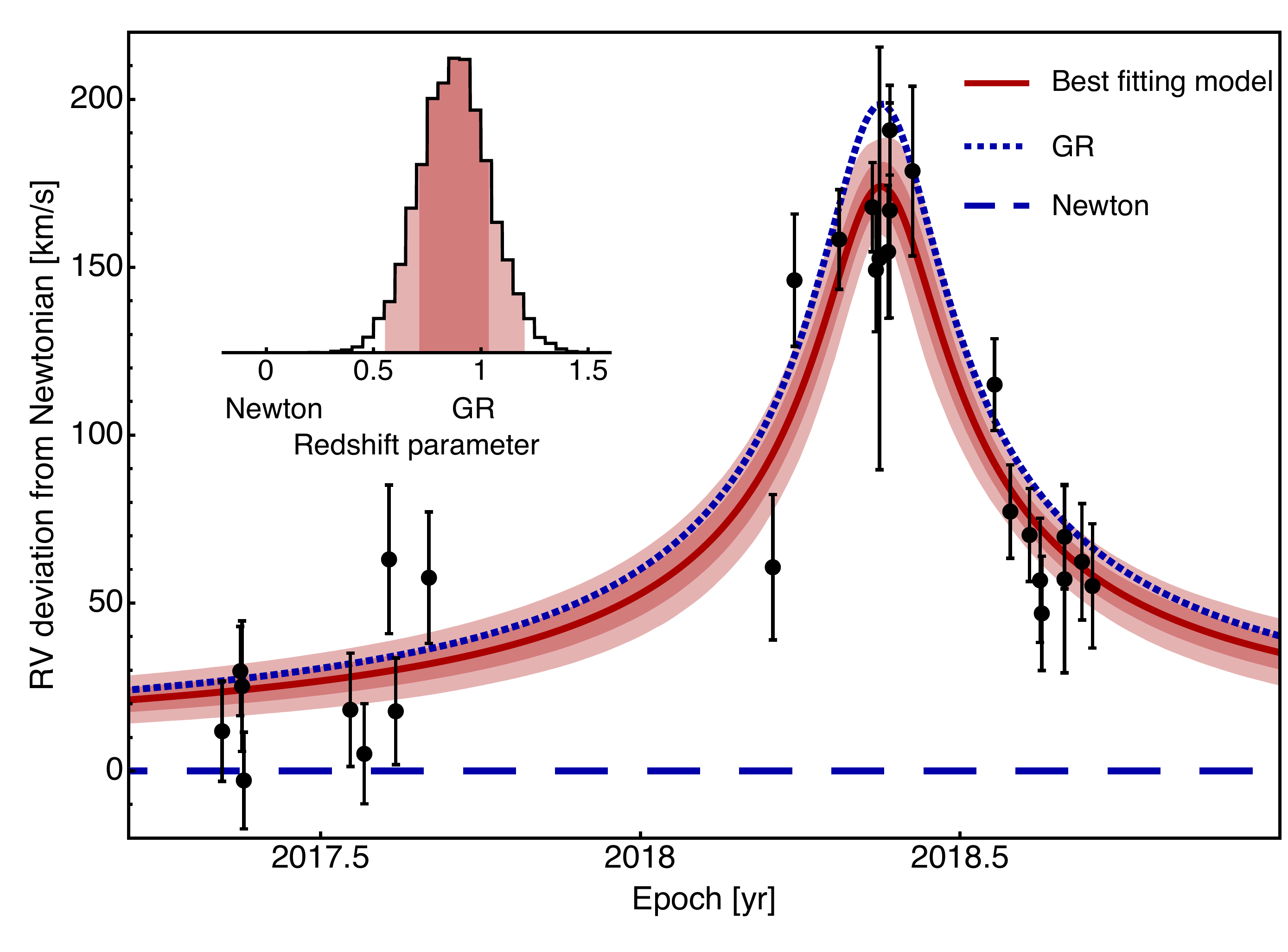}
  \label{subfig:grav_redshift_keck}
\end{subfigure}\begin{subfigure}{.40\textwidth}
  \centering
  \caption{}
  \includegraphics[width=\linewidth]{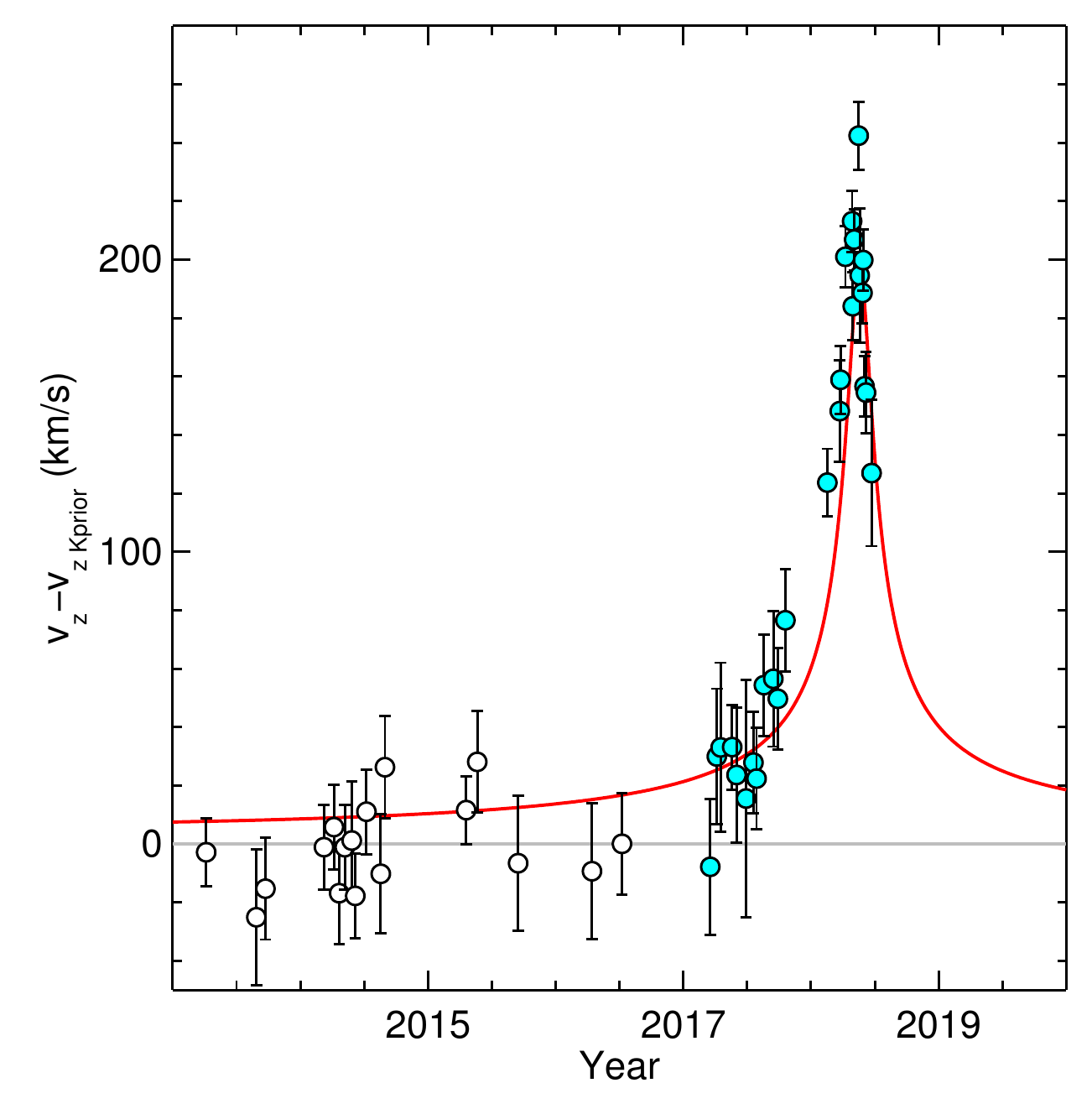}
  \label{subfig:grav_redshift_gravity}
\end{subfigure}
\caption{Contribution of the additional special + general relativistic redshift for the S2 star around its pericenter, which has a maximum magnitude of $\sim 200$ km/s at the point of highest proximity in $\sim$2018.38, \textbf{(a)} from the UCLA group, observed at the Keck telescope \cite{Do2019a}, and \textbf{(b)} with the VLT/SINFONI spectrographer by the MPE/Gravity Collaboration \cite{GravityCollaboration2018a}. In particular the curves report the difference between the relativistic model (blue dotted line in \textbf{(a)} and red solid line in \textbf{(b)}) and a corresponding Keplerian orbit with same orbital parameters (represented here as the horizontal 0-level lines). Recorded radial velocities and corresponding errors show how, starting in $\sim2017$, the line-of-sigh velocity began departing from the Newtonian prediction and followed the expected profile from GR. Images are courtesy of \cite{Do2019a} and \cite{GravityCollaboration2018a}, respectively.}
\label{fig:grav_redshift}
\end{figure}

If we compute the value of $\beta^2$ for the S2 star at its pericenter, where it reaches its maximum orbital speed of $v \simeq 7650$ km/s, one obtains $\beta^2 = (v/c)^2 = 6.6\times 10^{-3}$. This means that spectroscopic observations of the S2 star near its pericenter are supposed to carry an additional $\beta^2c \simeq 200$ km/s of reconstructed line-of-sight velocity that are ascribable to the combination of the special + general relativistic effects. The AO-assisted spectrographer employed by both the UCLA and the MPE groups (see Section \ref{sec:s_stars_obs}), provide with radial velocities with uncertainties on the order of a few 10 km/s and are hence capable of detecting this effect on the orbit of S2. This motivated both groups to intensively monitor S2 throughout 2018 (which is the year when S2 passed at its pericenter for the last time) in order to produce datasets in which the gravitational redshift signal could be successfully detected. Since measuring PN deviations from a Newtonian orbit heavily relies on a precise knowledge of the orbital Keplerian elements\footnote{Radial velocities for a single star on a ${\cal {O}}(\beta^{2})$ PN orbit can always be expressed as a Newtonian orbit (including the classical Roømer effect) with suitably modified orbital parameters \cite{Alexander2005}.}, spectroscopic observations in 2018 had to be complemented with proper motion measurements spanning more than 20 years in order to accurately fit simultaneously for the orbital parameters and for the redshift effect. In particular, both groups introduced an extra parameter ($\Gamma$ in UCLA notation and $f$ in MPE notation, here this parameter is denoted as $\zeta_{\rm redshift}$) in their orbital models which would uniformly map between a Newtonian orbit and a 1-PN orbit. In particular for $\zeta_{\rm redshift} = 0$ one obtains an orbit compatible with Newtonian gravity, while for $\zeta_{\rm redshift} = 1$ one obtains an orbit consistent with the first PN limit of GR.
By leaving this parameter free to vary in the orbital fits, it is possible to quantitatively establish the validity of GR against Newtonian gravity at the 1PN order. Both groups successfully detected a value of $\zeta_{\rm redshift}$ compatible with unity, thus confirming the detection of the additional redshift component. In Figure \ref{fig:grav_redshift} we report the reconstructed line-of-sight velocity from the analyses performed by the two groups \cite{Do2019a, GravityCollaboration2018a}. In particular, both plots represent the additional redshift contribution from the 1PN terms relative to the predictions for a Newtonian orbit. Around the 2018.38 pericenter passage, the expected 200 km/s ‘‘kick'' in radial velocity has been successfully observed and the two orbital fits yield
\begin{align}
    &\textrm{UCLA:}& & \zeta_{\rm redshift} = 0.88\pm0.16, \\ \cline{1-4}
    &\textrm{MPE:}& & \zeta_{\rm redshift} = 0.901\pm0.090.
\end{align}
Moreover, by simultaneously fitting a Keplerian orbital model to the data and a 1PN model containing the additional redshift component, it is possible to compare the two models via Bayesian evidence estimation. The differences in the Bayesian Information Criteria (BIC) and the Akaike Information Criteria (AIC) computed by the Gravity Collaboration are of order $\sim87$, thus providing ‘‘decisive'' evidence for the $\zeta_{\rm redshift} = 1$ model. A similar result was obtained by the UCLA group who obtained a difference in the log-evidence of order $\sim10.68$ indicating that the relativistic model has a Bayes factor of 10.68, implying that it is 43 650 times more likely, given the measurements, than the purely Newtonian model \cite{Do2019a}.

The redshift modeling was stopped at the second order in $\beta$ (Equation \eqref{eq:beta_r}) because with current instrumentation it has not been possible to detect higher order effects on the orbits of the S-stars. Third order effects on the spectroscopic observable would include frame dragging, gravitational lensing, the Shapiro time delay and effects of the relativistic orbital precession. We will discuss the latter (which is a $\mathcal{O}(\beta^{3})$ effect on the redshift, while being a $\mathcal{O}(\beta^{2})$ effect in the astrometric measurements) in the next section and the other higher order effects in \ref{sec:higher_order_effects}. 

\subsubsection{The orbital precession}
\label{sec:orbital_precession}

\begin{figure}[!t]
\centering
\begin{subfigure}{.33\textwidth}
  \centering
  \caption{}
  \includegraphics[width=\linewidth]{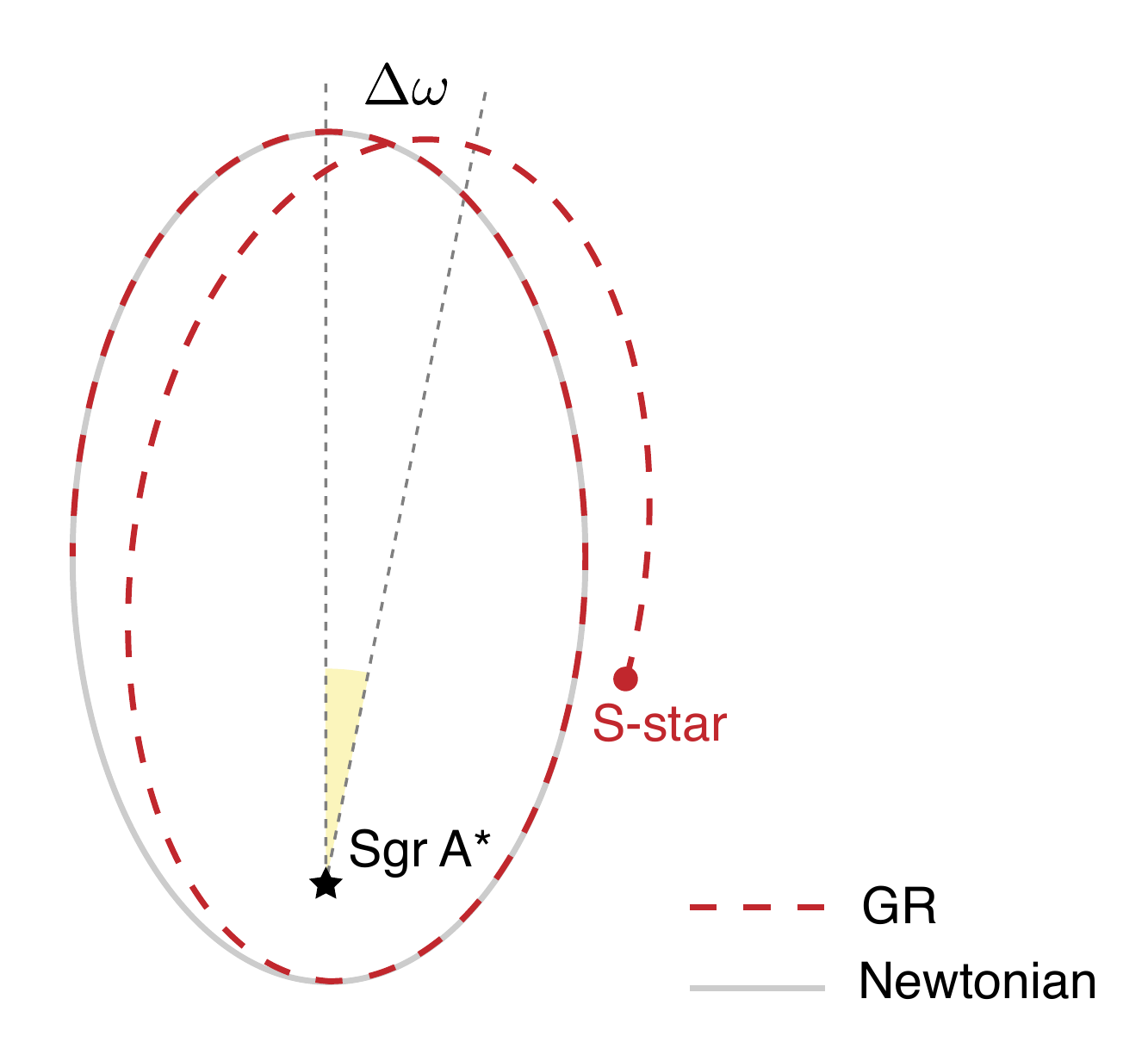}
  \label{subfig:orb_precession_illustration}
\end{subfigure}\begin{subfigure}{.67\textwidth}
  \centering
  \caption{}
  \includegraphics[width=\linewidth]{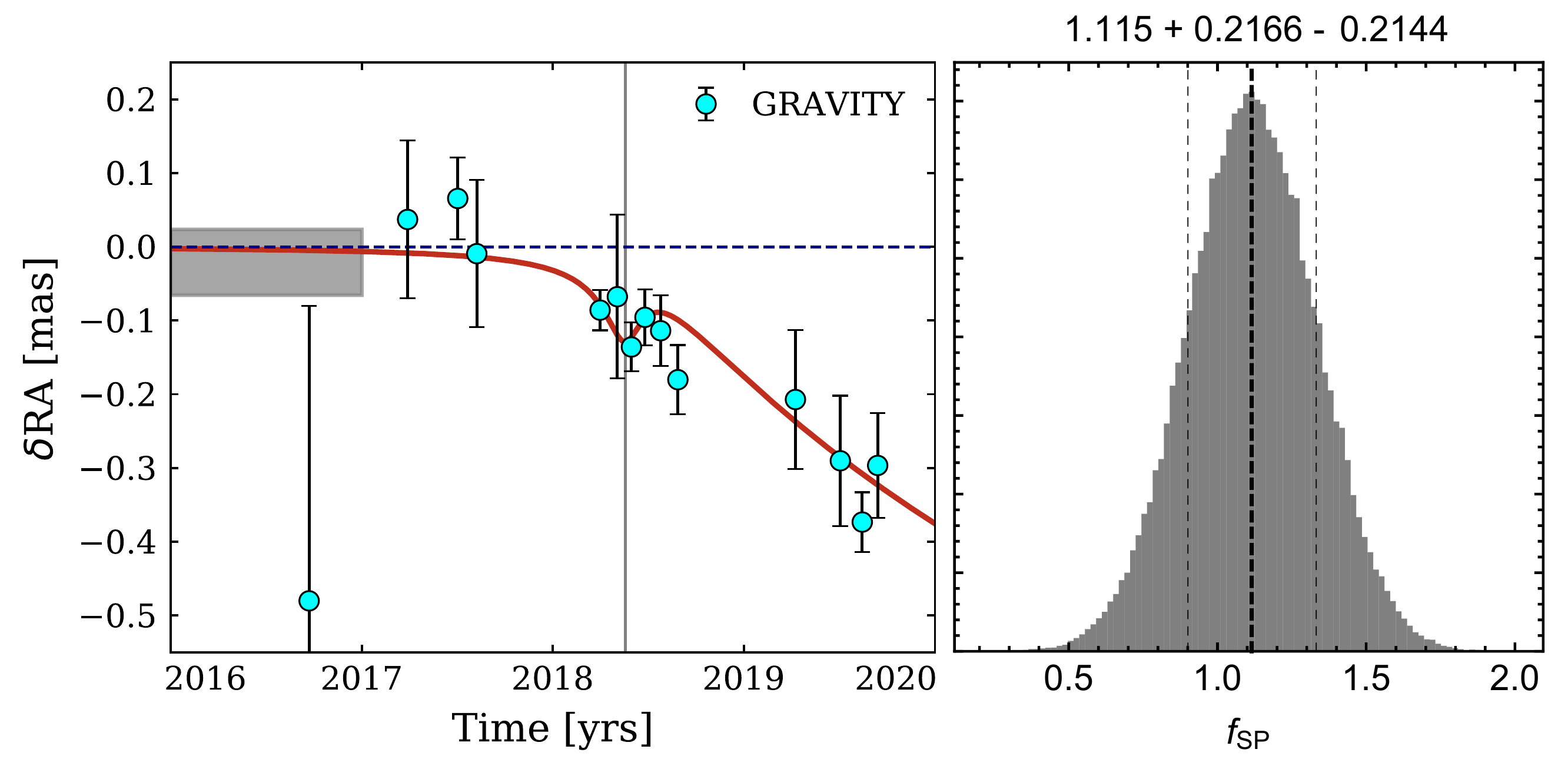}
  \label{subfig:orb_precession_gravity}
\end{subfigure}
\caption{\textbf{(a)} Illustration of the orbital precession around Sgr A*. The dashed red line is the rosette-like preceding orbit from GR, as compared to the solid gray Keplerian ellipse. The rate of precession per orbital period is the angle $\Delta\omega$ spanned by the pericenter/apocenter over one revolution; \textbf{(b)} Adapted from \cite{GravityCollaboration2020c}. The first plot shows the departure in the RA direction of the 1PN orbit from the Keplerian ellipse (dashed black line at 0 level) and the measured positions with the GRAVITY interferometer as cyan dots. After the 2018 pericenter passage the star appears to perfectly follow the 1PN line. This allowed the Gravity Collaboration to put a constraint on the parameter $f_{\rm SP}$, whose posterior probability distribution is depicted on the right, which confirms GR predictions ($f_{\rm SP} = 1$) against Newtonian gravity ($f_{\rm SP} = 0$) at more than $5\sigma$.}
\label{fig:orb_precession}
\end{figure}

The relativistic orbital precession (also known as periastron/ pericenter advance or shift) is regarded as one of the classical tests of GR and it represents a dynamical effect that naturally arises in the geodesic approach to the dynamics of test particles. While in Newtonian dynamics, the trajectory of a test particle in a central potential would trace out a perfectly closed ellipse with the source of the gravitational field in the focus, the deviation of GR from the Newtonian $1/r$ potential of a point mass modifies the resulting trajectory. In particular, GR predicts that the orbits are affected by a precession, i.e a rotation of the major axis of the ellipse about the center of mass, on the orbital plane in a prograde sense with motion (Figure \ref{subfig:orb_precession_illustration}). This results in rosette-like orbits which can be described as an ellipse whose argument of the pericenter $\omega$ linearly drifts with time, with a rate per orbital period that, for a non-rotating BH,
is given by \cite{Weinberg1972}
\begin{equation}
    \Delta\omega=\frac{6\pi GM}{c^2a(1-e^{2})}.
    \label{eq:precession}
\end{equation}
Here, $M$ is the mass of the central object, $a$ is the semi-major axis of the orbiting body and $e$ its eccentricity. Since the pericenter distance can be expressed as $r_p = a(1-e)$, the rate of orbital precession is directly proportional to the relativistic parameter $\Gamma = r_s/r_p$, which makes the pericenter advance an $\mathcal{O}(\beta^{2})$ effect. Moreover, the closer the orbiting body gets to the central object (and hence the greater its eccentricity is), the stronger this effects is. In particular, since $\Delta\omega$ represents an angular shift on the orbital plane, the effect is not directly measurable at pericenter (differently from the the gravitational redshift) but is accumulated at pericenter and becomes noticeable only once the body gets further away from the central object. In particular, at apocenter (the furthest point from the center along the orbit), $r_a = a(1+e)$, the precession of the ellipse on the orbital plane yields a spatial shift of the apocenter of about (assuming no inclination of the orbit)
\begin{equation}
    \Delta p_{a}\sim\Delta\omega r_a=\frac{6\pi GM}{c^2(1-e)},
\end{equation}
which ultimately depends only on the eccentricity of the orbit itself. Plugging the orbital parameters of the S2 star into Equation \eqref{eq:precession}, the orbital precession  amounts to an angle of $\Delta \omega \simeq 12.1'$ for each revolution around Sgr A*. Converting this number into the observable astrometric shift at pericenter, one obtains   the shift is $\Delta p_{a} \simeq 1$
mas per orbital period, which makes it observable with current instruments \cite{Zakharov2007, Heissel2022}. See columns 10 and 11 in Table \ref{tab:sstars} for an estimate of this effect for some of the S-stars for which it is more prominent.

In 2019, the Gravity Collaboration successfully detected \cite{GravityCollaboration2020c} the periastron advance of the S2 star, only a few months after the pericenter passage in 2018, thanks to the exquisite astrometric precision delivered by the GRAVITY interferometer\footnote{If near-infrared NACO/Keck measurements of the astrometric position of S2 would have not been affected by systematic effects in the construction of the infrared-to-radio references frame it would have been possible to detect the orbital precession of the S2 long before 2019. However, the drift of the orbit on the orbital plane due to the orbital precession of S2 is totally degenerate with a drift of the mass centroid itself on the reference frame and only with the ability of GRAVITY of observing directly the S2-Sgr A* separation it was possible to detect the relativistic precession in a matter of months \cite{GravityCollaboration2020c}.} (see Sect\ref{sec:s_stars_obs}), revealing a departure from a Keplerian orbit of a few $\sim 0.1$ mas (especially in the RA direction, due to the particular inclination of the orbit) which tightly follow predictions from GR (Figure \ref{subfig:orb_precession_gravity}). As already done for the gravitational redshift, in order to establish quantitatively the agreement of the observed data with the relativistic prediction against Newtonian gravity, a new parameter $f_{SP}$, which would continuously span between Newtonian and relativistic dynamics, was considered. In particular, this parameter was introduced in front of the Schwarzschild related terms into the relativistic equations of motion for a test particle in a central potential at 1PN order \cite{Will2008}:
\begin{equation}
    \vec{a} = -\frac{G M}{r^3} \vec{r}  + f_\mathrm{SP} \frac{G M}{c^2 r^2}\left[\left(4 \frac{G M}{r} - v^2\right)\frac{\vec{r}}{r} +4 \dot{r}\vec{v}\right].
\end{equation}
Here, $\vec{r}$ is the position of the star with respect to the central object and $\vec{v}$ its velocity, {while $\dot{r}$ is the derivative of the radial coordinate, corresponding to the radial component of $\vec{v}$}. The acceleration thus reduces to that of Newtonian gravity if $f_\mathrm{SP} = 0$, while if $f_{\rm SP} = 1$ the additional terms encode the relativistic dynamics at 1PN order which leads to the orbital precession (Equation \eqref{eq:precession}). The parameter $f_{SP}$ has been left free to vary in the posterior analysis performed in \cite{GravityCollaboration2020c}, along with all other orbital elements, including the mass and distance of the central object, in order to simultaneously fit the 1PN orbital model to the astrometric and spectroscopic data for S2. The results of the fit yield
\begin{equation}
    f_{\rm SP} = 1.10 \pm 0.19,
    \label{eq:precession_gravity}
\end{equation}
thus confirming the validity of GR predictions over Newtonian gravity at more than 5$\sigma$ \cite{GravityCollaboration2020c}.

The successful detection of the orbital precession of the S2 star at the GC center, not only provides with yet further evidence that Sgr A* behaves exactly as a BH described by GR would (albeit at 1PN order), but also places very tight constraints on a possible extended mass component around Sgr A*. In particular, from Newtonian gravity it is a known result that an extended mass component would yield retrograde precession \cite{Fragile2000}. The fact that observations detect unambiguously prograde precession allows to place an upper limit on the mass of this extended component which cannot exceed 0.1\% of the mass of the central object (about 4000 $M_\odot$) \cite{Heissel2022}.

\subsubsection{Higher-order effects}
\label{sec:higher_order_effects}

First order PN effects that have been successfully detected on the orbit of the S2 star \cite{Do2019a, GravityCollaboration2018a, GravityCollaboration2020c} provide with stringent test of GR in a regime that had not been explored before (Figure \ref{fig:eht_gravitational_probes}). With increasing observational capabilities and the potential discovery of a population of fainter/closer stars, higher order relativistic effects could be tested with the orbits of the S-stars. These effects include the frame dragging, effects related to the BH quadrupole moment and gravitational lensing by the SMBH. Detection of such effects (especially the first two) would not be merely the detection of yet another prediction of GR, but would rather provide evidence for the rotation of Sgr A* on one hand, and the first test of much more fundamental theoretical predictions, like the BH ''no-hair'' or uniqueness theorems of GR, among the others \cite{Misner1973}. In particular, this theorems predict that the space-time (and hence the dynamics of stars orbiting) around a non-charged rotating BH is completely characterized by its mass $M$ and angular momentum $J$.  Consequently, all the higher multipole moments of space-time have to be functions of $M$ and $J$. More specifically, the quadrupole moment $Q_2$ has to satisfy the relation
\begin{equation}
    Q_2 = -\frac{J^2}{M}
    \label{eq:no-hair}
\end{equation}
assuming ($G=c=1$).

The dynamical effects on a test particle related to the rotation of the central object and the presence of a quadrupole moment are secular changes in its trajectory through gravitomagnetic GR effects known as ''frame dragging''
or Lense-Thirring precession \cite{Lense1918}. The PN equations of motion of a test particle around such an object are \cite{Will2008}
\begin{align}
    \vec{a} =& -\frac{M}{r^3}\vec{r} + \frac{M}{r^2} \left[\left(4 \frac{M}{r} - v^2 \right )\vec{n} + 4\dot{r}\vec{v}\right]\vec{n}- \frac{2J}{r^3} \left[ 2\vec{v}\times \vec{{j}}-3 {\dot r} \vec{n}\times \vec{j} - 3\frac{\vec{n}(\vec{h} \cdot
    \vec{j})}{r} \right]+
    \nonumber \\
    \nonumber \\
    & + \frac{3}{2} \frac{Q_2}{r^4} \left[5\vec{n}(\vec{n} \cdot \vec{j})^2
    - 2 (\vec{n} \cdot \vec{{\vec j}})\vec{j} - \vec{n}\right] \,,
    \label{eq:pn_eom}
\end{align}
where $\vec{r}$ and $\vec{v}$ are the position and velocity of the
body, $\vec{n} = \vec{r}/r$, $\vec{h} = \vec{r}\times\vec{v}$, 
and $\hat{J} = \vec{J}/J$ being $\vec{J}$ the BH angular momentum. The first and second terms of Equation \eqref{eq:pn_eom} correspond to the PN acceleration related to the Schwarzschild part of the metric, already illustrated in Section \ref{sec:orbital_precession}. The third  and fourth terms represent the frame-dragging  and quadrupole moment effects, respectively. The combination of these effects results not only in an additional pericenter precession, but also to a rotation of the orbital plane itself, given by a precession of the orbital angular momentum vector around the SMBH spin vector $\vec{J}$. The two effects thus determine a secular change in the longitude of the ascending
node $\Omega$ and an additional change in the argument of pericenter, $\omega$, given by \cite{Alexander2005}:
\begin{align}
    \Delta\omega_{\mathrm{LT}}&=-3\Delta\Omega_{\mathrm{LT}}\cos i\,\label{eq:lensethirring_omega}\,,
\end{align}
with 
\begin{align}
    \Delta\Omega_{\mathrm{LT}} &=\sqrt{2}\pi\frac{a_0}{(1-e^{2})^{3/2}}\left(\frac{r_{S}}{a}\right)^{3/2}\,,
    \label{eq:lensethirring_Omega}
\end{align}
where $a_0 \equiv J/(GM/c)$ is BH spin parameter ($0\leq a_0 \leq 1$), $i$ is the angle between the BH angular momentum and the orbital angular momentum of the star. Here, the ‘‘no-hair'' theorem (Equation \eqref{eq:no-hair}) is assumed to be satisfied. Being both terms proportional to the 3/2-power of the relativistic parameter $\Gamma \sim \beta^2$ (at pericenter), the two effects are of order $\mathcal{O}(\beta^{3}$).

\begin{figure}[!t]
    \centering
        \begin{subfigure}{.32\linewidth}
          \centering
          \includegraphics[width=\linewidth]{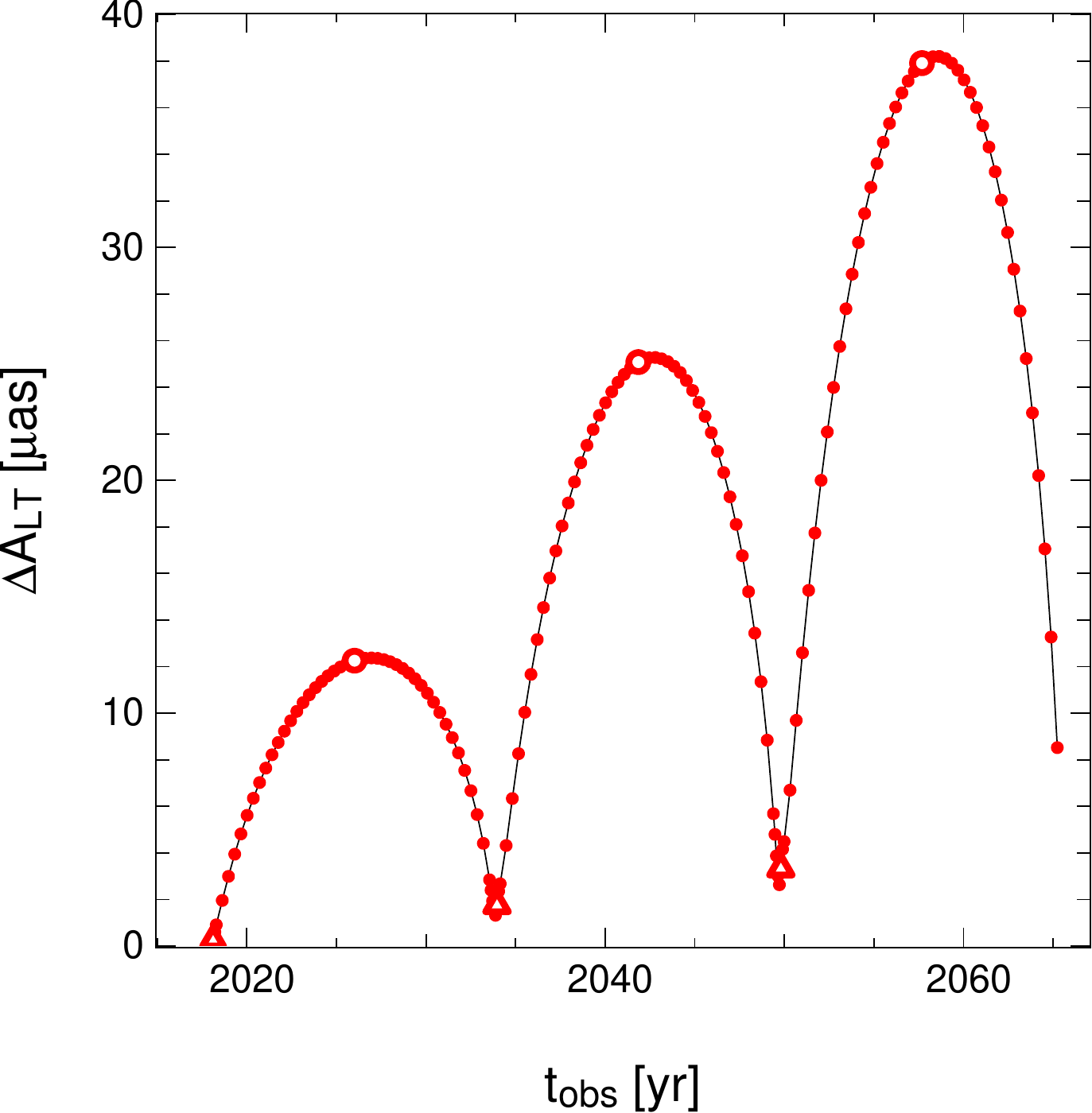}
          \caption{}
          \label{subfig:grould_lt}
        \end{subfigure}
        \begin{subfigure}{.32\linewidth}
          \centering
          \includegraphics[width=\linewidth]{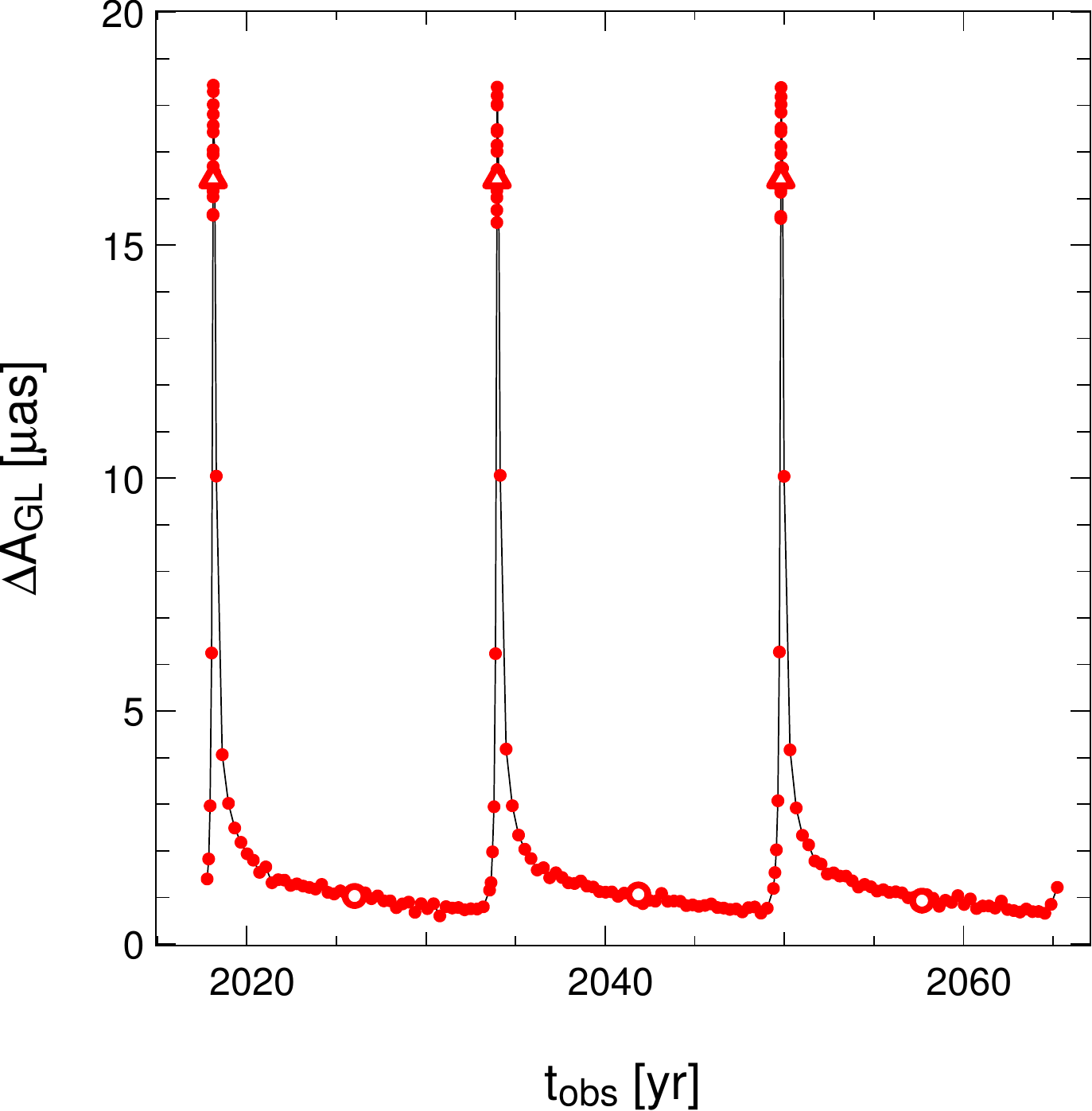}
          \caption{}
          \label{subfig:grould_lensing}
        \end{subfigure}
        \begin{subfigure}{.32\linewidth}
          \centering
          \includegraphics[width=\linewidth]{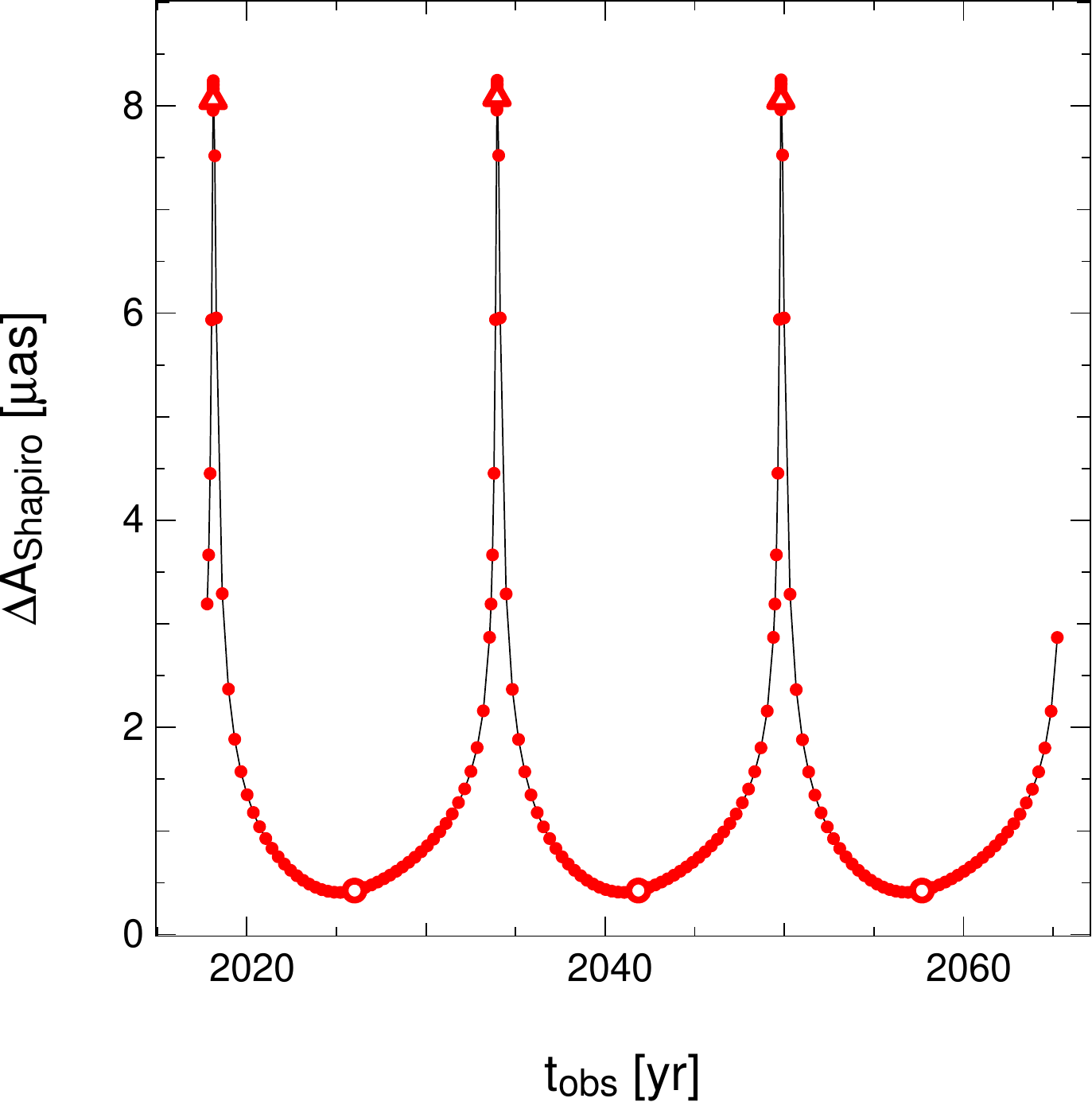}
          \caption{}
          \label{subfig:grould_shapiro}
        \end{subfigure}
    \caption{Astrometric impact of different higher-order relativistic effects over three orbits of the S2 star in the GC from the numerical studies of Grould and collaborators \cite{Grould2017b}. In particular: \textbf{(a)} reports the astrometric shift induced by Lense-Thirring precession on the star's trajectory due to the rotation of the central SMBH (which for this plot is assumed to have a spin of $a = 0.99$ inclined with angles $i'=45^\circ$ and $\Omega'=160^\circ$  - inclination and longitude of ascending node -  with respect to the observer reference frame). The astrometric shift is computed by comparing two relativistic orbital models for the star's trajectory, one assuming a Schwarzschild (non-rotating) BH and the other assuming a Kerr (rotating) BH. Since the Lense-Thirring effect induces secular perturbations on the orbit of S2, its magnitude grows over time, and is always maximum at apocenter; \textbf{(b)} reports the shift in the observed position of S2 due to gravitational lensing on the photons. The effect is always maximum at pericenter ($\sim 20\;\mu$as) and it induces a periodic modulation of the astrometric position, as it does not actually perturb the orbit. This shift is estimated by comparing the observed positions of S2 in the case of a straight-line propagation of the photons emitted by the star, with those computed assuming a fully relativistic geodesic propagation; \textbf{(c)} same as (b), but for the Shapiro time delay. The periodic modulation in this case, never goes above $\sim 8\;\mu$as, rendering this effect impossible to detect with GRAVITY observations. The plots are courtesy of \cite{Grould2017b}.}
    \label{fig:grould_s2_higher_order}
\end{figure}

The predicted magnitude of the effect in the GC is very small \cite{Jaroszynski1998}, and it is unlikely that it will be measured directly from one single orbit of the S-stars even with the proposed TMT \cite{Weinberg2005a} or the ELT. However, over the lifespan of one of the young S-stars near the SMBH, $\sim$10 Myr, the accumulated precession can be substantial if the orbit is eccentric enough and following multiple orbits\footnote{As a matter of facts, the Lense-Thirring precession is one the proposed mechanisms by which the 0.88-eccentric S2 could have been dragged out by the thin stellar disk (where it might gave originated) up to the position where we observe it today, which appears to be totally uncorrelated with the disk itself \cite{Eisenhauer2005}. This is one of the possible formation scenario for the S-stars cluster, as mentioned in Section \ref{sec:s_stars}, even though the orbital parameters of other stars in the cluster seem not to support this idea, since for many of them the time required in order to complete a Lense-Thirring induced revolution is larger than the age of the stars.} with high-precision astrometry can lead to the detection of the effect. In particular, the precession of the orbital plane around the BH spin axis for the S2 star amounts to $\Delta\Omega_{\rm LT} \sim 10^{-5}$ rad/orbital period which correspond to a $2\pi$ angle spanned on a period of $\sim 10^7$ yr \cite{Levin2003}, which is comparable to the lifetime of the star itself. Grould and collaborators \cite{Grould2017b} performed a numerical study of all the relativistic effects on the orbit of the S2 star (using a fully-relativistic stellar orbital model including both the computation of null and time-like geodesics for a Kerr space-time) and how long would it take for the GRAVITY interferometer and the SINFONI spectrographer (for which they assume a nominal astrometric precision of $10\;\mu$as and a radial velocity uncertainty of 10 km/s, respectively) to detect such effects on the orbit of S2 (Figure \ref{subfig:grould_lt}). In particular, the Lense-Thirring effect would shift the orbit of S2 by $\sim 12\;\mu$as for each orbital period (this difference is evaluated at apocenter where this effect is maximized) with respect to a non-preceding orbit, and hence GRAVITY observations of S2 should be able to detect it after one or two full orbits \cite{Grould2017b}. The effect is much less noticeable on the radial velocity, where it amounts to only $\sim -0.2$ km/s at pericenter per orbital period and an accuracy of $\sim$10 m/s is needed to constrain the Lense-Thirring effect in only one orbital period \cite{Angelil2010a}. Moreover, while the astrometric shift induced by the effect is expected to be above the sensitivity threshold of GRAVITY, using this effect in order to constrain the BH angular momentum (its magnitude $a_0$, and position angle) has been demonstrated to be quite difficult using only the orbital motion of the S2 \cite{Broderick2005, Zhang2015, Yu2016, Grould2017b} unless for very particular combination of BH inclination and spin (\emph{e.g.} quasi-extremal angular momentum $\gtrsim 0.99$ and inclination $i\lesssim 50^\circ$) and for a long (and thus impractical) monitoring period (at least three S2-orbits) \cite{Yu2016}. However, the BH spin could be constrained with a high precision (\emph{e.g.} with $1\sigma$ uncertainty of $\lesssim0.1$, or even $\lesssim0.02$) if a closer star ($a\lesssim 300$ AU) is discovered and its orbit is tracked with GRAVITY-like precision for $\sim$10 yr.
As for the effects related to the quadrupole moment of the BH, C. M. Will demonstrated in 2008 \cite{Will2008}, using the PN formalism, that several stars were to be found with orbital periods of fractions of a year and with sufficiently large orbital eccentricities, then the quadrupole-induced precessions could be as large as 10 $\mu$as/year. In particular, observing astrometric positions of at least two stars with eccentricity $e>0.9$ and period $T<0.1$ yr, could lead to constraining the BH quadrupole moment.

\begin{table}
        \renewcommand{\arraystretch}{1.5}
        \resizebox{.45\linewidth}{!}{
        \centering
        \begin{tabular}{lccr}
            \hline
             Effects  & $1^{\mathrm{st}}$ period  & $2^{\mathrm{nd}}$ period & $3^{\mathrm{rd}}$ period \\   
            \hline
            PA & 3000 (Pe) &  8000 (Pe) & $16 000$ (Pe) \\  
            Shapiro & 8 (Pe) & 8 (Pe) & 8 (Pe) \\
            LTS & 10 (Ap) & 25 (Ap) & 40 (Ap) \\
            LTP & $\ll 1$ & $\ll 1$ & $\ll1$ \\
            GL & 20 (Pe) & 20 (Pe) & 20 (Pe)\\
            \hline 
        \end{tabular}
        }
        \hspace{1cm}
        \resizebox{.45\linewidth}{!}{
        \centering
        \begin{tabular}{lccr}
            \hline
             Effects  & $1^{\mathrm{st}}$ period  & $2^{\mathrm{nd}}$ period & $3^{\mathrm{rd}}$ period \\   
            \hline
            PA & 140 (Pe) & 1520 (Pe) & 2800 (Pe) \\
            TD, GZ & 200 (Pe) & 200 (Pe) & 200 (Pe) \\
            LTS & 0.2 (Pe) & 0.5 (Pe) & 0.9 (Pe) \\
            LTP & $10^{-2}$ & $10^{-2}$ & $10^{-2}$ \\
            HOPC & 5 (Pe) & 5 (Pe) & 5 (Pe) \\
            \hline
        \end{tabular}
        }
        \caption{The maximum impact of relativistic effects on the astrometric and spectroscopic observables for the S2 star, as studied by Grould and collaborators \cite{Grould2017b}. In particular, in the left table the maximum astrometric shift for each effect is reported in units of $\mu$as for three consecutive orbital periods. The right table, on the other hand, reports the corresponding radial velocity shift in km/s induced by each effect. The following abbreviations have been adopted: PA stands for pericenter advance (\emph{i.e.} the orbital precession); TD is the transverse Doppler effect of special relativity, while GZ stands for the gravitational redshift; LTS and LTP report the Lense-Thirring effect on the star's trajectory and on photon paths, respectively; GL is the gravitational lensing, while HOPC, which stands for high-order photon curvature, reports the sum of all higher-order relativistic effects on the spectroscopic observable due to the curved photon paths and gravitational time dilation experienced by photons. Moreover, Pe and Ap denote whether the maximum shift for each effect is achieved at Pericenter or Apocenter, respectively. Both tables are adapted from \cite{Grould2017b}.}
        \label{tab:rel_effects}
\end{table}

Another relativistic effect that can potentially be detected at the GC is gravitational lensing (GL) \cite{Misner1973} of stars by the SMBH strong gravitational field. This process differs from the other secular perturbations to the stellar orbits that we have discussed, as it does not involve modification of the actual trajectory of a test particle, but only the apparent position due to the deflection of the light emitted by it and that can eventually generate secondary images. For this reason it is not possible to attribute a PN order to GL, as its magnitude is only related to how close on the sky (not necessarily in the 3D space) a photon path goes to the SMBH \cite{Wardle1992, Jaroszynski1998, Alexander1999b, Chaname2001, DePaolis2003, Nusser2004, Bozza2005}. This effect has been studied for the S-stars \cite{Bozza2012} and it has been demonstrated that new generation instruments like GRAVITY could potentially detect the effect of GL on their astrometric position. In particular, while the secondary image is always confused with Sgr A*, the astrometric shift of the primary image is appreciable by GRAVITY for many known S-stars in a cone with aperture $\sim40^\circ$. The star S17, in particular, while not being on a particularly eccentric orbit ($e\sim 0.3$) and thus not having a particularly close pericenter passage (if compared for example with S2 or  S62, Table \ref{tab:sstars}), will have the highest astrometric shift for several years, inducing an astrometric shift of about 30 $\mu$as in 2018 that surpasses the sensitivity threshold of the GRAVITY interferometer. Moreover, since photon paths are perturbed by the rotation of the central object (null geodesic around a Kerr BH \cite{Chandrasekhar1983}), the GL effect depends on the spin of the BH and can also help in providing and independent constraint on the Lense-Thirring effect \cite{Bozza2004, Bozza2005, Jorgensen2016}. Finally, Grould and collaborators \cite{Grould2017b} have investigate the astrometric impact of GL on the orbit of S2, by numerically tracking null geodesics in Kerr space-time from the star to a distant observer. They found that the greatest astrometric shift occurs at pericenter and it amounts to $\sim 20\;\mu$as, which is barely above the sensitivity threshold of GRAVITY. Moreover, since this effect does not perturb the orbit, it does not accumulate over time, and it thus induces a periodic shift on the orbit (as can be seen in Figure \ref{subfig:grould_lensing}).

Finally, an higher-order relativistic effect that one can take into account is Shapiro time delay. This effect, as for GL, alters photon trajectories from the emitter to the observer, due to the presence of a massive object \cite{Shapiro1973}. In particular, photons passing in the gravitational potential well of a compact mass, experience a time dilation, which results in a delay for the observer in receiving the photon. The effect of Shapiro time delay, is thus not accumulated over orbital periods, but it induces a transient modulation in the photon time of arrival. For S-stars on highly-eccentric orbits this effect peaks at pericenter \cite{Angelil2014}, where the photons emitted by the star graze closer to Sgr A* before reaching the observer, and it can be potentially measured by pulse-arrival data. In their numerical study, Grould and collaborators \cite{Grould2017b}, have translated the Shapiro time delay into an observed astrometric shift. In particular, they estimated the magnitude of the effect on the recovered astrometric position of the S2 star, by considering two relativistic orbital models: the first only considers relativistic star motion and light propagation happens on straight lines; the second is a fully relativistic model where both the star and photons follow geodesic paths in a Kerr space-time. This means that at a given moment $t_{\rm obs}$ for the observer the two models predict a slightly different position for S2 as seen by the observer, the second one carrying the additional travel-time of photons due to the massive object. In this way, they demonstrated that Shapiro time delay induces a periodic modulation (Figure \ref{subfig:grould_shapiro}) in the astrometry of S2 whose maximum value of $\sim8\;\mu$as is reached at pericenter and is thus impossible to detect only with astrometric GRAVITY observations of the GC.
The maximum impact of relativistic effects on the astrometric and spectroscopic observables for the S2 star, as studied in \cite{Grould2017b}, is summarized in Table \ref{tab:rel_effects}.

\subsubsection{The shadow of Sgr A*}
\label{sec:sgra_shadow}

Horizon scale images of SMBHs provide a conceptually new avenue for testing GR \cite{EventHorizonTelescopeCollaboration2022f}. Due to the fact that the observed photons originate in the near-horizon deep gravitation field, they carry information on the space-time properties in the strong-field regime \cite{Falcke2000a}. As mentioned in Section \ref{sec:eht}, the most prominent effect on direct images of BHs is their shadow \cite{Falcke2000a}. The BH shadow is by definition inherently related to the geometric properties of space-time and does not depend on astrophysical effects \cite{Johannsen2010, Psaltis2015, Narayan2019}. The shape and size of its boundary are strongly dependent on the mass-over-distance ($M/R$) ratio of the BH and exhibit only a very weak dependence on the BH spin and the observer's inclination. In particular, the shape is perfectly circular for a non rotating BH, its diameter being given by Equation \eqref{eq:shadow_diameter}. For a rotating BH, the shape gets increasingly asymmetric (assuming a D-shaped configuration when observed edge-on, \emph{i.e.} on a line-of-sight perpendicular to the BH spin axis) as the spin parameter $a_0$ increases, but due to a fortunate cancellation between the effects of frame dragging and the quadrupole moment, the mean radius of the shadow boundary is only weakly affected, ranging between  $\sim4.8\;GM/c^2$ to $\sim5.2\,GM/c^2$ \cite{Zakharov2005, Johannsen2010}.

When considering a spherically symmetric space-time, the radial areal coordinate $r_{\rm ph}$ of the photon sphere (\emph{i.e.} the locus of points where photons move onto circular orbits of constant radius $r_{\rm ph}$) can be related to the $g_{00}$ component of the metric tensor via the implicit relation \cite{Psaltis2020a}
\begin{equation}
    r_{\rm ph} = \sqrt{-g_{00}}\left(\left.\frac{d\sqrt{-g_{00}}}{dr}\right|_{r_{\rm ph}}\right).
\end{equation}
The BH shadow as observed at infinity is the lensed image of the photon sphere \cite{Psaltis2008b} and its diameter can be computed as
\begin{equation}
    d_{\rm sh} = 2r_{\rm sh} = \frac{2r_{\rm ph}}{\sqrt{-g_{00}(r_{\rm ph})}}.
    \label{eq:diameter_shadow_th}
\end{equation}
which thus depends only on the $00$-component of the space-time metric and, when particularized for a Schwarzschild space-time, reduces to Equation \eqref{eq:shadow_diameter}. In order to connect gravity tests based on the BH shadows with other tests, one can perform a PN expansion of the $g_{00}$ component of a generic spherically symmetric metric tensor in powers of $r_s/r$, where $r_s$ is the gravitational radius of the central object (Equation \eqref{eq:sch_radius}):
\begin{equation}
    -g_{00} = 1-\frac{r_s}{r}+2(\beta_{\rm PPN}-\gamma_{\rm PPN})\left(\frac{r_s}{r}\right)^2-2\zeta_{\rm PPN}\left(\frac{r_s}{r}\right)^3+\mathcal{O}(r_s/r)^4.
\end{equation}
Here, the coefficients $\beta_{\rm PPN}$ and $\gamma_{\rm PPN}$ are the usual first order parametrized-post-Newtonian (PPN) parameters \cite{Will2014}, while the $\zeta_{\rm PPN}$ encodes the 2PN contribution in the expansion. If $\beta_{\rm PPN} =\gamma_{\rm PPN} = 1$ and $\zeta_{\rm PPN} = 0$, the $g_{00}$ corresponds exactly to Schwarzschild metric. Weak-field tests of gravity in the Solar System have placed stringent constraints on $\beta_{\rm PPN}$ and $\gamma_{\rm PPN}$ at around $1\pm 10^{-5}$ \cite{Will2014}. Assuming these constraints valid at the GC, the 1PN term in $g_{00}$ vanishes, and the diameter of the BH shadow can be written as
\begin{equation}
    d_{\rm sh} = 6\sqrt{3}\left(1+\frac{1}{9}\zeta_{\rm PPN}\right)\frac{GM}{c^2D}.
    \label{eq:shadow_2pn}
\end{equation}
This expression reduces to Equation \eqref{eq:shadow_diameter} when $\zeta_{\rm PPN} = 0$ but quantitatively demonstrates that tests with the BH shadow probe the properties of space-time at least at the 2PN order, with the size of the shadow depending linearly on the magnitude of the 2PN term \cite{Psaltis2020a}.

The accurate knowledge of the $M/R$-ratio for Sgr A* from the stellar dynamics around the GC (see Section \ref{sec:s_stars_obs}) provides with stringent priors on the expected shadow diameter, making the direct imaging of Sgr A* a powerful probe for metric properties \cite{Psaltis2015}.
However, as mentioned in Section \ref{sec:eht}, what VLBI techniques detect is the light emitted at a specific frequency by the accreting flow, which results in the bright emission ring that surrounds the shadow. This means that the quantity that can be effectively measured (albeit with an uncertainty given by the instrumental resolution that blurs the emission features with a Gaussian beam of FWHM $\sim15\;\mu$as \cite{EventHorizonTelescopeCollaboration2022b,EventHorizonTelescopeCollaboration2019f}) is the diameter of the bright emission ring and not that of the shadow itself. Nevertheless, with an appropriate calibration, the ring diameter can be employed as a proxy for the shadow diameter \cite{EventHorizonTelescopeCollaboration2019b, EventHorizonTelescopeCollaboration2019c, Narayan2019, Ozel2021, Younsi2021, Kocherlakota2022}. In particular, suppose that $\hat{d}_m$ is measured diameter of the ring (that for Sgr A* amounts to $51.8\pm2.3\;\mu$as as estimated from the 2022 observation by the EHT collaboration) and $d_{\rm sh}$ is the true shadow diameter for Sgr A*. In order to use the measurement of $\hat{d}_m$ as a proxy for $d_{\rm sh}$ one has to establish a quantitative relation, in the form of a calibration factor, between the two diameters. This relation can be expressed as \cite{EventHorizonTelescopeCollaboration2022f}
\begin{equation}
    \alpha = \frac{\hat{d}_m}{d_{\rm sh}} \qquad\Rightarrow\qquad \hat{d}_m = \alpha d_{\rm sh} = \alpha (1+\delta) 6\sqrt{3}\frac{GM}{c^2 D}.
    \label{eq:calibration_factor}
\end{equation}

Here, $\alpha$ is the calibration factor, which addresses the extent to which the ring diameter can be used as a proxy for the shadow diameter \cite{EventHorizonTelescopeCollaboration2022f}. While $d_{\rm sh}$ is determined only by the metric and its properties, the calibration factor $\alpha$ is connected to the physics of image formation in the near-horizon regime. This factor, thus, involves both features strictly related to theoretical arguments, quantifying the degree to which the ring diameter tracks that of the shadow, for a given space-time metric and a given model of accretion for the plasma, and potential measurement biases. These two components are multiplicative in nature and are generally independent of each other, requiring them to be quantified separately \cite{EventHorizonTelescopeCollaboration2022f}. One can thus write $\alpha = \alpha_1\times\alpha_2$, where $\alpha_1 \equiv d_m/d_{\rm sh}$  is the ratio between the true diameter $d_m$ (not the actual measured value) of the bright ring and that of the shadow $d_{\rm sh}$, given a specific metric and plasma model, and $\alpha_2\equiv \hat{d}_m/d_m$ is the ratio between the measured ring diameter and its true value. In order to estimate $\alpha_1$, the EHT collaboration carried on a large suits of numerical General Relativistic Magnetohydrodynmics (GRMHD) simulations of BH shadows, employing several different models to explore a range of effects related to the characteristic properties of the accretion flow, and space-time metric \cite{EventHorizonTelescopeCollaboration2022f, Ozel2021, Younsi2021}. This allowed to derive a distribution of $\alpha_1$ for a great variety of different physical conditions that one can use as a prior on this parameter. In particular, the distribution for $\alpha_1 - 1$ peaks at small positive values ($\sim 0.05$), meaning that the bright ring tends to peak at slightly larger radii than that of the BH shadow. On the other hand, in order to compute the probability distribution for $\alpha_2$, segments from the aforementioned large suit of simulations can extracted based on image morphology and size, degree of variability, space-time metric, and plasma model, for each of them synthetic VLBI EHT data can be generated \cite{Janssen2019, Roelofs2020a, Natarajan2022}, taking into account interstellar scattering through the Galactic disk, realistic atmospheric, instrumental, and calibration effects. This allows to estimate the ratio between the actual ring diameter and that of the reconstructed image for each simulation. The resulting distribution on $\alpha_2-1$ always encompasses zero, implying that, on average, the imaging and fitting procedures in EHT data tend recover the actual size of the bright ring. Small differences in the inferred diameters between different algorithms are thus ascribable to the different methodologies, priors, and optimization techniques used \cite{EventHorizonTelescopeCollaboration2022f}.

The parameter $\delta$ in Equation \eqref{eq:calibration_factor}, on the other hand, is a purely geometrical factor that quantifies how much the theoretically predicted shadow size for a specific space-time metric differs from that of a Schwarzschild BH (\emph{e.g.} $\delta = \zeta_{\rm PPN}/9$ for a spherically symmetric space-time expanded at 2PPN order as shown in Equation \eqref{eq:shadow_2pn}). In fact, for $\delta = 0$, $d_{\rm sh}$ corresponds to the diameter of a Schwarzschild BH, a value $\delta \neq 0$  would encode deviations from Schwarzschild. For example, the Kerr metric predicts a shadow whose diameter is always less or equal (when $a_0 = 0$ or for an observer along the spin axis \cite{Johannsen2010}) than that of Schwarzschild. Depending on the spin and the inclination, the shadow size can get up to $\sim7.5\%$ smaller \cite{Takahashi2004, Chan2013}, implying a value of $\delta$ in the range $[-0.075, 0]$. A value outside this range can be considered to be in tension with it \cite{EventHorizonTelescopeCollaboration2022f}.

The parameter $\delta$ is the parameter of interest when performing tests of gravity with the BH shadow. However, extracting $\delta$ from the measurement of the bright ring diameter $\hat{d}_m$ requires to take into account four sources of uncertainty, relate to all the various factors in Equation \eqref{eq:calibration_factor}:

\begin{enumerate}
    \item the uncertainty on the $M/R$-ratio of Sgr A* from stellar dynamics as reported in Section  \ref{sec:s_stars_obs};
    \item the observational uncertainty on the bright ring diameter $\hat{d}_m$;
    \item the theoretical uncertainties in the ratio $\alpha_1$ between the true diameter $d_m$ (not the actual measured value) of the bright ring and that of the shadow $d_{\rm sh}$, given a specific metric and plasma model;
    \item the uncertainties in the ratio $\alpha_2$ between the measured ring diameter and its true value, due to the instrumental and fitting techniques in order to retrieve the bright ring diameter.
\end{enumerate}

\begin{figure}
    \centering
    \includegraphics[width = 0.75\textwidth]{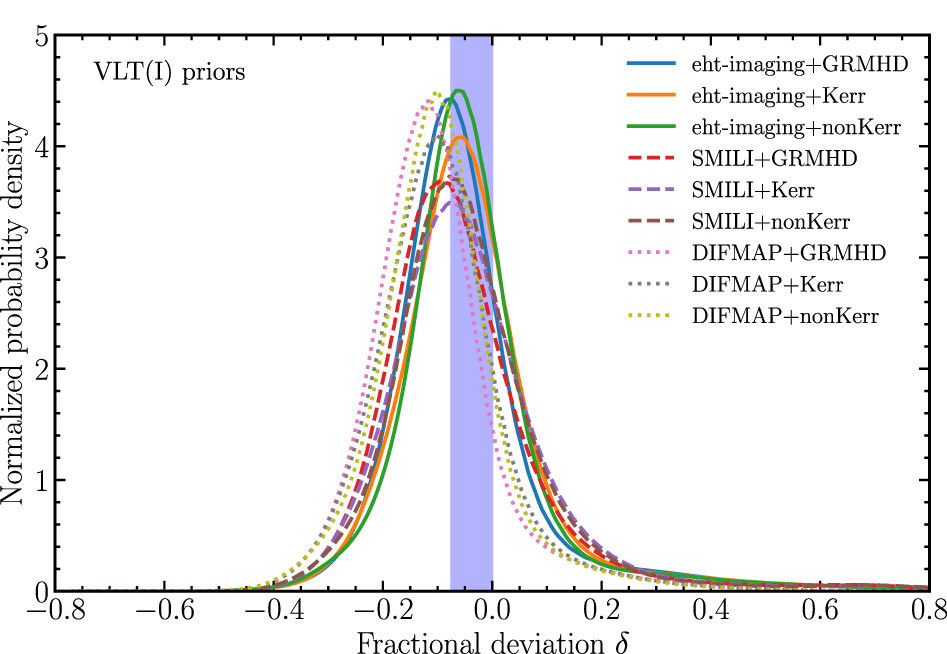}
    \caption{Posterior distributions on the deviation parameter $\delta$ between the shadow size obtained for EHT observations of Sgr A* and theoretical predictions for a Schwarzschild BH. The $M/R$-ratio considered here is the one coming from the MPE group in 2020 using the orbit of the S2 star \cite{GravityCollaboration2020c}. As reported in the legend, the various curves correspond to different sets of theoretical models used for deriving the $\alpha_1$ calibration factor (either GRMHD simulations \cite{Prather2021, Porth2017, Moscibrodzka2018, Younsi2012, Younsi2016}, using a set of covariant plasma models in the Kerr metric \cite{Ozel2021}, or using analytic models exploring a range of BH metrics parametrically different from the Kerr, such as Johannsen–Psaltis - JP - metric \cite{Johannsen2011, Johannsen2013b, Younsi2021}) and and different imaging techniques (either eht-imaging , SMILI or DIFMAP \cite{EventHorizonTelescopeCollaboration2022d}) for the shadow diameter. The purple shaded area shows the $\sim7.5\%$ range consistent with a Kerr metric, for different values of the BH spin and observer inclination.}
    \label{fig:posterior_delta}
\end{figure}

Taking all these uncertainties into account, one can express the posterior probability on the deviation parameter $\delta$, given an observed diameter $\hat{d}$ for the bright ring, as
\begin{equation}
    P(\delta|\hat{d}) = C\int d\alpha_1\int d\alpha_2 \int d(M/R) \,\mathcal{L}[\hat{d}|M/R, \alpha_1, \alpha_2,  \delta]P(\delta)P(M/R)P(\alpha_1)P(\alpha_2).
    \label{eq:posterior_delta}
\end{equation}
In this expression, $C$ is a normalization constant, $\mathcal{L}[\hat{d}|M/R, \alpha_1, \alpha_2, \delta]$ is the likelihood of obtaining a measurement $\delta$, given a specific set of parameters $(M/R, \alpha_1, \alpha_2, \delta)$ for the model \eqref{eq:calibration_factor}. The multiplying functions $P(\dots)$ are the prior probability distributions on each parameter, derived as described above. The integral in Equation \eqref{eq:posterior_delta} can be computed numerically \cite{EventHorizonTelescopeCollaboration2022f} and results (taking the prior for the $M/R$-ratio from the Gravity collaboration analysis of the S2 star orbit \cite{GravityCollaboration2020c}) are shown in Figure \ref{fig:posterior_delta}. They indicate that even though the current determination by the EHT Collaboration of the shadow diameter for Sgr A* is perfectly compatible with the theoretical expectation for a Kerr BH (whose diameter range is reported as a purple vertical strip), the posterior for the deviation parameter $\delta$ results in a larger credible interval (reported in Table 2 of \cite{EventHorizonTelescopeCollaboration2022f} for all the combinations of accretion models and imaging reconstruction techniques) that could in principle allow for deviations of the shadow diameter beyond the Kerr metric.

\subsection{The nature of Sgr A* under probe}
\label{sec:nature_Sgr}

The evidence accumulated over the years that Sgr A* contains a large mass confined into a compact volume is compelling. Analysis of gas dynamics measurement (Section \ref{sec:gas_dynamics}), stellar orbital motion (Section \ref{sec:stellar_dynamics}) and the successful observation of a horizon-scale emission (Section \ref{sec:eht}), consistent with accreting plasma, and a brightness depression in Sgr A* all point towards a $\sim4\times10^6$ solar masses object confined into an angular diameter of $\sim 50\;\mu$as. However, whether this object is \textit{exactly} a BH as described by GR or not still remains unclear \cite{EventHorizonTelescopeCollaboration2022f} and its intrinsic nature is still debated.

The orbital motion of the S-stars, with the successful detection of relativistic effects in the astrometric and spectroscopic observables, has been a powerful tool to test the validity of GR against Newtonian gravity with unprecedented accuracy (Section \ref{sec:test_GR}). It is, hence, a natural question whether the motion of such stars could provide hints about the intrinsic nature of the central object, whether BH or an alternative object. With the term \emph{alternative}, we refer to other classes of exotic compact object that, still remaining in the framework of GR (\emph{i.e.} not considering extensions or modifications to the underlying theory of gravity, these will be discussed separately in Section \ref{sec:alternative_theories}) could behave as BH-mimickers (see \cite{Cardoso2019} for an extensive review of such objects). A BH mimicker is by definition a possible alternative to a BH, that would look observationally almost identical to a BH but would have no horizon \cite{Lemos2008, Danielsson2021, Chirenti2007, Morris1988a, Visser1989a, Brito2015a, Visinelli2021}. The properties in the strong-field near-horizon region can differ substantially between a BH and a mimicker, but at infinity the space-times produced by the two kinds of objects tend to approach to each other. An alternative to the SMBH paradigm that has been widely investigated for Sgr A* is the wormhole, a singularity-free solution of Einstein's field equations that connects two different regions of space-time through a geometrical throat \cite{Morris1988a, Bambi2021}. In particular, the effects on the motion of the S-stars of replacing Sgr A* with a wormhole have been studied in \cite{Dai2019, Simonetti2020}, showing peculiar secular effects on the orbit of the S2 star. In particular, they show that a wormhole connects two different space-times, not only casually, but also in terms of field propagating from one side to the other. These are the cases of the either scalar, electromagnetic, or gravitational fields and, thus, particles that move in the vicinity of a wormhole in one space must feel the influence of objects on the other side of the throat. It is using this argument that they demonstrate that if a star of a few solar masses exists orbiting around Sgr A* on the other side of the wormhole, at the distance of a few gravitational radii from the central object, it would interact gravitationally with S2 on our side, leaving a detectable imprint on its orbit that could be detected by observations of the GC with sufficient precision (such that could detect accelerations of S2 of order $10^{-6}$ m/s$^2$).

A more quantitative approach on constraining a BH-mimicker wormhole model with the orbits of the S-stars has been performed in \cite{DellaMonica2022b}. Della Monica and De Martino used the proper motion data for S2 up to 2016 \cite{Gillessen2017}, complemented with the orbital precession measurement in 2020 by the Gravity Collaboration \cite{GravityCollaboration2020c}, to statistically constrain a Black Bounce model \cite{Simpson2019}. These kind of model is described by the space-time line element
\begin{equation}
    ds^2 = -\left(1-\frac{2M}{\sqrt{r^2+\alpha^2}}\right)dt^2+\left(1-\frac{2M}{\sqrt{r^2+\alpha^2}}\right)^{-1}dr^2+(r^2+\alpha^2)(d\theta^2+\sin^2\theta d\phi^2),
    \label{eq:bb_metric}
\end{equation}
which presents an additional parameter $\alpha$ with respect to the Schwarzschild BH. This parameter allows to uniformly map between a BH for $\alpha = 0$; a regular BH (free of singularities in the center, but endowed with an horizon) for 
$0 < \alpha < 2M$, a one-way transversable wormhole (\emph{i.e.} with its throat coinciding with the horizon) for $\alpha = 2M$, and a two-way transversable wormhole (no singularity and no horizon) for $\alpha > 2M$ \cite{Morris1988a, Morris1988b, Visser1989a, Visser1989b}. The constraints on $\alpha$ derived from this analysis \cite{DellaMonica2022b}, not only showed that current observational data (limited to the public availability of S2 proper motion datasets) don't have enough constraining power to say anything conclusive on the nature of Sgr A* (being only able to place an upper limit on the size of the throat $\alpha \lesssim 70 $ at $99.7\%$ confidence level), but also that neither high-precision observations of S2 with the GRAVITY interferometer (using a mock catalogue for this analysis) could shed light on the nature of Sgr A*. Finally, it was estimated that only the discovery of a much closer and faster star (with a period of $\sim$5 years and a pericentre distance of $\sim5$ AU) observed with a uniform phase coverage by an instrument with GRAVITY-like accuracy could identify the nature of the central object at almost $5\sigma$ confidence.

{
Recently, Cadoni et al. \cite{Cadoni2023} have investigated the possibility to test a model of nonsingular BH in the GC using the motion of the S2 star. Contrary to classical BHs, presenting a singularity at their core and subsequently implying a breakdown of the classical theory, these objects are completely regular everywhere, retaining a GR framework. An intriguing feature of such objects is the presence of a hair appearing in the form of a new scale length $\ell$ responsible for the smearing of the classical singularity, which can be hierarchically larger than the Planck scale. The model considered in this work is described in the usual Schwarzschild coordinates by the space-time metric
\begin{equation}
    ds^2 = -f(r)dt^2+\frac{dr^2}{f(r)}+r^2d\Omega^2,
\end{equation}
where $d\Omega^2 = d\theta^2+\sin^2\theta d\phi^2$ and
\begin{equation}
    f(r) = 1-\frac{2GMr^2}{(r+\ell)^3}.
\end{equation}
The Schwarzschild space-time is recovered in the limit $\ell\to0$, and the model is particularly interesting for two reasons: first, it represents the most powerful deformation of the classical BH picture that is still compatible with Schwarzschild asymptotically. This leads to strong deviations from the standard GR phenomenology, having a clear observable experimental signature when the hair $\ell$ is super-Planckian. Secondly, it has a nice astrophysical analogy. In fact, the mass density associated with the solution reproduces that of some density profiles of dark matter in elliptical and spherical galaxies. Interestingly, in \cite{Cadoni2023} a first order analytical expression for the orbital precession of massive test particles is derived
\begin{equation}
    \Delta\omega = \Delta\omega_{\rm GR}\left(1-\frac{\ell}{GM}\right)
\end{equation}
where $\Delta\omega_{\rm GR}$ is the precession for a Schwarzschild BH (Equation \eqref{eq:precession}). When $\ell = GM$ the precession in this model is predicted to be 0, while for a hair greater than the gravitational radius $\ell > GM$ it becomes retrograde ($\Delta\Omega < 0$), providing a clear phenomenological signature of this model. Observational data for S2, along with the measurement of its orbital precession, result in a constraint on the hair of $\ell \lesssim 0.47GM$ at 95\% level, which, while tightly restricting the parameter space of the model, is not able to completely rule out the possible existence of a regular BH with a super-Planckian hair at the GC.
}

\begin{figure}
    \centering
    \includegraphics[width = \textwidth]{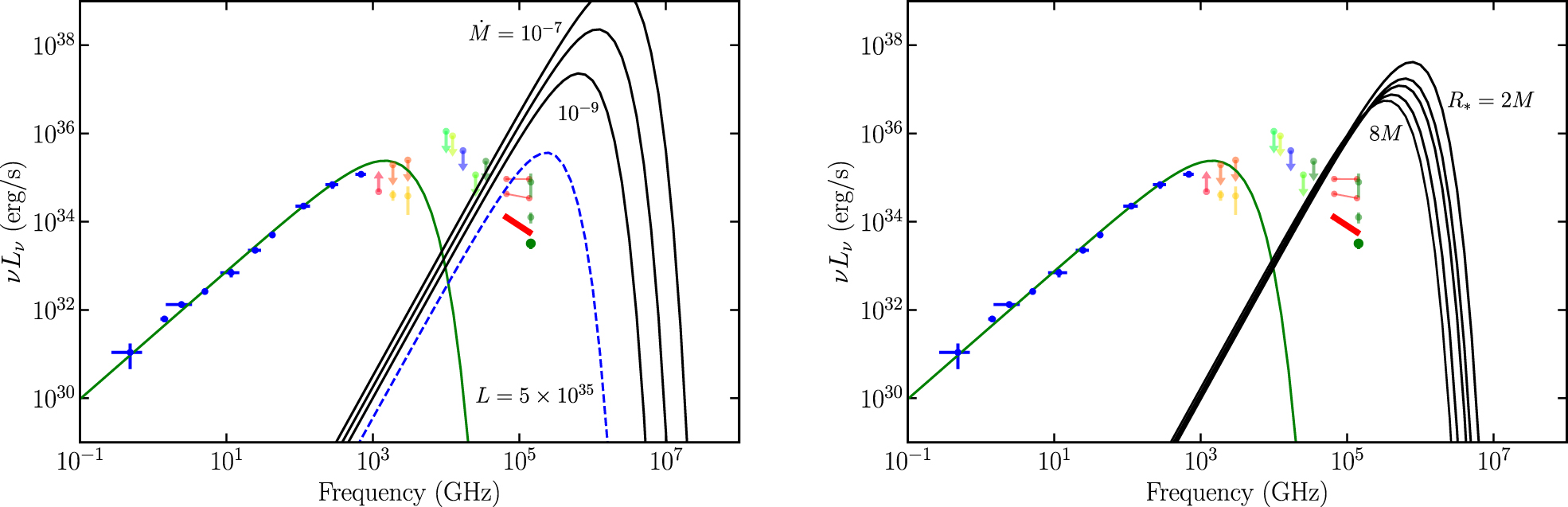}
    \caption{Predicted SEDs, of Sgr A* resulting from the synchrotron-radiation-emitting accretion flow (green solid line) in the radio and sub-millimeter part of the spectrum and the presence of a thermalizing surface (black solid lines). In particular, on the left plot the three solid lines represent, given a fiducial radius for the surface $R_* = 2.5GM/c^2$, the black-body thermal emission assuming three different accretion rates of $\dot{M} = 10^{-7},\,10^{-8},\,10^{-9}\,M_\odot/$yr; on the right, assuming $\dot{M} = 10^{-9}M_\odot/$yr and five different thermalizing surface radii $R_* = 2,\,3,\,4,\,6,\,8\,GM/c^2$. The dashed blue line corresponds to a conservative lower bound on on the surface luminosity of $5\times10^{35}$ erg/s. Filled dots report data and upper limits on the infrared luminosity of Sgr A* in various bands, along with a red solid line representing the 5th percentile of the flaring infrared emission \cite{Witzel2018, GravityCollaboration2020b}. Image is courtesy of \cite{EventHorizonTelescopeCollaboration2022f}.}
    \label{fig:eht_therm_surface}
\end{figure}

On a much smaller observational scale, confrontation between EHT observed images of Sgr A* and the numerical simulation shows that the observed brightness distribution is consistent with the presence of an event horizon. However, conclusive evidence that all other possible alternatives are ruled out is still missing \cite{EventHorizonTelescopeCollaboration2022f}. Among these objects, in particular, observations of the BH shadow can provide constraints, or completely rule out, models of compact objects endowed with surface (either a thermalizing or a reflecting one), {exotic (boson and Proca) stars (see \cite{Cardoso2019} for an extensive review of such objects and \cite{LuisRosa2022a,LuisRosa2022b} for the computations of the shadow and the observational signatures of hot spots orbiting such object) or naked singularities.}
The presence of a thermalizing surface, would mean that the thermal and kinetic energy carried by the plasma in the accretion flow\footnote{Hot gas with near-virial temperature, located external to the central object, is expected to either be accreting towards the center, flowing out in a jet, or a combination of the two effects \cite{Rees1982, Falcke1993, Narayan1995a, Falcke2013, Yuan2014}. Since the accreting plasma is expected to be radiatively efficient, a lot of the energy $\dot{M}c^2$ (where $\dot{M}$ is the accretion rate), reaches the surface of the central object.}, would be radiated back to infinity once the system reaches a steady state. This, on turn, would result into a large surface luminosity that should be visible to a distant observer \cite{Broderick2006b}.
On the other hand, in the case of a BH this energy would disappear through the event horizon. This process happens under three ‘‘natural'' minimal assumptions \cite{EventHorizonTelescopeCollaboration2022f}: \emph{(i)} energy conservation is satisfied in the compact object at the center of Sgr A*; \emph{(ii)} laws of thermodynamics are applied to such object and, in particular, the system tends to approach statistical equilibrium in steady state; \emph{(iii)} the compact object couples and emits in all electromagnetic modes. Some exotic models, \emph{e.g.} a shell-like BH mimicker \cite{Danielsson2021} or gravastars \cite{Chirenti2007}, could radiate only in specific electromagnetic modes or not radiate at all, violating assumption (iii). Such models cannot be constrained using astronomical observations, not even the horizon-scale EHT observations, and only complementary analyses, \emph{e.g.} with gravitational waves, could potentially distinguish them \cite{Abbott2021}. Matter that accretes via a radiatively inefficient process from infinity onto a spherically symmetric object endowed with a thermalizing surface at radius $R_*$, releases a thermal energy that is a fraction $\eta$ of the rest-mass energy of the gas. Such fraction for a Schwarzschild space-time is given by \cite{Broderick2006b}
\begin{equation}
    \eta = 1-\left(1-\frac{2GM}{c^2R_*}\right)^{1/2}\,.
\end{equation}
Once the object reaches steady state, the released energy is radiated back to infinity, providing an extra luminosity from the thermalizing surface, that is much larger than the luminosity radiated by the hot plasma itself during accretion\footnote{For stellar-mass BHs, this phenomenon provides an effective mechanism to distinguish BHs and neutron stars
\cite{Narayan1995b, Narayan1997, Garcia2001, Done2003, McClintock2004}.}. For SMBHs, on the other hand, the radiation observed in the submillimeter band provides a lower limit on $\dot{M}$ (whether radiation is produced by the in-flowing plasma or by the out-flowing jet) that, for a given surface radius $R_*$, can be converted into a minimum luminosity of the thermalizing surface that can be observed at infinity \cite{EventHorizonTelescopeCollaboration2022f},
\begin{equation}
    L_\infty > \eta \dot{M}_{\rm min}c^2.
\end{equation}
EHT observations allowed for the first time to place an upper limit on the radius of the putative emitting surface, $R_* \leq  8GM/c^2$ at 95\% confidence \cite{EventHorizonTelescopeCollaboration2022f}. Most importantly, thermal radiation from the surface is expected to produce an infrared peak in the Spectral Energy Distribution (SED), $\nu L_\nu$ as a function of $\nu$ (where $L_\nu$ is the specific luminosity at frequency $\nu$), of Sgr A* in the form of black-body radiation, due to the optical thickness of the gas \cite{McClintock2004, Broderick2006c, Broderick2007}, in addition to the sub-millimeter peak from the accretion flow. Observations in the near-infrared \cite{Eckart2004, Eisenhauer2005, Hora2014, Witzel2018}, however, have provided with very stringent upper limits on the infrared emission of Sgr A* that lie far below the predicted minimum surface luminosity, thus implying that Sgr A* does not have a radiating surface (Figure \ref{fig:eht_therm_surface}).

Summarizing, the EHT observation of Sgr A* allowed to constrain the surface radius at $R_* < 8GM/c^2$, whereas earlier works could only be sensible to much larger scales (\emph{e.g.} the upper limit on $R_*\lesssim1000 GM/c^2$ in 2002 \cite{Narayan2002b} or $R_*\lesssim100 GM/c^2$ in 2009 \cite{Broderick2009b}) and the analysis on the SED and the infrared constraints of Sgr A* allowed to strengthen even further the evidence in disfavour of a thermalizing surface. However, some phenomena could potentially weaken the arguments in favour of an event horizon and it is important to take them into account. In \cite{EventHorizonTelescopeCollaboration2022f} all the various possibilities are considered. For example, the presence of dust could potentially obscure radiation emitted by the surface; the radiation from the surface, being emitted from deeper within the potential well of Sgr A*, could be affected by a curvature-induced time delay and thus it could not yet be reaching a distant observer; similarly, gravitational redshift could dim the observed luminosity, making the thermalizing surface impossible to detect; the surface itself could have not yet reached steady state equilibrium with the accretion flow; moreover, the surface has to grow as the horizon radius radius grows due to mass accretion, limiting further the radius of the surface. Nevertheless, strong arguments against each of these possibility have been proposed, leaving only a very narrow window of models with thermalizing surfaces that are between $10^{-21}$\% and $10^{-12}$\% larger than the event horizon, that are currently allowed to have a thermalizing surface and are not yet ruled out by infrared constraints \cite{EventHorizonTelescopeCollaboration2022f}.
Moreover, shortly after the 2022 publication of the EHT results for Sgr A*, criticism emerged \cite{CarballoRubio2022} on the reported arguments against a thermalizing surface, as it relies on the strong assumption of perfect balance in the energy exchange between the accretion disk and the central object. It has been shown that if this assumption is violated, which happens if the surface has a non-zero absorption coefficient, the very low infrared emission by Sgr A* could be justified even with the presence of a thermalizing surface. The conclusions by the EHT collaboration, on the other hand, would hold only for unnaturally small values of the absorption coefficient ($\ll10^{-14}$).

A different possibility is that the putative surface of Sgr A* reflects incident radiation with a given albedo $A$ (\emph{i.e.} the ratio between the intensity of reflected v.s. incident radiation). A series of simulated images have been produced by the EHT collaboration \cite{EventHorizonTelescopeCollaboration2022f}, demonstrating that models in which Sgr A* has a reflecting surface with perfect albedo, A = 1, can potentially be distinguished by the EHT 2017 observations due to a drastically different radius of the central brightness depression. However, models with only partial albedo ($A\lesssim0.3$), are harder to distinguish from the case of a BH. In the cases where, $A < 1$, due to the conservation of energy, a fraction $(1 - A)$ of the incident radiation on the reflecting surface has to be absorbed and will radiated back to infinity, through the steady state thermalization process already discussed.

Moreover, EHT observations of Sgr A* have successfully ruled out (mini-) boson star models \cite{Cardoso2019, Olivares2020} for Sgr A*. As a matter of fact, GRMHD simulations of radiatively inefficient accretion flows on such kind of horizonless BH mimickers \cite{Olivares2020} have shown that in the case of a stable boson star there total lack of a central brightness depression, with the inner portion of the image being very bright. Unstable boson stars simulated observations, on the other hand, despite their space-times not possessing an unstable photon orbit (and thus an actual shadow), have been demonstrated to posses a central brightness depression, ascribable to a centrifugal barrier \cite{Olivares2020}. However, this brightness depression is expected to have a significantly smaller intrinsic size than the counterpart for a BH. EHT observations, thus allow to exclude these models, since image morphologies would be generally too compact and lack both the ring-like feature and the central shadow-related brightness depression \cite{EventHorizonTelescopeCollaboration2022f}. However, in order to conclusively rule out boson stars as alternatives to the BH model for Sgr A*, a more extensive study of their spin, compactness, and astrophysical properties is required \cite{Vincent2021}.

Other models, however, are excluded right away by only by considering presence of a central brightness depression in the EHT images of Sgr A*. For example, a non-spinning Joshi–Malafarina–Narayan-2 \cite{Joshi2014}
naked singularity\footnote{This model is an exact solution of the Tolman–Oppenheimer–Volkoff equations of GR \cite{Misner1973}, formed as the non-empty end state of the gravitational collapse of a non-thermal perfect fluid from regular Cauchy data.} is expected not to possess a shadow at all \cite{Shaikh2019}, and thus no argument could make this model compatible with actual observations of Sgr A* \cite{EventHorizonTelescopeCollaboration2022f}.

\subsection{Testing Extended Theories of Gravity at the Galactic Center}
\label{sec:alternative_theories}

\begin{figure}[!t]
    \centering
    \includegraphics[width = \textwidth]{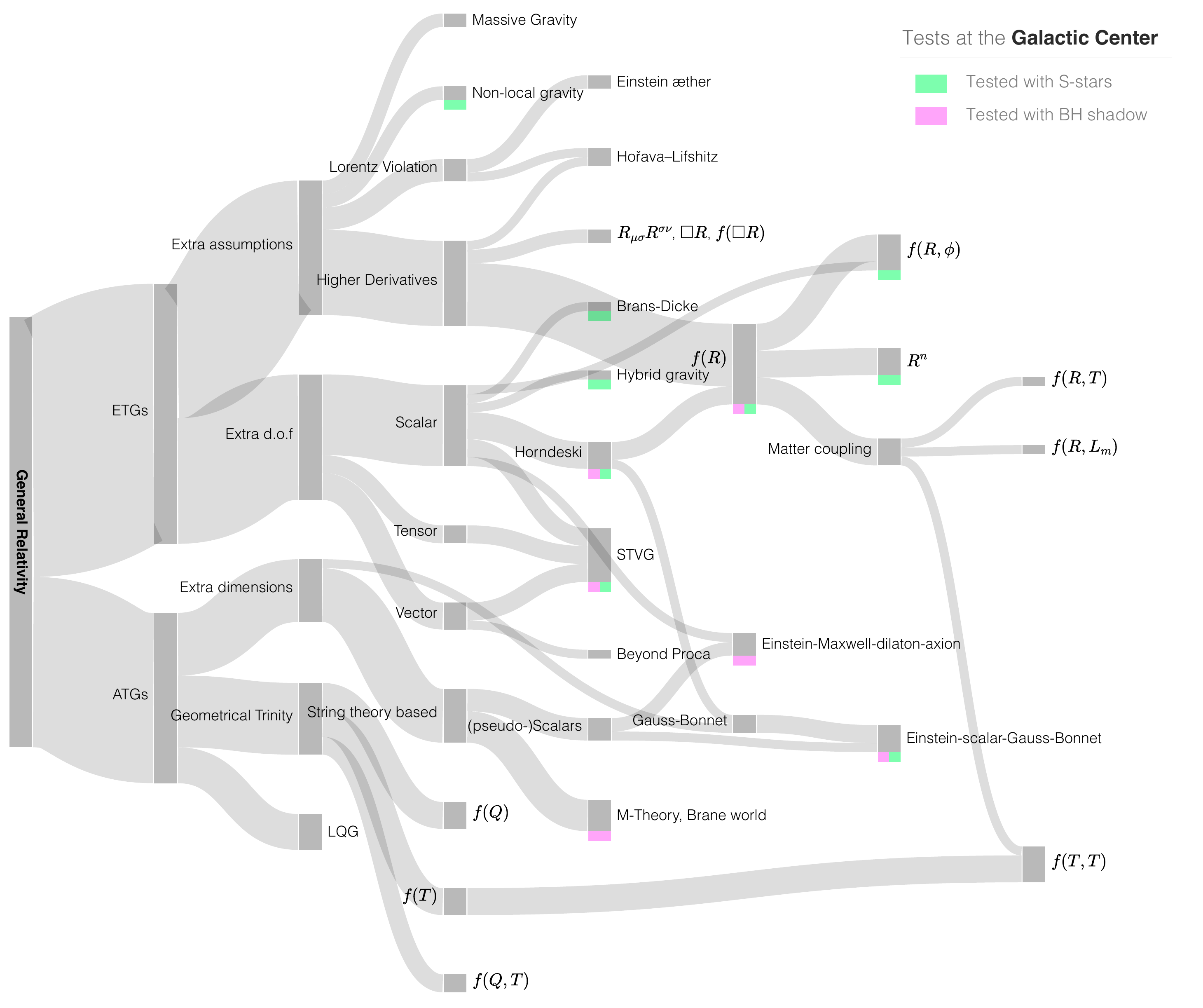}
    \caption{Diagram of families of modifications/extensions/alternatives to general relativistic theory of gravity and the assumptions from which they arise. Theories for which tests at the GC, either using the S-stars orbits or the shadow of Sgr A*, are present in literature have been highlighted with a green or a pink square, respectively.}
    \label{fig:etg}
\end{figure}

Extended Theories of Gravity (ETGs) are usually regarded as a new paradigm to overcome shortcomings of GR at infrared and ultraviolet scales \cite{Capozziello2011}. As a matter of fact, their formulation is strongly motivated by the possibility to naturally explain cosmological and astrophysical data at different scales without the need to introduce exotic forms of energy and matter. For example, GR fails at reproducing the rotation curves of nearby galaxies \cite{Freeman1970} and to account for the velocity dispersion \cite{Zwicky1933} and for the strong and weak gravitational lensing of galaxy clusters \cite{Clowe2004, Ellis2010, Salucci2021}. Finally, on cosmological scales GR cannot explain the emergence of Large Scale Structures in the observable Universe, it doesn't account for the accelerated expansion of the Universe and it breaks down in the description of the first stages of the Universe in the current Big Bang theory. The introduction in Einstein's Field Equations of a cosmological constant $\Lambda$, representing a dark energy \cite{Frieman2008} permeating the Universe, can solve the problems of the theory on cosmological scales \cite{Adam2016, Abbott2019b}, while the existence of dark non-baryonic matter is postulated to solve the problems of the theory on the scales of galaxies and galaxy clusters. At the present time, though, no direct evidence of dark matter or energy has been experimentally achieved \cite{DeMartino2020}. The phenomenology that emerges when additional degrees of freedom of the gravitational field are taken into account can provide with a natural explanation for all this range of observations, encompassing, in a self-consistent scheme, flaws of GR at all scales. All extensions to GR have to preserve the positive results of Einstein's Theory in weak field limit, \emph{i.e.} on the scales of the Solar System, where we have stringent tests validating it \cite{Will2014}.
{
    However, except for this obvious observational restriction, there are many ways one can construct a self-consistent theory of gravity extending GR. In Figure \ref{fig:etg} we report a diagram in which we aim at giving a compact (yet not exhaustive) overview of the plethora of possible extensions to GR (for more comprehensive summaries we refer to \cite{Capozziello2011,Nojiri2017,Saridakis2021}). One possible alternative is altering the founding hypotheses of GR, either by considering extra assumptions or by violating preexisting-ones. It is the case for example of theories where the tensorial degree of freedom of gravity is endowed with a non-zero mass (the so called massive gravity \cite{deRham2014}), or where the gravitational interaction is considered as non-local \cite{Capozziello2022}. Violation of the Lorentz invariance \cite{Mattingly2005} gives rise to theories in which a preferred reference frame exists like Einstein aether \cite{Jacobson2008} or Hořava–Lifshitz gravity \cite{Mukohyama2010}. In the latter case, the theory is regulated by field equations presenting spatial derivatives higher than the second order \cite{Blas2018}. More in general, one can consider theories that, while preserving all the other basic assumptions of GR (locality, Lorentz invariance, ecc.) present higher-curvature terms in the action and thus naturally give rise to higher order field equations. This is the case of $f(R)$ gravity (see Section \ref{sec:fR}) or theories whose action presents terms of the kind of $R_{\mu\nu}R^{\mu\nu}$, $\Box R$, etc. \cite{Capozziello2011}. In the metric formalism, $f(R)$ can be regarded as GR plus an additional scalar degree of freedom \cite{Magnano1994}. In general, considering extra scalar-tensorial-vectorial degrees of freedom in the action is a viable way of formulating meaningful extensions to GR. In 1974 G.W. Horndeski derived the most general scalar-tensor thoery giving rise to second-order field equations \cite{Horndeski1974}. This theory, Horndeski-gravity, due to its generality, incorporates a broad variety of families of ETGs, among which sub-classes of $f(R)$ and Gauss-Bonnet gravity \cite{Fernandes2022}. More recently, a theory in which all possible additional degrees of freedom are taken into account, namely Scalar-Tensor-Vector Gravity (STVG, see Section \ref{sec:STVG}) has been formulated \cite{Moffat2006}. Finally, it is worth to notice that for all the mentioned ETGs the modifications to the theory are introduced in the form of extra degrees of freedom (extra fields, higher-order terms in the action, etc.) that do not alter the basic geometrical (metric + connection + curvature) formulation of GR. However, gravity theories which arise as distinct geometrical formulations have been proposed. These are based on the fact that, while in the classical GR picture gravity is described in terms of curvature on a torsion-free ($T=0$, where $T$ is the torsion tensor) metric ($Q=0$, where $Q$ is the non-metricity tensor) manifold, the very same underlying theory can be equivalently described in terms of either torsion ($T\neq 0$) or non-metricity ($Q\neq 0$) while considering the manifold as flat ($R=0$) \cite{Beltran2019}. While at GR level this \emph{geometrical trinity} \cite{Beltran2019} provides with a fully equivalent description of gravity, Alternative Theories of Gravity (ATGs) have been proposed to extend the theory starting from these equivalent description. It is the case, for example, of \emph{teleparallel} ATGs which, applying the same idea behind $f(R)$, consider a modification of the form $f(T)$ to the teleparallel equivalent of the GR action \cite{Cai2016}, $f(Q)$ theories that build upon the non-metricity equivalent \cite{Jimenez2020}, or even $f(Q,T)$ theories where both entities are modified \cite{Xu2019}. All such theories, along with other ATGs like string theory Brane-World gravity \cite{Maartens2010}, do present different phenomenology that, especially in cosmological scenarios, have gained much interest in the last years \cite{Khyllep2023, Paliathanasis2021, Najera2022, Anagnostopoulos2021, Ayuso2021, Jimenez2020}. However, these ATGs have been applied only in a very limited number of cases to non-cosmological settings and only a few tests exists in the GC for such ATGs \cite{Jusufi2022,Jusufi2022b} (in which it is demonstrated that within the current sensitivity observable features in these models cannot be distinguished from that of a Schwarzschild BH), so in this review we will preferentially focus on the much broader literature that is present for ETGs.
}

\begin{table*}
    \small
    \caption{A summary of works in literature considering the orbits of the S-stars in ETGs or  models where Sgr A* is described as an exotic object or is surrounded by a given dark matter distribution. The main results/constraints are reported along with references for each class of models considered. The table is adapted from \cite{GravityCollaboration2019b}.}\label{tab:eGR_literature}
        \renewcommand{\arraystretch}{1.5}
        \footnotesize
        \resizebox{\textwidth}{!}{
        \parbox{1.05\linewidth}{
        \begin{tabularx}{\linewidth}{|l|XXr|} 
            \hline
            \multicolumn{2}{|c}{Model} & Content &  Ref. \\\hline
                        \multirow{5}{*}[-5em]{\rotatebox{90}{Exotic objects}} & Charged non-rotating BH &  Upper limit on th BH charge from the upper limit on the precession of S2 & \cite{DeLaurentis2018a, Iorio2012a, Zakharov2018b} \\
                        & Nonsingular BH with super-Planckian hair & Motion of the S2 star strongly constrains the value of the hair but does not rule out the existence of such objects & \cite{Cadoni2023} \\
                        & Charged rotating BH and plasma effects & Upper limits from BH mass, spin and local magnetic field & \cite{Zajavcek2018}\\
                        &  Fermion ball & Ruled out by \cite{Ghez2005b} and \cite{GravityCollaboration2018a}& \cite{Munyaneza2002}\\
                        &  Boson star & Effects on S2 orbit are much smaller than those of GR, and are only relevant at a few tens of gravitational radii & \cite{AmaroSeoane2010,Boshkayev2019,Grould2017a} \\
            
            & Wormhole & The orbit of S2 with current sensitivity cannot detect the BH/wormhole nature of Sgr A*. Closer S-stars are required & \cite{Dai2019,DellaMonica2022b} \\ \hline
                        \multirow{11}{*}[-13em]{\rotatebox{90}{Extended theories of gravity}} & Yukawa potential& Upper limits on potential parameters and graviton mass from S-stars orbits and precession upper limit &\cite{Borka2013,Hees2017,Zakharov2016,Zakharov2018a,DAddio2021,Borka2021} \\
                        & Einstein-Maxwell-Dilaton-Axion gravity & Effects smaller than $10^{-3}$ of GR for S2, need pulsars or inner stars for further test & \cite{DeLaurentis2018a}\\
                        & Brans-Dicke theory & Effects smaller than $10^{-3}$ of GR for S2, need pulsars or inner stars for further tests. & \cite{DeLaurentis2018a,Kalita2018}\\
                        & $f(R)$ gravity & Effects smaller than $10^{-3}$ of GR for S2, however orbital precession demonstrated to have great constraining power on the parameters of the theory.&\cite{Capozziello2014,DeLaurentis2018a,DeLaurentis2018b,Kalita2018,DeMartino2021,Borka2021,Lalremruati2022a,Debojit2023}\\
                        & STVG & Orbital precession scales linearly with the extra parameter of the theory and this allows a constraint of the model & \cite{DellaMonica2022a,DellaMonica2023b} \\
                        & Non-local gravity & Precession compatible with observational upper limit, of the order of GR prediction & \cite{Dialektopoulos2019}\\
                        & Scalar-tensor gravity & Precession is $13\times$ GR value, and hence ruled out by orbital precession measurement & \cite{Borka2021}\\
                        & $f(R,\phi)$ gravity & Best fit precession prediction for S2 is $20\times$ larger than GR value, and hence ruled out by orbital precession measurement & \cite{Capozziello2014}\\
                        & Hybrid Gravity & Best fit precession prediction too high, ruled out by orbital precession measurement & \cite{Borka2016}\\
                        & $R^n$ gravity & When compared with \cite{Hees2017} upper value, the GR value ($n=1$) is recovered to  $<1\%$, or smaller if extended mass distributions are present & \cite{Borka2012,Zakharov2014}\\
                        & Quadratic Einstein-Gauss-Bonnet gravity & Derive expressions for gravitational redshift in function of theory coupling parameters (scalar/matter \& scalar/Gauss-Bonnet invariant).&\cite{Hees2019a}\\ 
                        & Horndeski gravity & Data for S2 are able to constrain two of the five parameters of the theory that emerge at Newtonian level & \cite{DellaMonica2023a} \\ \hline
                        \multirow{3}{*}[-4em]{\rotatebox{90}{Dark matter}} & Dark matter profiles & Dark matter mass required to explain TeV emission compatible with orbital upper limits. Limits on spatial distribution of non-annihilating dark matter.&\cite{dePaolis2011, Zakharov2009,Dokuchaev2015,Hall2006,Iorio2013,Lacroix2018,Zakharov2007,Shen2023}\\
                        & Scalar fields and ultralight dark matter & Upper limits on scalar field mass (1\% of BH) for particles of  mass $4\times 10^{-19}~\mathrm{eV/c}^2$&\cite{NitsanBar2019,Yuan2022,DellaMonica2022c,Lalremruati2022a,Lalremruati2022b,Debojit2023}\\
            & Fermionic dark matter, RAR model & Fermionic dark matter concentration can offer alternative to SMBH paradigm that can lead to the same orbital motion for the S-stars & \cite{Ruffini2015,BecerraVergara2020}\\
                        \hline
        \end{tabularx} 
        }}
\end{table*}

Even though the evidence supporting a SMBH scenario as described in GR at the GC is extremely strong (and this hypothesis has passed much more tests than other alternatives \cite{Eckart2017}), extensions to GR have been widely studied in the context of S-stars orbits. As shown in Figure \ref{fig:eht_gravitational_probes}, the S-stars explore a regime of gravity where no other probes are available, providing with a test bench for theories of gravity that had never been considered before. Moreover, the possibility to track single orbits for the S-stars with exquisite precision (Section \ref{sec:s_stars_obs}) and the ability to detect relativistic signatures on the orbital motion of these object, allows to exhibit features of their dynamics directly related to the geodesic nature of their motion. This offers a great advantage compared to other tests of gravity in the (Newtonian) weak field limit. As a matter of fact, the relativistic geodesic approach in the description of the motion of test particles has been demonstrated to represents the best strategy for tests of modified theories of gravity \cite{DeLaurentis2018c}. Among the almost countless extensions to GR that have been proposed over the years (see Figure \ref{fig:etg}), in Table \ref{tab:eGR_literature} we report a summary of the models whose astrometric signatures have been studied for the S-stars and those for which experimental constraints have allowed to narrow the parameter space of (or completely rule out) the theory. In particular we consider three major groups of models: \emph{(i)} exotic objects, \emph{i.e.} modifications to the BH paradigm, that do not require a modification of the underlying theory (these have been reviewed in Section \ref{sec:nature_Sgr}); \emph{(ii)} ETGs, \emph{i.e.} fundamental extensions to GR; \emph{(iii)} several models of dark matter concentration around Sgr A* that leave detectable imprints on the S-stars orbits. 

A phenomenological framework that has been extensively used to describe deviations from GR is the fifth force formalism \cite{Fischbach1986, Talmadge1988, Fischbach1992, Adelberger2003, Adelberger2009}. In this context, deviations from standard gravity are introduced in the form of a Yukawa-like potential in addition to Newton's classical potential
\begin{equation}
    V(r) = -\frac{GM}{r}(1+\alpha e^{-r/\lambda}).
    \label{eq:yukawa_deviation}
\end{equation}
Here $G$ is Newton’s constant, $M$ is the mass of the central object, and $r$ is the distance of the test particle from the central mass. This potential is characterized by two extra parameters: a scale-length $\lambda$ that defines the spatial range where the additional interaction is not-negligible and a multiplicative factor $\alpha$ that modulates the strength of the interaction. Moreover, the introduction of a Yukawa potential is a common feature to several modified theories of gravity when analyzed in the weak-field limit \cite{Quandt1991}. For instance,  it naturally arises in massive-gravity theories \cite{Hinterbichler2012} that predict new fundamental interactions with a gauge boson of mass $m_\psi$ (which is related to the scale length of the interaction, $m_\psi = \hbar/\lambda c$) \cite{Fischbach1992, Fujii1971}, in theories with higher dimensions \cite{Bars1986, Floratos1999, Goldberger1999, Kehagias2000, Hoyle2001}, in braneworld scenarios \cite{ArkaniHamed1998, ArkaniHamed1999, Dvali2000, Kogan2001}, certain models of dark matter \cite{Frieman1991, Gradwohl1992, Stubbs1993, Carroll2009, Carroll2010}, massive Brans-Dicke theories \cite{Perivolaropoulos2010, Alsing2012}, $f(R)$ gravity \cite{Capozziello2012, Capozziello2015}, Scalar-Tensor-Vector Theory (STVG) \cite{Moffat2006}. In 2017, the UCLA group studied the impact of a modified potential like the one in Equation \eqref{eq:yukawa_deviation} on the orbit of the S2 star \cite{Hees2017}. In particular, they used 19 years of astrometric and spectroscopic observation of S2 in order to constrain the two extra parameters ($\alpha$, $\lambda$) of the Yukawa-like potential along with all the other orbital parameters of the star that they left free to vary. This study provided a fully self-consistent test of gravity using orbital dynamics in the strong field regime of a SMBH, and resulted in an interesting constraint on $|\alpha| < 0.016$ on a scale-length of $\lambda = 150$ AU, at 95\% confidence. This result demonstrated for the first time that the short-period S-stars at the GC can effectively be used as probes of the gravitational theory, thus opening a new avenue to test ETGs. 

Several studies focused on the impact of the modification of the underlying theory on secular orbital changes like the pericenter advance. As a matter of facts, since it is a relativistic effect, the pericenter advance is a strong falsifier of theories and provides a crucial tool to test their validity \cite{GravityCollaboration2019b}. For some models the modification of the orbital precession induced by a modification of the theory itself is very weak and S-stars do not carry any signature of such theories. This is the case, for example, of Einstein-Maxwell-Dilaton-Axion gravity \cite{DeLaurentis2018a} or Brans-Dicke theories \cite{DeLaurentis2018a,Kalita2018}, for which the orbital precession is perturbed only by $0.3\%$ and thus constraints on the theory from current observations of the orbital precession in the GC (see Section \ref{sec:orbital_precession}) are not possible\footnote{However, closer objects like inner stars or pulsars might be used to constrain such theories.}. For other models, on the other hand, a much larger orbital precession than in GR is predicted and thus observations of the periastron advance for S2 already provide evidence to rule out the whole theory. Examples of such models are $f(R,\phi)$ gravity \cite{Capozziello2014} (in which the periastron advance is expected to be larger by a factor 13) or hybrid gravity \cite{Borka2016} (in which S2 should precess 20 times faster). {Recently, a preliminary study focused on a possible test of Horndeski-gravity at the Newtonian level using the pre-pericenter orbital data of the S2 star at the GC, giving new constraints on some of the parameters of this model \cite{DellaMonica2023b}.} More quantitative results in the form of constraints that the S2 star allows to place on ETGs, were possible after the successful measurement of the rate of orbital precession by the Gravity Collaboration \cite{GravityCollaboration2020c}. In particular, it was possible to show that once the measurement of the orbital precession (see Equation \eqref{eq:precession_gravity}) is taken into account, it is possible to considerably narrow down the parameter space of ETGs. The orbital precession measurement has been successfully used to constrain $f(R)$-gravity \cite{DeMartino2021}, STVG \cite{DellaMonica2022a} (see Sections \ref{sec:fR} and \ref{sec:STVG} for a more detailed description) and, recently, the parameters of power-law $f(R)$, general Yukawa-like corrections, scalar-tensor gravity and non-local-gravity have been estimated making use of the Schwarzschild precession of the S2 star \cite{Borka2021}.

\subsubsection{f(R)-gravity}
\label{sec:fR}
{Among different attempts to extend GR, historically $f(R)$-gravity has covered a particularly important role \cite{Sotiriou2010} and we devote this section to a brief overview of its characteristics and of its tests at the GC. The idea behind $f(R)$ gravity, as its name suggests, is a straightforward generalization of the Lagrangian in the Einstein-Hilbert action of GR
\begin{equation}
    \mathcal{S}_{GR} = \frac{1}{2\kappa}\int d^4x\sqrt{-g}R\qquad\to\qquad\mathcal{S}_{f(R)} = \frac{1}{2\kappa}\int d^4x\sqrt{-g}f(R),
    \label{eq:GR_fR_action}
\end{equation}
where $\kappa\equiv8\pi G$, $G$ is the gravitational constant, $g$ is the determinant of the space-time metric $g_{\mu\nu}$ and $R$ is the Ricci scalar built upon $g_{\mu\nu}$ (here we only deal with the metric formalism of $f(R)$-gravity. For a much more detailed review of other formalisms, \emph{e.g.} Palatini or metric-affine, we refer to \cite{Sotiriou2010}). An important consequence of such modification is that the appearance of a non-linear dependence of the lagrangian on $R$ leads to field equations of the fourth order, whereas GR is characterized by second-order equations (which are recovered in the special case $f(R) = R$). Due to the higher order, these field equations prescribe a much richer phenomenology than GR solutions \cite{Capozziello2011} and the propagation, in metric formalism, of an additional scalar degree of freedom with respect to to GR \cite{Magnano1994}. Such features have been shown to be able to overcome drawbacks of GR in a variety of astrophysical \cite{Banerjee2017, Naik2018, Leizerovich2022, deMartino2023,Capozziello2012b,DeLaurentis2013,DeMartino2014,DeLaurentis2015,DeMartino2016,DeMartino2018,DeLaurentis2018c} and cosmological contexts \cite{Starobinsky1980, Li2007, Starobinsky2007, Hu2007, Tsujikawa2008, Miranda2009, Lazkoz2018,DeMartino2015}. Moreover the propagation of gravitational waves in $f(R)$-gravity is significantly modified from that in GR, as a consequence of the extra degree of freedom in the theory \cite{Gogoi2020,Bora2022}.}
In the weak field limit, that is derived by expanding the Lagrangian $f(R)$ in powers of $R$, the space-time around a point-like source of mass $M$ is described as in spherical coordinates \cite{Capozziello2012, Capozziello2015, DeLaurentis2018b} by the line element
\begin{equation}
    ds^2 = [1+\Phi(r)]dt^2+[1-\Phi(r)]dr^2-r^2d\Omega^2\,,
\end{equation}
where the function $\Phi(r)$ is defined as
\begin{equation}
    \Phi(r) = -\frac{2GM(\delta e^{-r/\lambda}+1)}{rc^2(\delta+1)}.
    \label{eq:fR_yukawa}
\end{equation}
and $d\Omega^2 = d\theta^2 + \sin^2\theta d\phi^2$ is the solid angle element in spherical coordinates.
Here, the modification to the standard weak-field spherically-symmetric metric appears in terms of a Yukawa interaction with a scale length $\lambda$, whose strength is modulated by the parameter $\delta$. These two parameters, on the other hand, are related to the coefficients of the Taylor expansion of the function $f(R)$ by
\begin{equation}
    \delta = f'_0-1\,;\qquad\qquad\lambda =\sqrt{-6f''_0/f'_0}.
\end{equation}
A value $\delta = 0$ or, equivalently, $\lambda \to \infty$ reduce the theory to GR. Interestingly, in such a weak field limit, it is possible to derive an analytic expression for the rate of orbital precession of a test particle \cite{DeLaurentis2018b}, given by \begin{align}
    \Delta\omega_{\rm{f(R)}} = \frac{\Delta\omega_{\rm GR} }{(\delta +1)}\biggl( 1 & +
  \frac{2 \delta  G^2 M^2 }{3a^2 c^4  \left(1-e^2\right)^2} -\frac{2 \pi \delta  G^2 M^2 }{a c^4  \left(1-e^2\right) \lambda }  \\&-
  \frac{3\delta  G M}{a c^2  \left(1-e^2\right)} -\frac{ \delta G^2 M^2 }{6c^4(\delta +1) \lambda ^2}+
     \frac{\delta G M }{3 \lambda c^2 }\biggr)\,,
     \label{eq:fR_precession}
\end{align}
where $\Delta\omega_{\rm GR}$ is the orbital precession predicted by GR (Equation \eqref{eq:precession}), $a$ and $e$ are the orbital semi-major axis and eccentricity respectively. The expression reduces to Equation \eqref{eq:precession} when $\delta = 0$. Starting from this result, in \cite{DeMartino2021} the orbit of the S2 around Sgr A* has been derived in $f(R)$-gravity by considering the geodesic equations related to the weak-field metric. These have been solved numerically and converted into the astrometric and spectroscopic mock observations of S2, which have then been compared to publicly available data from the MPE group, up to 2016, and constraints on $\delta$ and $\lambda$ (as well as on all the orbital parameters) have been derived. An important result was the great improvement on the constraints established in the parameter space of $f(R)$-gravity when the measurement of the orbital precession by the Gravity collaboration (Equation \eqref{eq:precession_gravity}) was taken into account and compared to the theoretical one in Equation \eqref{eq:fR_precession} (see Figure \ref{fig:fR_STVG_improvement}). This allowed to considerably narrow the allowed region for these parameters and set a 95\% confidence interval on $\delta = -0.01^{+0.61}_{-0.14} $ and lower limit on $\lambda \gtrsim 6300$ AU. These values, on one hand, represent a factor 100 improvement over previous analysis of the same theory on the same system \cite{Capozziello2014}, which reduce the parameter space of $f(R)$ in a much narrower region around the GR limit of the theory; on the other hand, they prove how the geodesic approach and the use of relativistic effects can efficiently help in constraining ETGs. Interestingly, a study on the same weak-field potential using a geodesic approach was performed in \cite{DAddio2021} but without considering the impact of the orbital precession. In this study, not only the star S2 was considered, but also S38 and S55 (among the brightest and shortest-period ones in the GC). Moreover, the only free parameters of the analysis were the two $f(R)$ parameters $\delta$ and $\lambda$ (as opposed to the study in \cite{DeMartino2021} where the entire set of orbital parameters was constrained along with $\delta$ and $\lambda$). Despite the lower dimensionality of the problem and the advantage of considering multiple stars (that in principle should provide better constraints on the gravitational field), the constraints obtained in \cite{DAddio2021} are consistent with the ones derived by \cite{DeMartino2021} using only the S2 star and without the preccesion information (pink contours in Figure \ref{fig:fR_STVG_improvement}). 
On the other hand, in a previous study by Borka and collaborators \cite{Borka2013}, the orbit of the S2 star was used to constrain the Yukawa-like potential in Equation \eqref{eq:fR_yukawa}. In that analysis, however only the parameters ($\delta$, $\lambda$) were varied, fixing the orbital elements of S2 to the best fit ones from \cite{Gillessen2009b} (derived by fitting a Keplerian orbit to the data) and orbits were derived integrating Newton’s Law instead of integrating the geodesic equations. Moreover, it is not clear whether all observational and relativistic effects have been taken into account. This study favours strongly positive values of $\delta$, while suggesting values of $\lambda$ in the range $5000-7000$ AU, a region of the parameter space that seems not to be compatible with GR.
\subsubsection{Scalar-Tensor-Vector Gravity}
\label{sec:STVG}
\begin{figure}[!t]
    \centering
      \centering
      \includegraphics[width=\linewidth]{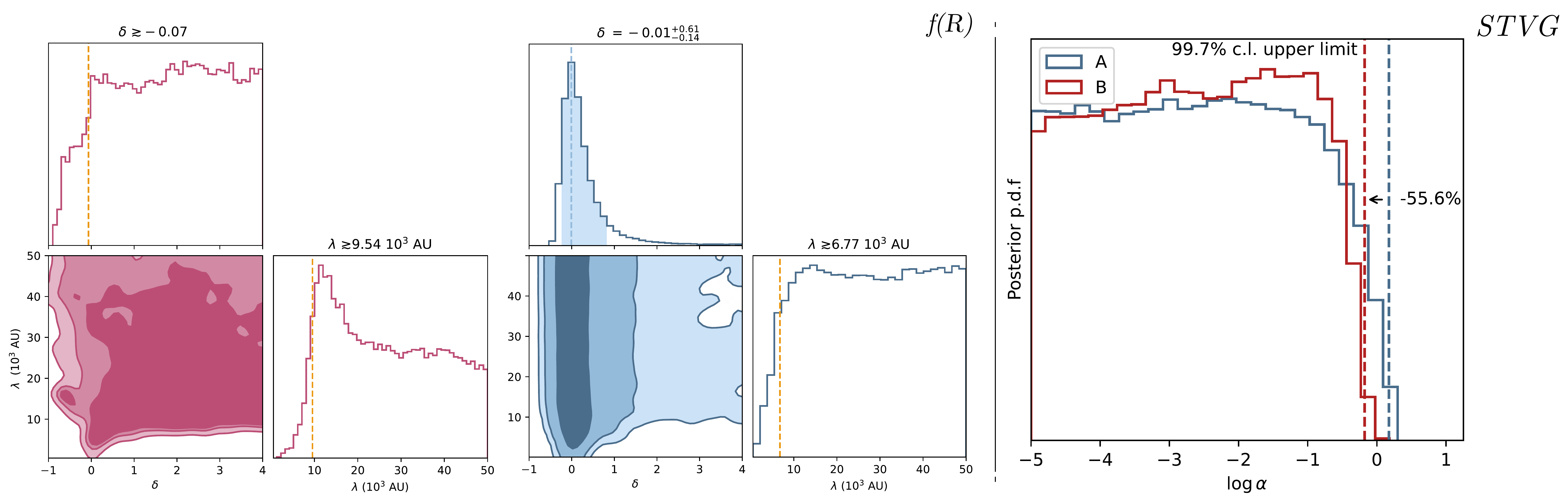}
    \caption{The two left plots show the posterior probability distribution for the parameters $(\delta, \lambda)$ from \cite{DeMartino2021} in the weak field limit of $f(R)$-gravity \cite{Capozziello2014} before (pink contours) and after (blue contours) adding the measurement of the orbital precession by the Gravity Collaboration \cite{GravityCollaboration2020c} in the analysis. Both the 68\% and the 95\% contours are considerably narrower in the latter case, allowing to place a constraint on the parameter $\delta= -0.01^{+0.61}_{-0.14}$ at 68\% confidence. The right plot reports the marginal posterior distribution in log-scale of the parameter $\alpha$ in STVG as derived in \cite{DellaMonica2022b} before (blue histogram) and after (red histogram) inserting the measurement of the orbital precession of S2 into the analysis. As in the previous case, the additional information carried by the pericenter advance allows to bring the upper limit on the parameter $\alpha$ from $\alpha\lesssim1.499$ (w/o precession) to $\alpha\lesssim0.662$ (w/ precession) at 99.7\% confidence, allowing to improve the constraint by $\sim$55.6\%.}
    \label{fig:fR_STVG_improvement}
\end{figure}

In a similar study \cite{DellaMonica2022b, DellaMonica2023b}, the same approach has been adopted to constrain the parameter space of the Scalar-Tensor-Vector Gravity (STVG), an ETG developed by J. Moffat in 2006 \cite{Moffat2006}. As the name suggests, this theory considers the introduction of extra degrees of freedom of gravity in the form of scalar fields and a massive vector field $\phi^\mu$ (along with metric tensor field). In particular, the gravitational constant $G$ and the mass of the vector field are elevated to the role of scalar fields. The vector field $\phi^\mu$, on the other hand, couples with massive test particles (of mass $m$), introducing a fifth-force interaction that modifies the geodesic equation
\begin{align}
    \left(\frac{d^2x^\alpha}{d\lambda^2}+\Gamma^\alpha_{\beta\rho}\frac{dx^\beta}{d\lambda}\frac{dx^\rho}{d\lambda} \right)=\frac{q}{m}{B^{\alpha}}_\beta\frac{dx^\beta}{d\lambda}.
    \label{eq:geodesic-equations_mog}
\end{align}
where $q$ is the charge of this additional interaction, and $B_{\alpha\beta}:= \nabla_\alpha\phi_\beta-\nabla_\beta \phi_\alpha $ is the Farady tensor associated to the vector field. {
In STVG, the generally covariant action can be written as \cite{Moffat2006}
\begin{equation}
    \mathcal{S} = \mathcal{S}_{GR}+\mathcal{S}_M+\mathcal{S}_\phi+\mathcal{S}_S.
    \label{eq:stvg-action}
\end{equation}
Here the first term is the classical Hilbert-Einstein action of GR and the second one is related to the ordinary matter energy-momentum tensor. On the other hand, the two additional terms, $\mathcal{S}_\phi$ and $\mathcal{S}_S$, encode the novel features of this theory
\begin{align}
    \mathcal{S}_\phi =& -\int d^4x\sqrt{-g}\left(\frac{1}{4}B^{\alpha\beta}B_{\alpha\beta}-\frac{1}{2}\mu^2\phi^\alpha\phi_\alpha+V(\phi)\right),\\
    \mathcal{S}_S =& \int d^4x\sqrt{-g}\frac{\omega_M}{G^3}\left(\frac{1}{2}g^{\alpha\beta}\nabla_\alpha G\nabla_\beta G-V(G)\right)+\nonumber\\
    &+\int d^4x\frac{1}{\mu^2G}\left(\frac{1}{2}g^{\alpha\beta}\nabla_\alpha\mu\nabla_\beta\mu-V(\mu) \right),
\end{align}
where $g$ denotes the determinant of the metric tensor $g_{\alpha\beta}$ and $R$ represents the Ricci scalar, $\nabla_\alpha$ is the covariant derivative related to the metric tensor $g_{\alpha\beta}$, $\omega_M$ is a constant, and $V(\phi)$, $V(G)$ and $V(\mu)$ are scalar potentials arising from the self-interaction associated with the vector field and the scalar fields, respectively.  Upon minimization of the action in Equation \eqref{eq:stvg-action}, the field equations in vacuum (\emph{i.e.} $T_{\alpha\beta}^M = 0$) read \cite{Moffat2021}
\begin{align}
    G_{\alpha\beta}=&-\frac{\omega_M}{\chi^2}\biggl(\nabla_\alpha\chi\nabla_\beta\chi -\frac{1}{2}g_{\alpha\beta}\nabla^\sigma\chi\nabla_\sigma\chi\biggr)+
\\&-\frac{1}{\chi}(\nabla_\alpha\chi\nabla_\beta\chi-g_{\alpha\beta}\Box\chi)+\frac{8\pi}{\chi}T^\phi_{\alpha\beta},
\end{align}
where the scalar field $\chi = 1/G$, and $T^\phi_{\alpha\beta}$ is the gravitational $\phi$-field energy momentum tensor given by
\begin{equation}
    T^\phi_{\alpha\beta} = -\left({B_\alpha}^\sigma B_{\sigma\beta}-\frac{1}{4}g_{\alpha\beta}B^{\sigma\rho}B_{\sigma\rho}\right).
\end{equation}}STVG has an exact BH solution \cite{Moffat2015} derived under the assumptions that the scalar field nature of $G$ can be regarded as a constant variation with respect to the Newtonian value, $G = G_{\rm Newton}(1+\alpha)$, where $\alpha$ is an extra parameter of the theory. Moreover, on the scales of compact objects, it is postulated that the mass of the vector field can be neglected and that the fifth-force charge of a test particle can be written as $q = m\sqrt{\alpha G_{\rm Newton}}$. This assumptions allow to derive a spherically symmetric space-time metric given by
\begin{equation}
    ds^2 = \frac{\Delta}{r^2}dt^2-\frac{r^2}{\Delta}dr^2-r^2d\Omega^2
\end{equation}
with $\Delta = r^2-2Mr+\alpha M\left((1+\alpha)M-2r\right)$ and $d\Omega^2 = d\theta^2 + \sin^2\theta d\phi^2$. This space-time is formally equivalent to that of a Reissner–Nordström charged BH \cite{Reissner1916} where in this case the charge is the fifth-force charge of the central object and reduces to the usual Schwarzschild metric when $\alpha = 0$. When studying the free-fall motion of massive (and thus fifth-force-charged) particles in STVG and the interaction with the vector field $\varphi^\mu$, it is possible to derive a first-order analytical expression for the rate of orbital precession, that is given by \cite{DellaMonica2023b}
\begin{equation}
    \Delta\omega_{\rm STVG} = \Delta\omega_{\rm GR}\left(1+\frac{5}{6}\alpha\right).
    \label{eq:precession_STVG}
\end{equation}
The precession scales linearly with the parameter $\alpha$ and reduces to the usual GR expression in Equation \eqref{eq:precession} when $\alpha = 0$. Given this result, it appears natural that, taking into consideration the measured rate of precession for the S2 star in Equation \eqref{eq:precession_gravity} \cite{GravityCollaboration2020c}, the parameter $\alpha$ for STVG theory can be constrained. As a matter of fact, as already done for the case of $f(R)$-gravity, an orbital model for the S2 star in STVG was developed in \cite{DellaMonica2022b} by solving numerically the modified geodesic equations in Equation \eqref{eq:geodesic-equations_mog}. Two separate posterior analyses were performed on the whole dimensional space (orbital parameters for S2 + the STVG parameter $\alpha$) deriving upper limits on $\alpha$. In the first one, only astrometric and spectroscopic datasets for S2 were used up to 2016, \emph{i.e.} including no information at all on the $\sim2018$ pericenter passage. This resulted in an upper limit on $\alpha\lesssim1.499$ at 99.7\% confidence. In the second run, the orbital precession measurement was added as a single data point, and compared with the analytical expression in STVG in Equation \eqref{eq:precession_STVG}. This allowed to narrow down the 99.7\% confidence region on $\alpha$ by almost 56\% (at $\alpha\lesssim0.662$), once again demonstrating the great constraining power of the rate of pericenter advance as a gravity theory falsifier. In Figure \ref{fig:fR_STVG_improvement} the marginalized posterior distribution for the parameter $\alpha$ in both cases are shown.

\subsection{Testing alternative Dark Matter models at the Galactic Center}
\label{sec:darkmatter}

Studying the orbits of the S-stars does not only provide with a powerful tool to test gravity at the GC but also serves as a probe of extended mass distributions around Sgr A* \cite{GravityCollaboration2022a}. As a matter of fact, the in-plane orbital precession of stars, besides being a prediction of GR, can also be influenced by other processes. In particular, a test particle moving through a an extended continuous mass distribution (that is assumed to be dark, in the sense that it cannot be directly observed and hence its density cannot be quantified by its luminous emission) feels an additional acceleration imposed by the extended mass component. This can cause an additional pericentre advance, even in the Newtonian framework \cite{Rubilar2001, Merritt2013}. Such a distribution can be composed of faint stars, stellar remnants and/or dark matter \cite{Heissel2022}. This effect is usually referred to as mass precession, and, depending on the specific mass profile and enclosed mass, it can be of the same order (or may even dominate) the relativistic precession in Equation \eqref{eq:precession} \cite{Rubilar2001, Merritt2013}.  Furthermore, for particularly steep mass profiles it can end up producing a retrograde precession (\emph{i.e.} opposite to sense of orbital revolution of the test particle) whereas the Schwarzschild precession is always prograde \cite{GravityCollaboration2022a}. Studying the effects of such putative mass distribution around the GC is of paramount importance to directly infer details on the the matter distribution in the immediate vicinity of the central SMBH, and can also allow to narrow down constraints on (or rule out) alternatives to the standard BH paradigm for the central object in the form of continuous dark matter concentrations present in literature (see the last section of Table \ref{tab:eGR_literature}) \cite{Heissel2022}.

Specifically, for the case of the GC, the precision of current S-stars data allows to establish upper bounds on the amount of mass present within the apocentre of S2. A recent study by the Gravity Collaboration \cite{GravityCollaboration2022a} has allowed to place limits on the mass of the putative extended component at  $\lesssim3000M_\odot$ (corresponding to $\sim0.1\%$ of the BH mass). For the sake of comparison, a precedent study by the same group \cite{Gillessen2017} (when they could not yet rely on the precise astrometry of the GRAVITY interferometer) was able to place much weaker constraint on the extended mass, at $\sim1\%$ of the mass of central SMBH. The 10-fold improvement in the upper limits of such a mass distribution is due to the great advancement in the observational facilities employed. However, with increasingly precise astrometry, it is important to be able to unambiguously distinguish signatures on the orbits of the S-star that are ascribable to the extended mass component, from that of other sources of precession \cite{Heissel2022}. On secular timescales, this may be of particular concern for the relativistic effects of Schwarzschild precession (Section \ref{sec:orbital_precession}) and Lense-Thirring frame-dragging (section \ref{sec:higher_order_effects}).
Recently, Heisßel and collaborators have proposed an in-depth study \cite{Heissel2022} of the dark mass signature on the orbit of S2, with the goal of exploring a strategy to separate the Schwarzschild precession from the mass precession, despite their secular interference. What they found is that \emph{(i)} the mass precession almost exclusively impacts the orbit in the apocentre half of the orbit. The Schwarzschild precession, on the other hand, almost exclusively impacts it in the pericentre half. This allows for a clear separation of the effects when a uniform orbital phase coverage is present; \emph{(ii)} Consequently, data that are limited to the pericentre half are insensitive to the dark mass component, by which data on the apocentre half are affected; \emph{(iii)} A full orbital-phase coverage of data is required to substantially constrain the amount of a extended mass; \emph{(iv)} When a full orbit of astrometric and spectroscopic data is available, the greatest constraining power belongs to astrometric data in the pericentre half. A forecast analysis in the same paper, suggests that with GRAVITY data with a $\sim50\;\mu$as precision collected over an entire orbital period (which will occur in $\sim2033$) the $1\sigma$ upper bound on the extended mass component can be brought down to $\sim1000\;M_\odot$.

In 2015 the possibility of dark matter composed by a self-gravitating system of fermions at finite temperature, in equilibrium, and distributed in phase-space according to the Fermi-Dirac statistics, known as the Ruffini-Argüelles-Rueda (RAR) model \cite{Ruffini2015, Arguelles2018}, has been proposed. Interestingly, in the RAR model, the dark matter distribution in galaxies (obtained from the general relativistic field equations) as a function of the radius, predicts three different physical regimes: \emph{(i)} a quantum-degenerate inner core of almost constant density; \emph{(ii)} an intermediate quantum-classical region where the density decreases sharply, followed by an extended plateau; \emph{(iii)} a power-law decreasing density distribution ($\rho\propto r^{-2}$ leading to flat rotation curves) described by the classical Boltzmann statistics \cite{Ruffini2015}. More specifically, the presence of a dense core that naturally arises in this model has been proposed as an alternative to the SMBH in the galactic center. This possibility has been investigated \cite{BecerraVergara2020, BecerraVergara2021} by looking at the orbital motion of the S-stars and of the G2 object \cite{Gillessen2012, Witzel2014}. The post-pericenter motion of the latter, in particular, exhibits an acceleration that cannot be explained only in the context of a freely falling test particle in the Sgr A* gravitational field \cite{Plewa2017}. A possible solution is that G2 is a gas cloud\footnote{As a matter of fact, the true nature of G2 is still debated, with some authors suggesting that it might be an evaporating circum-stellar disc \cite{Miralda2012, Murray2012} or proto-planet \cite{Mapelli2015}, a stellar source endowed with a gaseous-dusty envelope \cite{Ballone2013, Eckart2013, Valencia2015, Ball2016}, the result of the merger of a binary system of stars \cite{Witzel2014}, or a set of two substructures that are remnants of a dissolved young stellar cluster that initially formed in the circumnuclear disk \cite{Peissker2021}.} and that the additional acceleration results from a drag force from the interaction with the SMBH accretion flow \cite{Burkert2012, Gillessen2019}. The authors of the RAR model, on the other hand, treat G2 as a test particle following geodesic motion in the space-time generated by the core-halo profile of fermionic dark matter and obtain a superior fit to the orbital data of G2 \cite{BecerraVergara2020}. In this context, the additional acceleration arises from the steep dark matter density gradient encountered by G2 on its trajectory, which goes in and out of the fermionic dense core. {While compelling for its ability of explaining the anomalous motion of the G2 object, it should be noted that the RAR model has been shown to exhibit some shortcomings in the description of conventional orbits of bright stars in the galactic center \cite{Zakharov2022c} and, due to the replacement of the central compact object with a continuous self-gravitating fermionic distribution, it is inherently impossible to reproduce the observed shadow for Sgr A* in this model.} 

Alternatively, an interesting model that in the last years has become very popular is the scenario of a Kerr BH endowed with a scalar 'hair'. These models arise from considering massive scalar fields\footnote{There are several example of candidate scalar fields from theoretical and particle physics. Among them the axions represent one of the most compelling alternatives. This kind of particles were first introduced introduced by Peccei and Quinn in 1977 \cite{Peccei1977} to solve the strong CP problem. High-energy theories, like string theory, predict that low-energy effective models contain a set of scalar fields with very small rest masses \cite{Svrcek2006}. Theoretical models predict that the mass of the scalar field can be as small as $10^{-33}$ eV leading to the putative existence a family of particles with a broad range of masses, called the ‘Axiverse’ \cite{Arvanitaki2010}. See \cite{Hui2017} for an analysis of several cosmological and astrophysical effects related to the potential existence of ultralight axions.} (of mass $m_\psi$) that tend to develop quasi-bound state solutions in the proximity of astrophysical BHs \cite{Detweiler1980, Dolan2007, Witek2013}. Such states correspond to profiles of the scalar field that vanish at infinity as a result of in-going waves at the event horizon. While these solutions tend to decay over time, there exists a regime in which growing modes of the scalar field are generated in the form of long-lived quasi-bound states supported by extraction of rotational energy from the BH. This happens whenever the Compton wavelength of the scalar field is comparable with the gravitational radius of the BH.
This phenomenon is regarded to as superradiance \cite{Brito2015b}. Considering the $\sim 4\times10^6M_\odot$ of Sgr A* in the GC, an upper limit on the mass of the scalar field that can lead to the formation of a superradiant state is estimated to be $m_\psi < 10^{-17}$ eV \cite{Kodama2012}. Superradiance can grow scalar field structures in the vicinity of BHs \cite{Dolan2013, Witek2013, Okawa2014} that are thought to have astrophysical effects on the surrounding environment \cite{Brito2015a, Cunha2015, Vincent2016b, Rosa2018} and, specifically, on the orbits of stars revolving around the BH \cite{Ferreira2017, Fujita2017}. In 2019, the Gravity collaboration, studied the secular changes that would incur in the orbit of the S2 star due to the presence of such a scalar field superradiant structure in the vicinity of Sgr A* \cite{GravityCollaboration2019b}. In particular, they consider a scalar field cloud with axial symmetry with respect to the BH's spin axis that reaches a peak at a certain distance from the BH (on its equatorial plane) that is proportional to the inverse-square of the mass $m_\psi$ of the field. For large-mass couplings the cloud is concentrated near the BH, while for small-mass couplings the cloud is located further away. Form the numerical solution of the equations of motion for S2 in such a configuration, it results that the strongest effects on the orbit take place when the star crosses the scalar field density peak (feeling a greater density gradient along its orbit). The effect on the argument of the pericentre can be competitive (ranging between -1.37 and -17.19 arcmin per orbital period) with the relativistic precession, and its contribution can either reinforce or reduce the GR value, depending on the BH spin \cite{GravityCollaboration2019b}. Moreover, due to the lack of spherical symmetry of the scalar-field structure, other orbital elements tend to incur into secular changes over time. This affects both the eccentricity $e$, the orbital inclination $i$ and longitude of the ascending node $\Omega$ whose variation depends on the relative inclination between the orbit and BH axis and amount to up to $\sim0.1$ arcmin in the case of a maximally rotating BH. Such estimates show that with the current ability of detecting secular changes on the orbit of the S2 star at the GC, the presence of a scalar field cloud in the vicinity of Sgr A* could in principle be revealed.

On a different line of investigation (\emph{i.e.} without considering a coupling at a fundamental level of the scalar field with gravity), self-gravitating halos of ultralight ($m_\psi < 10^{-18}$) dark matter are known to naturally form cored solitons in their inner regions \cite{Hui2017, Schive2014a}. Such ultralight particles possess a de Broglie wavelength that can reach the kpc scale, leading to the appearance of quantum behavior over macroscopic astrophysical scales. For this reason, this dark matter model is usualy regarded to as Fuzzy Dark Matter (FDM). As a matter of fact a self-gravitating halo of FDM is supported by an additional internal quantum-pressure \cite{Hui2017} that arises from Heisenberg's uncertainty principle and that ultimately prevents collapse of the halo below the de Broglie scale, forming the solitonic core. The radial profile of such cores is well approximated by \cite{Schive2014a, Mocz2017}
\begin{equation}
    \rho_s(r) = \frac{\rho_0}{(1+Ar^2)^{8}},
    \label{eq:soliton}
\end{equation}
where $A$ is related to the core radius $r_c$ via $A = 9.1\times 10^{-2}/r_c^2$, and $\rho_0$ corresponds to the central density of the halo:
\begin{equation}
    \rho_0 = 1.9\left(\frac{m_\psi}{10^{-23}\;\textrm{eV}}\right)^{-2}\left(\frac{r_c}{\rm kpc}\right)^{-4} \frac{M_\odot}{\rm pc^3}\,.
\end{equation}
The core radius, on the other hand, can be related to the virial mass of the entire halo via a scaling relation \cite{Schive2014b}
\begin{equation}
    r_c = 1.6\left(\frac{m_\psi}{10^{-22}\textrm{ eV}}\right)^{-1}\left(\frac{M_{\rm halo}}{10^9\,M_\odot}\right)^{-1/3}\textrm{ kpc}.
\end{equation}
This means that, depending on the inverse mass of the boson, the solitonic cores may be compact enough to add a non-negligible amount of dark matter mass around the SMBH. Recently, the impact that the presence of such a distribution of dark matter around Sgr A* could have on the orbit of the S2 star has been investigated \cite{DellaMonica2022c}. In particular, using a weak-field approach, the combined presence of the SMBH plus a distribution of dark matter, has been considered by adding to the 1PN acceleration related to the presence of the SMBH (Equation \eqref{eq:pn_eom}) an acceleration term due to the presence of the cored halo (derived from Poisson's equation applied to the solitonic radial profile in Equation \eqref{eq:soliton}). In particular, it was demonstrated, using mock astrometric observation of S2, that future analyses with the precision of the GRAVITY interferometer could provide with an upper limit on the mass of the ultralight boson $m_\psi\lesssim10^{-19}$ eV at 95\% confidence level. This result is particularly interesting, because, when compared with other studies of the same model, it allows to narrow down by several orders of magnitude the allowed region for the mass of the dark matter particle. This study provides with yet another demonstration that the GC and the S-stars orbiting Sgr A* can help to provide independent (usually perpendicular) constraints on theories or models that are usually tested on different astrophysical scales, thanks to the particular regime of gravity explored by such systems (Figure \ref{fig:eht_gravitational_probes}).

\subsection{Testing Extended Theories of Gravity at the horizon scale}

In Section \ref{sec:sgra_shadow} we have described the procedure by which a prior information on the mass-to-distance ratio of Sgr A* and an appropriate calibration factor can convert the bright ring diameter measurement by EHT into a measurement of its shadow and of the deviation parameter $\delta$ from that of a Schwarzschild BH. The analysis in \cite{EventHorizonTelescopeCollaboration2022f} found that there is no evidence of violations of GR as the inferred values of $\delta = -0.04^{+0.09}_{-0.10}$ (from the UCLA priors) and $\delta = -0.08^{+0.09}_{-0.09}$ (from the MPE priors) are both compatible with predictions for a Kerr BH. Such constraints on the deviation parameter $\delta$ can be converted into bounds on the parameter space of ETGs. Given a specific metric for which the diameter of the shadow is predicted to be a certain value $\tilde{d}_{\rm metric}$ (that depends on the parameters of the specific model and that for the case of a non circular orbit is defined as the median shadow diameter) the fractional diameter deviation $\delta_{\rm metric}$ that one compares with the value of $\delta$ derived by EHT analysis is defined as
\begin{equation}
    \delta_{\rm metric} = \frac{\tilde{d}_{\rm metric}}{6\sqrt{3}}-1.
\end{equation}
In particular, two different approaches have been adopted, (1) constrain the parameters of stationary metrics with a specific parametrization, that are agnostic to the underlying physical theory; (2) constrain the parameters of stationary metrics that are generated by specific modifications to GR that depend on additional generalized charges\footnote{These represent only a particular class of deviations from GR, but it is important to constrain this kind of metrics as they allow us to translate directly the constraints from the EHT images to bounds on physical parameters \cite{EventHorizonTelescopeCollaboration2022f}.} \cite{Kocherlakota2021}.

\begin{figure}[t]
    \centering
    \includegraphics[width = 0.7\textwidth]{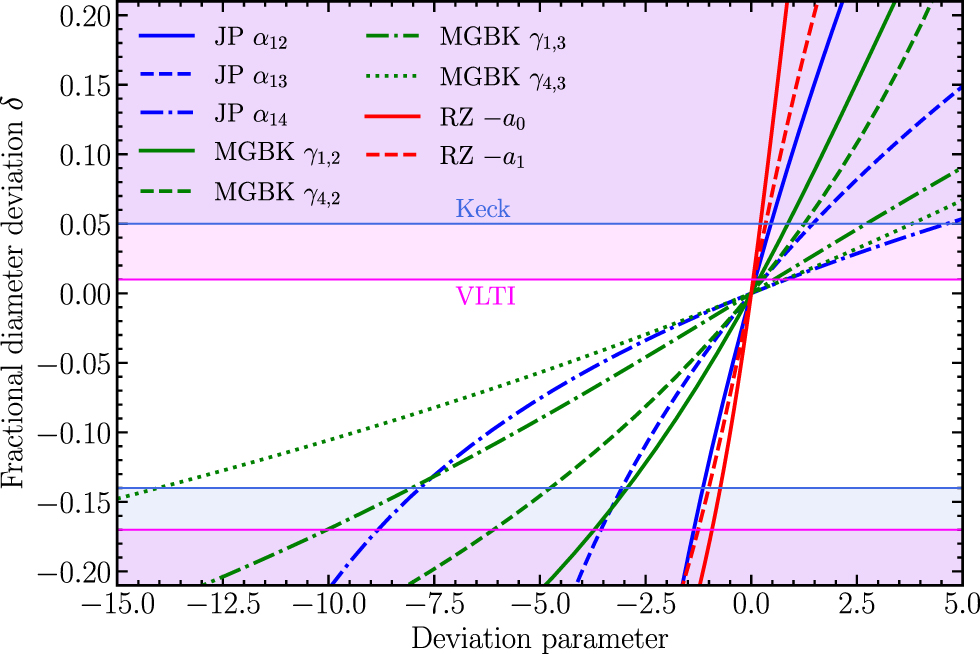}
    \caption{Analytically computed constraints on the JP (blue), MGBK (green) and RZ (red) deviation parameters derived in \cite{EventHorizonTelescopeCollaboration2022f} by setting, for each metric, all deviations parameters to zero and leaving only one free to vary in the shadow diameter determination. Solid lines represent the lowest-order deviation parameters, which are also the best constrained. Higher order parameters (with different dashes) are increasingly less constrained.}
    \label{fig:eht_alternative}
\end{figure}

In the former case, while the used parametrizations do not arise from any particular modification to gravity, one has the advantage of constrain phenomenologically a broad spectrum of models, that can be mapped later to the parameters of a fundamental theory \cite{Psaltis2020a}. Several different metric parametrization that are agnostic to the underlying physical theory have been formulated, allowing for general deviations from the Kerr space-time while minimizing pathologies (\emph{i.e.} not satisfying the no-hair theorem) due to a relaxation of the Ricci flatness hypothesis \cite{Psaltis2020a, Suvorov2021}. In \cite{EventHorizonTelescopeCollaboration2022f}, the shadow for three specific metric parametrization has been studied, converting the bounds on $\delta$ into constraints on the deviation parameters. In particular, the analyzed parametrizations are the Johannsen-Psaltis (JP) metric \cite{Johannsen2011, Johannsen2013a}; the Modified Gravity Bumpy Kerr (MGBK) metric \cite{Vigeland2011} and the Rezzolla-Zhidenko (RZ) metric \cite{Rezzolla2014, Konoplya2016}.
As shown in Section \ref{sec:sgra_shadow}, the diameter of the shadow for a given spherically-symmetric metric, depends solely on the $g_{00}$ component (Equation \ref{eq:diameter_shadow_th}). For the three parametrization analyzed the $g_{00}$ components of the metric have the expressions:
\begin{equation}
    g_{00} = \left\{\begin{array}{lr}
         \displaystyle-\left(1-\frac{2}{r}\right)\left(1+\sum^\infty_{i=2}\frac{\alpha_{1i}}{r^i}\right)^{-2}&  JP\\
         \\
         \displaystyle-\left(1-\frac{2}{r}\right)\left(1-\sum^\infty_{i=2}\frac{\gamma_{1,i}}{r^i}-2\sum_{i=0}^\infty\frac{\gamma_{4,i}}{r^i}\left(1-\frac{2}{r}\right)\right)^{-2}&  MGBK\\
         \\
         \displaystyle -x\left[1-\epsilon(1-x)+(a_0-\epsilon)(1-x^2)+\frac{a_1}{1+\frac{a_2x}{1+\frac{a_3x}{\dots}}}(1-x)^3\right]&  RZ\\
    \end{array}\right.
    \label{eq:metric_parametrizaion}
\end{equation}
where $r$ is the aerial radius, $a_{1i}$ are the deviation parameters for the JP metric, $\gamma_{1,i}$ and $\gamma_{4,i}$ are those of the MGBK parametrization and $\epsilon$ and $a_i$ are the deviation parameters for the RZ parametrization. Moreover, in the last case, $x \equiv 1-r_0/r$, where $r_0$ is the coordinate radius of the infinite redshift surface (that corresponds with the horizon in case no pathology exists). The measurement of the diameter of the shadow allows to place constraints of order unity mainly on parameters such as $\alpha_{13}$ and $\gamma_{1,2}$ that depend weakly on the BH spin \cite{Psaltis2020a}. Using the analytic expressions for the shadow diameter derived by the $g_{00}$ components of the metric in Equations \ref{eq:metric_parametrizaion}, the bounds on the shadow size from EHT images of Sgr A* can be converted into constrains on the several deviation parameters. More specifically, since only one quantity is measured from the EHT images (the shadow diameter), but the $g_{00}$ components in Equation \ref{eq:metric_parametrizaion}, it can only be possible place actual constraints on given combinations of the parameters. One possible solution to avoid this and derive directly upper and lower bounds on each parameter is setting alternatively to zero all deviations parameter except one. The resulting constraints are shown in Figure \ref{fig:eht_alternative}. As it appears, the deviation parameters that correspond to higher-order corrections have a smaller impact on the shadow diameter, thus resulting in larger bounds with respect to lower order parameters.

\begin{figure}[!t]
    \centering
        \begin{subfigure}{.48\linewidth}
          \centering
          \includegraphics[width=\linewidth]{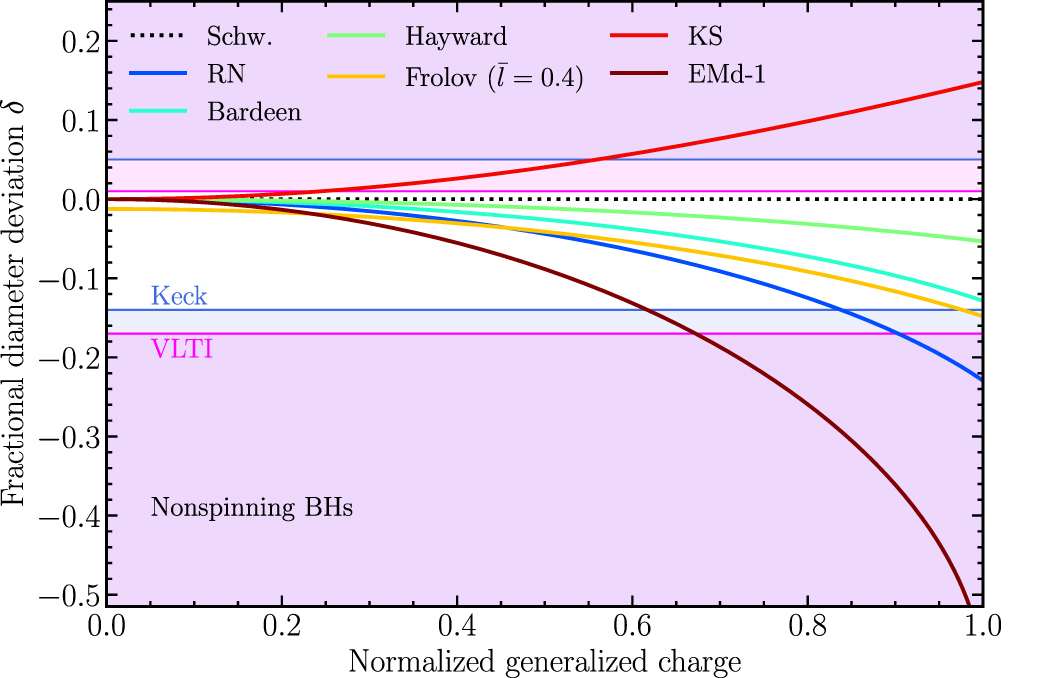}
          \caption{}
          \label{subfig:eht_alt_1}
        \end{subfigure}
        \begin{subfigure}{.48\linewidth}
          \centering
          \includegraphics[width=\linewidth]{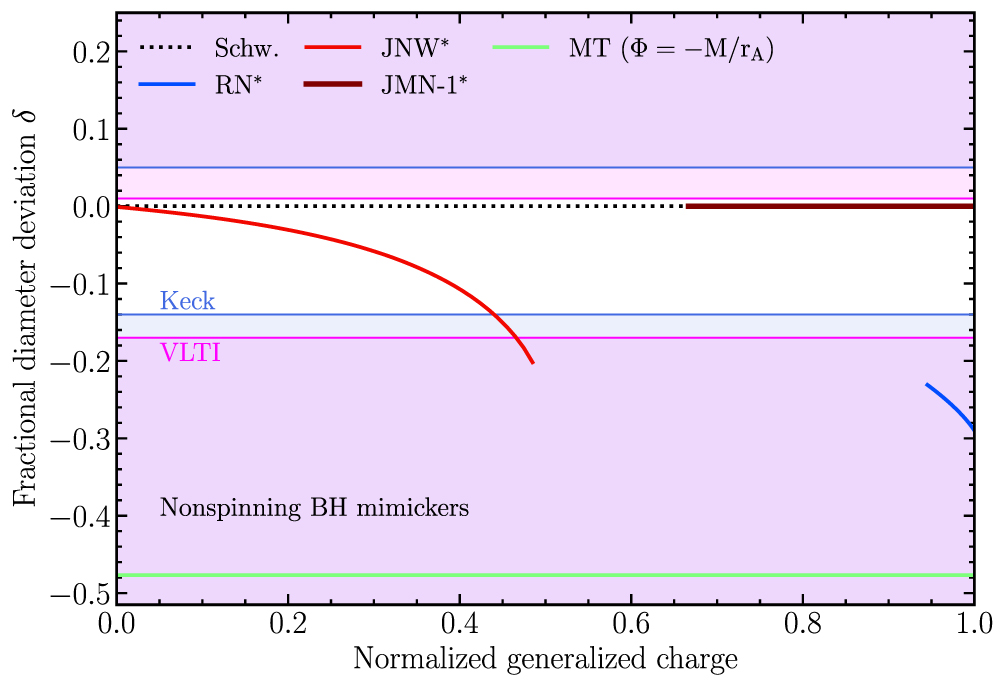}
          \caption{}
          \label{subfig:eht_alt_2}
        \end{subfigure}
    \caption{Dependence of the shadow diameter deviation, $\delta$ on the normalized charge for \textbf{(a)} non-spinning BH metrics and \textbf{(b)} for naked singularity and wormholes. The white regions report shadow sizes corresponding to the 68\% level with the 2017 EHT observations for Sgr A* \cite{EventHorizonTelescopeCollaboration2022a}. The EHT observations rule out values of the physical charges of this space-times  comparable to their maximum theoretically allowed values. The used abbreviations are: RN (Reissner-Nordstroöm), KS (Kerr-Sen), EMd-1 (Gibbons–Maeda–Garfinkle–Horowitz–Strominger BH in the context of  Einstein–Maxwell-dilaton-axion theory), JNW (Janis–Newman–Winicour), JMN-1 (Joshi–Malafarina–Narayan-1), MT (Morris-Thorne). An asterisk corresponds to a naked singularity solution. Except for the RN$^*$ metric and the MT wormhole, whose shadows are significantly out of the confidence region for $\delta$, it is clear that the EHT observation alone cannot rule out  the naked singularity nature for Sgr A*. Images are courtesy of \cite{EventHorizonTelescopeCollaboration2022f}.
    }
    \label{fig:eht_alternatives}
\end{figure}

Metrics arising from specific modifications of the theory of gravity have been considered in \cite{EventHorizonTelescopeCollaboration2022f}, as well. For example, Einstein–Maxwell-dilaton-axion (EMda) theory \cite{Gibbons1988, Garfinkle1991, Kallosh1992, Sen1992, Garcia1995, Kocherlakota2020}, emerging as the low-energy effective descriptions of the heterotic string, violates the weak equivalence principle (WEP) and the local positional invariance (LPI). Two metrics have been investigated in this framework: the Gibbons–Maeda–Garfinkle–Horowitz–Strominger BH \cite{Gibbons1988, Garfinkle1991, Mizuno2018}, describing a charged, static BH in an EMda theory (labeled EMd-1 in Figure \ref{fig:eht_alternatives}); and (Kerr-)Sen BHs \cite{Sen1992, Younsi2021, Ozel2021}, which are the spinning generalizations of the EMd-1 BHs. Moreover, the the Reissner–Nordström \cite{Reissner1916, Nordstrom1918} and the Kerr–Newman \cite{Newman1965} solutions, describe charged static and rotating BHs in GR, respectively, have been considered for comparison. {Useful investigation on the strong-lensing and shadow features of such space-times have been discussed in \cite{Zakharov2005b} based on the approach studied in \cite{Zakharov1994}.}
Additionally, solutions arising from by regularizing the central singularities of classical BH within general relativity. In particular, solutions by Bardeen \cite{Bardeen1968}\footnote{This solution was originally presented in 1968 at the 5th International Conference on Gravitation and the Theory of Relativity, in Tblisi (USSR). The proceedings of the conference \cite{Bardeen1968} are not readily available. However, a discussion of the model can be found in \cite{Borde1994} and the regular black hole presented by Bardeen back in 1968 is commonly agreed upon as the first example of such regular solutions \cite{Ansoldi2008}.}, Hayward \cite{Hayward2006}, and Frolov \cite{Frolov2016} have been considered\footnote{These are usually nonempty solutions to the non-vacuum field equations in the presence of a stationary configuration of an exotic matter violating one or more energy conditions \cite{Hawking1973}.}, along with a rotating counterpart and the static BH solution by Kazakov and Solodukhin \cite{Kazakov1994} solution, in which the singularity is onto a surface.\par

Similarly to the case of the Kerr and the parametric metrics, it results \cite{EventHorizonTelescopeCollaboration2022f} that the BH spin introduces minor corrections to the size of the shadow. Moreover, the current bounds imposed by the EHT images of Sgr A* place constraints of order unity to the charges of several of the space-times, comparable to their maximum theoretical values. In Figure \ref{subfig:eht_alt_1} the constraints that can be placed on the parameter spaces of the considered space-times are shown. As it appears, no evidence of violations of the equivalence principle or of the presence of energy-conditions violating matter within the present context is found \cite{EventHorizonTelescopeCollaboration2022f}. 

{Interestingly, the constraints imposed on the charge for a Reissner–Nordström space-time in the GR framework can be directly converted into constraints on BH solutions in theories with extra dimensions \cite{Zakharov2012, Zakharov2022b}. In particular, it has been shown that in a Randall–Sundrum II braneworld scenario \cite{Randall1999b, Randall1999a,BinNun2010}, the Reissner–Nordström metric would appear as a BH solution in the model. Such solution is interpreted as a BH without an electric charge but a tidal charge emerging as a consequence of gravitational effects from the fifth dimension. The 2022 estimation by the EHT of the shadow size for Sgr A* implies a bound on such a tidal charge of $-0.27<q<0.25$, as shown in \cite{Zakharov2014, Zakharov2022a}.}

Finally the possibilities that the space-time in the vicinity of Sgr A* can be described by that of a naked singularity with a photon sphere or a wormhole have been considered in \cite{EventHorizonTelescopeCollaboration2022f}. For example, a naked singularity space-time can be found in the Reissner–Nordström metric \cite{Reissner1916, Nordstrom1918}, whenever the electromagnetic charge is $q > 1$. Additionally, it admits a photon sphere only if $1<q<\sqrt{9/8}$. Another example is the Janis–Newman–Winicour \cite{Janis1968} naked singularity space-time, a solution of the Einstein–Maxwell-scalar theory with a scalar charge $\hat{\nu}$. While this charge can theoretically span in the range $0<\hat{\nu}<1$, this space-time casts a shadow only if $0<\hat{\nu}<1/2$. Finally, a last class of naked singularities within GR have been considered, \emph{i.e.} the Joshi–Malafarina–Narayan-1 \cite{Joshi2011} space-time. This represents a uniparametric class of static space-times containing a compact region filled with an anisotropic fluid, which admits a photon sphere when $M_0\geq 2/3$, $M_0$ being the additional parameter in the theory. The shadow produced by this naked singularity has been demonstrated to be basically identical to that of a Schwarzschild BH \cite{Shaikh2019}, along with the spectra produced by the accreting matter in these space-times. This indicates that a JMN-1 naked singularity with a photon sphere may represent one of the best candidate BH mimicker.
Moreover, an example of wormhole, \emph{i.e.} a non-spinning, transversable Morris–Thorne wormhole \cite{Morris1988a} has been considered. The location of the circular null geodesic in this space-time \cite{Bambi2013a} implies a shadow radius of $\sim  2.72$ M, which corresponds to $\delta \approx -0.48$ , which is immediately ruled out by the EHT measurement \cite{EventHorizonTelescopeCollaboration2022f}.
Figure \ref{subfig:eht_alt_2} reports the bounds imposed by the EHT observations \cite{EventHorizonTelescopeCollaboration2022f} on the physical charges of these metrics (opportunely normalized to unity). Except for the Reissner–Nordström metric {for a naked singularity} and the Morris-Thorne wormhole, whose shadows are significantly out of the confidence region for $\delta$, it is clear that the EHT observation alone cannot rule out  the naked singularity nature for Sgr A*. \par
{Shortly after the publication of the 2022 EHT observation for Sgr A*, Vagnozzi et al. \cite{Vagnozzi2022} published an interesting collection of constraints for a wide variety of well-motivated space-time solutions arising from the measurement of the BH shadow diameter for Sgr A*. In this study various solutions for BH-mimicker space-times (including wormholes and naked singularities) and BH-like solution from both ETGs ($f(R)$, STVG, Horndeski, etc.) and ATGs (string-inspired space-times, violations of the no-hair theorem driven by additional fields) are considered. Remarkably, particularly stringent constraints on models in which the shadow size  is predicted to be larger than that of a Schwarzschild BH emerge from this study, with limits that in some cases have a better constraining power than cosmological tests. This study clearly demonstrates the ability to test fundamental physics models with the shadow of Sgr A*, with a number of well-motivated alternative scenarios (including BH mimickers) that are far from being ruled out at present.}

\section{Conclusions}
\label{sec:conclusions}

In 2010, R. Genzel would conclude his review article \cite{Genzel2010}, in collaboration with F. Eisenhauer and S. Gillessen, with the following sentence: \emph{‘‘[In the next decade] the Galactic Center may then become a test bed for probing general relativity in the strong-field limit''}. As we have illustrated thoroughly through the sections of this review article, twelve years later (up to the date of this publication) this prediction has found confirmation. Thanks to the technological advancement in the near-infrared astronomy of the GC \cite{Eisenhauer2003, Wizinowich2006, Larkin2006, GravityCollaboration2017} occurred over the last decades, we can now confidently measure the PN effects of general relativistic gravitational redshift \cite{GravityCollaboration2018a, Do2019a} and Schwarzschild precession \cite{GravityCollaboration2020a} on the orbits of the S-stars. These observations allowed to check the validity of predictions from GR in a totally different regime than previous tests (Figure \ref{fig:eht_gravitational_probes}), representing, in fact, a redefinition of the classical tests of GR; at the same time, the fact itself of being able to test gravity at the GC, opens a new way of testing models that go \emph{beyond} the standard general relativistic BH paradigm, either by a modification of the underlying theory of gravity \cite{Hees2017, DeMartino2021, DellaMonica2022a, Capozziello2014}, by the presence of a BH mimicker \cite{Grould2017a, DellaMonica2022b, Munyaneza2002} or due to a concentration of DM around the compact object \cite{BecerraVergara2020, DellaMonica2022c, Lacroix2018}. Similarly, the breakthrough event-horizon scale observation of the strongly-lensed emission features in Sgr A* by the EHT \cite{EventHorizonTelescopeCollaboration2022a}, that was only possible due to a leap forward in both theoretical, computational and technological fields, allowed to validate the predictions of a BH shadow on one hand, and to provide with yet another tool to test gravity in the strong field regime.

In spite of the breakthrough discoveries done so far at the GC, these do not represent the endpoint of this line of investigation. Rather, they provide source of motivation to invest more in theoretical and observational efforts to reach the next frontiers of our understanding of this peculiar environment. Over the next decade, great new observational facilities will allow us to peer into the GC with new eyes. Future upgrades of the VLT/GRAVITY instrument (the GRAVITY+ project \cite{GravityCollaboration2022b}), will result in deeper observations of the GC that will potentially lead to the identification of a new population of closer, fainter stars \cite{GravityCollaboration2021a}, that are expected to yield more accurate measurements of the SMBH properties like mass, spin and quadrupole moment \cite{Alexander2005}. Further progress is expected to occur when greater observing facilities like the TMT on Hawaii \cite{Weinberg2004} or the 39-meter ELT at ESO \cite{Davies2021} will become online. Additionally, radio-pulsars on relativistic orbits could be detected in a few years with the next generation millimeter observatories (the LMT \cite{OrtizLeon2016}, the phased-array configuration of the Northern Extended Millimeter Array, NOEMA \cite{Neri2022}, and of ALMA \cite{Matthews2018}) and especially with the radio Square Kilometer Array (SKA) \cite{Eatough2015}. Such objects are often regarded as an \emph{holy grail} for astrophysics \cite{Faucher2011} and, thanks to the intrinsic regularity of their pulses \cite{Cordes1997}, they are expected to provide unparalleled tests of gravity theories and BH physics, potentially enabling the most stringent tests of gravity to date \cite{Wex1999, Kramer2004, Pfahl2004, Liu2012, Torne2021}. As for the event-horizon scale observation of Sgr A*, the proposed Next Generation EHT \cite{Doeleman2019, Raymond2021} would bring greater sensitivity, thanks to larger array of stations and grater bandwidth. Nevertheless, a substantial improvement to the angular resolution, will only be achieved by expanding the telescope array into space, sending one (or more) radio dishes into large-radius orbits around Earth the \cite{Zakharov2005, Palumbo2019, Fish2020, Roelofs2020a, Fromm2021, Gurvits2021, Kudriashov2021}. Such an expanded array is expected to provide images with sufficient angular resolution to distinguish the details of the accretion flow and to directly probe the presence of a surface in Sgr A*.

Looking in retrospective to the astounding progresses made in the last decades in the astronomical facilities, in the quality of the resulting observations and in the theoretical modeling of the GC, we can anticipate even more exciting advancement in the years to come, both from new object that will be found orbiting Sgr A* and from the direct observation of Sgr A* itself.

\section*{Acknowledgments}
RDM acknowledges support from Consejeria de Educación de la Junta de Castilla y León, from the Fondo Social Europeo and from grant PID2021-122938NB-I00 funded by MCIN/AEI/10.13039/501100011033 and by “ERDF A way of making Europe”. IDM acknowledges support from Grant IJCI2018-036198-I  funded by MCIN/AEI/ 10.13039/501100011033 and, as appropriate, by “ESF Investing in your future” or by “European Union NextGenerationEU/PRTR”; and from the  grant PID2021-122938NB-I00  funded by MCIN/AEI/ 10.13039/501100011033 and, as appropriate, by “ERDF A way of making Europe”, by the “European Union” or by the “European Union NextGenerationEU/PRTR”. Finally, IDM acknowledges support from grant SA096P20 funded by Junta de Castilla y León.

\section*{Abbreviations}

The following abbreviations are used in this manuscript:\\
\noindent 
\footnotesize
\begin{tabular}{@{}ll}
AGB & Asymptotic Giant Branch\\
ALMA & Atacama Large Millimeter/submillimeter Array\\
APEX & Atacama Pathfinder Experiment telescope \\
AO & Adaptive Optics\\
ARO/SMT & Radio Observatory 10-m Sub-millimetre Telescope\\
BH & Black hole \\
CARMA & Combined Array for Research in Millimeter-wave Astronomy\\
EHT & Event Horizon Telescope\\
ELT & Extremely Large Telescope\\
ETG & Extended Theories of Gravity\\
ESO & European Southern Observatory \\
GC & Galactic Center \\
GL & Gravitational Lensing\\
GPS & Global Positioning System\\
GRMHD & General Reltivistic Magnetohydrodynamics\\
JCMT & 15-m James Clerk Maxwell Telescope\\
IMBH & Intermediate Mass Black Hole\\
ISCO & Innermost Stable Circular Orbit\\ 
LLAGNs & Low-Luminosity Active Galactic Nuclei\\
LMT & Large Millimeter Telescope Alfonso Serrano\\
MPE & Max Planck Institute for Extraterrestrial Physics\\ 
MW & Milky Way\\
NOEMA & Northern Extended Millimeter Array \\
NSC & Nuclear Star Cluster\\
NTT & New Technology Telescope \\
PN & Post-Newtonian\\
PPN & Parametrized Post-Newtonian\\
PSF & Point-spred-function\\
Sgr A* & Sagittarius A*\\
SMBH & Super Massive Black Hole\\
SKA & Square Kilometer Array \\
SMA & Submillimeter Array \\
SMT & Submillimeter Telescope Observatory\\
SPT & South Pole Telescope \\
TMT & Thirty Meter Telescope\\
UCLA &  University of California, Los Angeles\\
VLA & Very Large Array\\
VLBA & Very Long Baseline Array\\
VLBI & Very-Long Baseline Interferometry\\ 
VLT & Very Large Telescope\\
WR & Wolf-Rayet\\
\end{tabular}

\clearpage

\clearpage

\bibliographystyle{apsrev4-2}
\bibliography{bibliography}

\end{document}